\documentstyle[epsfig,12pt]{article}
\newtheorem{theorem}{Theorem}
\begin{document}
\hyphenation{ana-lyz-ed any-more ap-pro-xi-ma-tion can-not 
   ei-gen-func-tion ei-gen-val-ue Ha-mil-to-ni-an 
   iso-tro-pi-cal-ly
   re-cog-ni-ze re-la-ti-vis-tic re-nor-ma-li-za-ti-on
   }
\title{Discretized light-cone quantization and\\
the effective interaction in hadrons}
\author{Hans-Christian Pauli \\ 
Max-Planck-Institut f\"ur Kernphysik \\ 
D-69029 Heidelberg}
\date{30 August 1999}
\maketitle
\begin{abstract}
Light-cone quantization of gauge theories is discussed 
from two perspectives: as a calculational tool for 
representing hadrons as QCD bound-states of 
relativistic quarks and gluons, and as a
novel method for simulating  quantum field theory 
on a computer. 
A general non-perturbative method for numerically 
solving quantum field theories,
`discretized light-cone quantization',
is outlined. 
Both the bound-state spectrum and the corresponding
relativistic wavefunctions can be obtained 
by matrix diagonalization and related techniques. 
Emphasis is put on the construction of the light-cone Fock 
basis and on how to reduce the many-body problem to an 
effective Hamiltonian. 
The usual divergences are avoided by cut-offs 
and subsequently removed by the renormalization group.
For the first time, this programme is carried out 
within a Hamiltonian approach, 
from the beginning to the end. 
Starting with the QCD-Lagrangian, a regularized
effective interaction is derived and renormalized,
ending up with an almost solvable integral equation.
Its eigenvalues yield the mass spectrum of physical mesons,
its eigenfunctions yield their wavefunctions including
the higher Fock-space components.
An approximate but analytic mass formula is derived for 
all physical mesons.
\end{abstract}
\vfill 
\noindent  
{Preprint MPIH-V19-1999}\\ 
60 pages, 13 figures, 11 tables, 41 references\hfill \\
PACS-Index: 11.10Ef, 11.15Tk, 12.38Lg, 12.40Yx\hfill \\
Better figure quality is obtained from .uu and .ps files from \\
ftp anonymous@franny.mpi-hd.mpg.de/pub/pauli/99korea/\\
\hrule\vskip 1ex
\noindent  
\textit {To be published in:}\\
\textit {New Directions in Quantum Chromodynamics},
C.R. Ji, \textit {Ed.,} \\
American Institute of Physics, New York, 1999.\\
(Proceedings of a Summer School in Seoul, Korea, 26 May - 18 June 1999.)\\
%
%
\noindent \hrule
\newpage \tableofcontents \newpage
\section{Introduction}
One of the outstanding problems in particle physics is the 
determination of 
the structure of hadrons such as the proton and neutron in 
terms of their fundamental quark and gluon degrees of freedom.  
Over the past twenty years two fundamentally different pictures
have developed. 
One, the constituent quark model 
is closely related to experimental observation. The
other,  quantum chromodynamics is based on a
covariant non-abelian quantum field theory.
The front form (also known as light-cone quantization)
appears to be the only hope of reconciling these two. 
This elegant approach to quantum field theory is a Hamiltonian 
gauge-fixed formulation that avoids many of the most difficult
problems in the conventional equal-time formulation of the theory.  

The natural gauge for light-cone Hamiltonian 
theories is the light-cone gauge $A^+=0$.  In this physical
gauge the gluons have only the two physical transverse degrees 
of freedom.  One imagines that there is an expansion in 
multi-particle occupation number Fock states. 
But even in the case of  the simpler abelian quantum theory
of electrodynamics very little is known about the nature of the 
bound state solutions in the strong-coupling domain. 
In the non-abelian quantum theory of chromodynamics 
a calculation of bound-state structure has to deal with many 
difficult aspects simultaneously.  
Confinement, vacuum structure and chiral symmetry 
inter-twine with the difficulties of describing a
(relativistic) many-body system 
and the non-perturbative renormalization of a Hamiltonian.
 
In the conventional approach based on equal-time 
quantization the Fock state expansion
becomes quickly intractable because of the 
complexity of the vacuum. 
Furthermore, boosting such a wavefunction from the hadron's 
rest frame to a moving frame is as complex a problem as 
solving the bound state problem itself. 
The presence of the square root operator in the equal-time 
Hamiltonian approach presents severe mathematical difficulties. 

Fortunately `light-cone quantization'
offers an elegant avenue of escape. 
It can be formulated independent of the Lorentz frame. 
The square root operator does not appear, 
and the vacuum structure is relatively simple. 
There is no spontaneous  creation of massive 
fermions in the light-cone quantized vacuum. 

In fact, there are many reasons to quantize relativistic field
theories at fixed light-cone time. 
Dirac \cite{dir49} showed, in 1949, that in this so called 
`front form'  of Hamiltonian dynamics a maximum number 
of Poincar\'e generators become independent of the 
interaction, including certain Lorentz boosts. 
In fact, unlike the traditional equal-time Hamiltonian 
formalism, quantization on a plane tangential to the 
light-cone, on the `null plane', can be formulated without 
reference to a specific frame. 
One can construct an operator whose eigenvalues are the invariant 
masses of the composite physical particles.
The eigenvectors describe bound states of arbitrary 
four-momentum and invariant mass, and allow  the
computation of scattering amplitudes and other dynamical 
quantities. 
In many field theories the vacuum state of the free Hamiltonian 
is also an eigenstate of the light-cone Hamiltonian. 
The Fock expansion built on this vacuum state  
provides a complete relativistic many-particle basis for 
diagonalizing the full theory. 

The main thrust of these lectures will be to discuss the
complexities that are unique to this formulation of 
QCD, in varying degrees of detail. 
The goal is to present a self-consistent framework
rather than trying to cover the subject exhaustively. 
A review all of the successes or applications will, however, 
not be undertaken. 

One of the reasons is, that the subject was reviewed recently
\cite{bpp97}.
Another is that other lecturers in this school
emphasize complementary aspects. 
Stan Brodsky shows how the knowledge of the the light-cone
wavefunction has impact on hadronic physics and 
exclusive processes. 
Steve Pinsky demonstrates how the method of 
discretized light-cone quantization is constructive 
for analyzing super-symmetric string and M(atrix) theories.
Last not least, Simon Dalley expands on the 
transverse-lattice calculations 
within the light-cone approach.

Comparatively little space will be devoted to 
canonical field theory, just so much as to plausibilize
that a light-cone Hamiltonian exists. 
This so called `naive Hamiltonian' will be derived 
and written down explicitly as a Fock-space operator 
in the light-cone gauge and disregarding zero modes.
For historical and paedagogical reasons 
these notes will be rather outspoken 
for one space and one time dimension,
mostly to show that periodic boundary conditions 
(discretized light-cone quantization)
are helpful, indeed, for solving the bound-state
problem in a relativistic theory.
The attempt is made to be complementary to the review \cite{bpp97}
and some new material on the Schwinger model is included.

The bulk of these notes deals with the many-body aspects 
of a gauge field theory in the real world of 3+1 dimensions.
A hadron not only contains the valence quarks
but also an infinite amount of gluons and sea-quarks.
Progress often comes with new technologies.
The presentation of the method of iterated resolvents
therefore takes broad room. 
It allows to derive a well-defined effective interaction.
Some thus far unpublished work is included,
in particular more instructive examples.
The presentation is separated into two parts.
In the first considerations are essentially exact.
In the second some approximations and simplifications are
admitted and well marked in the text.
One arrives at the effective interaction 
in the form of a tractable and solvable integral equation.
Its solutions allow to construct explicitly
the many-body amplitudes corresponding to see-quarks and gluons
by comparatively simple quadratures, 
without the need of solving another bound-state problem.
Explicit equations for that are given.
Moreover, some new  research work will be included 
in the section on renormalization.

What is not included in these notes, however, is a complete survey
of the literature. Mostly for the reasons of space,
it is refered to the some 469 items of \cite{bpp97}.
I apologize with my colleagues whose work is not mentioned. 
But I will be careful to cite the work
which I need for the present discussion and presentation.  
Some selected monographies which I found useful to consult are
\cite{bas91,bjd65,mof50,mut87,wei95},
some selected review articles or conference proceedings
might be \cite{brp91,coe92,gla95,gra97}.

\section{Forms of Hamiltonian Dynamics}

Dirac defines the Hamiltonian $ H $ of a closed system as that
operator whose action on the state vector  
$\vert\ t \ \rangle $ has the same 
effect as taking the partial derivative with respect to time 
$ t $, {\it i.e.} 
\begin{equation}
    H \ \vert\ t \ \rangle = i {\partial \over \partial t} 
      \ \vert\ t \ \rangle 
.\label{eq:1}\end{equation}
The concept of an Hamiltonian is applicable irrespective of 
whether one deals 
with the motion of a non-relativistic particle in classical
mechanics or  with a non-relativistic wave function in the
Schr\"odinger equation,  and it  generalizes almost 
unchanged to a relativistic and covariant field theory. 

In a covariant theory, the very notion of `time' becomes, however,  
questionable since the time is only one component
of four-dimensional space-time.
But the concept of usual space and of usual time can be generalized.
One can define `space' as that hypersphere in four-\-space 
on which one chooses the initial conditions.
The remaining fourth coordinate can be understood as `time'.

\begin{figure}
 \epsfxsize=148mm\epsfbox{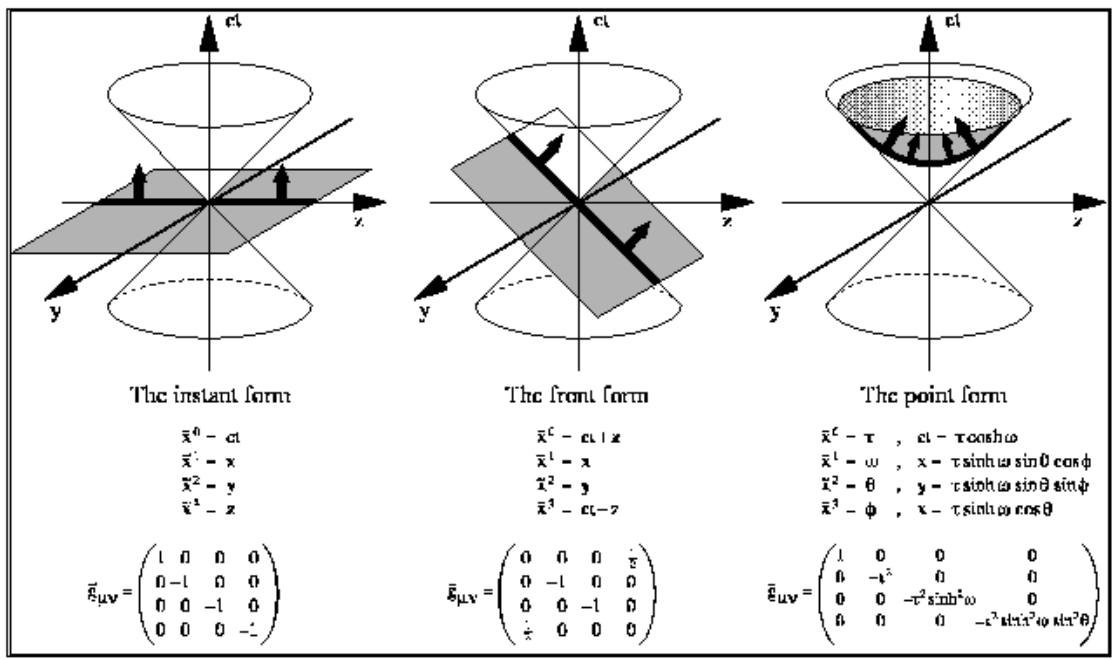}
\caption{\label{fig:bir2}
    Dirac's three forms of Hamiltonian dynamics.
}  \end{figure}

More formally, one conveniently 
introduces generalized coordinates $ \widetilde x  ^\nu $. 
Starting from a baseline parametrization of space-time 
like the instant form in Figure~\ref{fig:bir2},
one parametrizes space-time by a coordinate transformation 
$\widetilde x  ^\nu = \widetilde x  ^\nu (x ^\mu)$.
The metric tensors for the two parametrizations are 
then related by 
\[ 
 \widetilde g _{\kappa \lambda} 
  = \left( {\partial x ^\mu \over
 \partial \widetilde x ^\kappa} \right) 
 g _{\mu \nu} \left( {\partial x ^\nu \over
 \partial \widetilde x ^\lambda} \right)  
.\nonumber 
\] 
The physical content of the theory can not depend on such a 
re-parametrization. 

But the raising and the lowering of Lorentz indices are 
then non-trivial operations. 
As an example consider the front form parametrization 
in Figure~\ref{fig:bir2}.
Interpret $\widetilde x ^3= x ^0 - x ^3$ as the new 
space coordinate and denote it by $x ^-= ct - z$.
Then $\widetilde x ^0= x ^0 + x ^3$ must be interpreted
as the new time coordinate denoted by $x ^+= ct + z$,
or by the `light-cone time' $\tau= t + z/c$.
Of course, one also could exchange the two.
Since the lowering operation is $x_\mu=g_{\mu\nu} x^\nu$,
both $x _+=\frac{1}{2}x ^-$ are space, 
and  $x _-=\frac{1}{2}x ^+$ are  time coordinates.  
The new space derivative is therefore 
$ \partial _- = {1\over2}\partial ^+ $,
while   
$ \partial _+ = {1\over2}\partial ^- $ 
is a time-derivative.
The Lorentz indices `+' and `-' 
have a different physical meaning, 
depending on whether they occur up- or down-stairs.
Co-variant and contra-variant vectors are different objects.
The Hamiltonian is only one component of a four-vector 
$P^\mu$, particularly its time-like component.
Taking the partial time derivative of the state vector
\[ 
   P _+ \ \vert\ x ^+ \ \rangle = i {\partial \over \partial x ^+} 
      \ \vert\ x ^+ \ \rangle 
,\nonumber\] 
defines then $ P _+ = {1\over2} P ^- $ as the Hamiltonian 
in the transformed coordinates,
in line with Eq.(\ref{eq:1}),
and $ P _- = {1\over2}P ^+ $ as the 
longitudinal momentum. 

Following Dirac \cite{dir49} there are no more than
three different parametrisations of space-time. They are 
illustrated in Figure~\ref{fig:bir2}, and 
cannot be mapped onto each other by a Lorentz transform. 
They differ by the hypersphere on which the fields are 
initialized. They have thus different `times' 
and different `Hamiltonians'.
Dirac \cite{dir49}
speaks of  the three {\em forms of Hamiltonian dynamics}:
The {\em instant form} is the familiar one, 
with its hypersphere given by $t=0$. 
In the {\em front form} the hypersphere is a tangent plane 
to the light cone.
In the {\em point form} the time-like coordinate is 
identified with the eigentime of a physical system
and the hypersphere has a shape of a hyperboloid.

Which of the three forms should be prefered, is an
ill-posed question, since all three forms must 
yield the same physical results.
Comparatively little work has been done in the point form. 
The bulk of research on field theory implicitly uses the 
instant form. 
Although it is the  conventional choice for quantizing field 
theory, it has many practical disadvantages. 
For example, given the wavefunctions of an $n$-electron 
atom at an initial time $t=0$,  $\psi_n(\vec x_i,t=0)$, 
one can use the Hamiltonian $H$ to 
evolve $\psi_n(\vec x_i,t)$ to later times $t$.
However, an experiment which specifies the initial 
wave function would require the simultaneous measurement 
of the positions of all of the bounded electrons.
In contrast, determining the initial wave function at fixed 
light-cone time $\tau=0$ only requires an experiment which 
scatters one plane-wave laser beam, since the signal reaching 
each of the $n$ electrons, along the light front, at the 
same light-cone time $\tau = t_i+z_i/c$.

Dirac's legacy had been forgotten and was re-invented  several times. 
The front form approach carries therefore
names as different as 
{\em Infinite-Momentum Frame}, 
{\em Null-Plane Quantization}, 
{\em Light-Cone Quantization}, or
most recently 
{\em Light-Front Quantization}.
In the essence they are all the same. 
The infinite-momentum frame is a misnomer
since  the total momentum is finite and since the front form 
is {\em frame-independent} and covariant.
Light-cone quantization is also unfortunate, 
since the inital data are set one a plane tangential to 
but not on the light cone, and since both equal-time or
equal light-cone-time quantization stand both for 
the same quantization, 
for a quantum as opposed to a classical theory.
We propose to stay with Dirac's different `forms of Hamiltonians'.

\subsection{The canonical Hamiltonian for gauge theory}

The prototype of a field theory is Faraday's and Maxwell's 
electrodynamics. 
Every field theory has its own canonical Hamiltonian
and is governed by the action density.
The Lagrangian is the subject of
the canonical calculus of variation, 
given in many text books \cite{bjd65}. 
Its essentials shall be recalled briefly.

The Lagrangian, in general, is a function of a finite number
of fields $ \phi _r (x)$ and their first space-time 
derivatives $ \partial _\mu \phi _r (x)$, thus  
${\cal L} = {\cal L} \left[\phi _r, \partial _\mu \phi _r \right]$.
Independent variation of ${\cal L}$ with respect to 
$ \phi _r$ and $\partial _\mu \phi _r $  
results in the equations of motion,
\[ 
   \partial _\kappa \pi _r ^\kappa
  - \delta {\cal L}  / \delta \phi _r = 0
   \ , \quad {\rm with } \quad
   \pi _r ^\kappa [\phi ] \equiv {\delta  {\cal L} \over \delta 
   \left(\partial _\kappa \phi _r \right)} 
,\nonumber 
\]
canonically refered to as the Euler equations. 
The canonical formalism is particularly suited for discussing
the symmetries of a field theory. 
Every continuous symmetry of the Lagrangian is associated 
with a vanishing four-divergence of a current and a conserved charge.
Since ${\cal L}$ does not explicitely depend on the coordinates,
every field theory in 3+1 dimensions has ten 
conserved four-currents.
The four-divergences of the
energy-\-momentum tensor $T ^{\lambda \nu}$ and of 
the boost\--angular-\-momentum stress tensor 
$J ^{\lambda , \mu \nu}$ vanish,
\[ 
    \partial _\lambda   T ^{\lambda \nu} = 0
,\quad 
    \partial _\lambda   J ^{\lambda , \mu \nu} = 0 
.\nonumber \] 
As a consequence the Lorentz group has ten `conserved 
charges', the 4 components total momentum $P^\nu$ and the 
6 components of boost-angular momentum $M^{\mu \nu}$,
\begin{eqnarray}
    P  ^\nu = \int _\Omega d \omega_0 T ^{0\nu} 
    \,,\quad  {\rm and} \qquad 
    M  ^{\mu \nu} = \int _\Omega d \omega_0 J ^{0, \mu \nu}
.\label{eq:2.17}\end{eqnarray}
With the totally antisymmetric tensor in four dimensions
$ \epsilon _{\lambda \mu \nu \rho}$
($ \epsilon _{0123} = 1$),
the three-dimensional surface elements of a hypersphere are
$ d \omega _\lambda = 
  \epsilon _{\lambda \mu \nu \rho} d x ^\mu d x ^\nu d x ^\rho / 3 !$.
A finite volume is thus
$ \Omega = \int d \omega_0 = \int d x ^1 d x ^2 d x ^3 $,
and correspondingly in the front form 
$d\omega_+=\int dx_+d^2x_{\!\perp}$.
The time-like components of $P_\mu$ is the Hamiltonian,
{\it i.e.} $P_0$ for the instant form and $P_+$ for the front form.
The transition from instant to front form 
is thus simple: substitute `0' by `+'.

Working out the canonical procedure for QED with its Lagrangian, 
\begin{equation}
     {\cal L}  = 
   - {1 \over 4}  F  ^{\mu \nu}  F _{\mu \nu} + {1 \over2} \left[ 
      \overline \Psi \left( i \gamma ^\mu D _\mu  -  m\right) \Psi  
   + \ {\rm h.c.} \ \right]
,\label{eq:2.7}\end{equation}
where $F ^{\mu \nu}$ is the electro-magnetic field tensor
and $D _\mu$ the covariant derivative,
\begin {eqnarray} 
   F ^{\mu \nu} = \partial ^\mu A ^\nu - \partial ^\nu A ^\mu
,\qquad\quad
   D _\mu = \partial _\mu - ig  A _\mu 
,\nonumber 
\end {eqnarray}
one ends up with a manifestly gauge-invariant total momentum,
\begin {eqnarray} 
    P_\nu & = & \int_\Omega \! d\omega _0 
    \ \biggl( F^{0\kappa}      F _{\kappa\nu} 
   +{1\over4} g^0_\nu F ^{\kappa\lambda} F _{\kappa\lambda} 
   +{1\over2} \Bigl[i\overline\Psi\gamma^0 D_\nu\Psi
   +\ {\rm h.c.} \Bigr] \biggr) 
\ ,\nonumber \\
   P_\nu & = & \int_\Omega \! d\omega _+ 
   \ \biggl( F^ {+\kappa} F _{\kappa\nu}
   +{1\over4}g^+_\nu F^ {\kappa\lambda} F_{\kappa\lambda}
   +{1\over2}\Bigl[i\overline\Psi \gamma^+ D _\nu \Psi 
   +\ {\rm h.c.} \Bigr] \biggr) 
,\nonumber 
\end {eqnarray}
in both the instant and the front form, respectively.
The boost angular momenta will not be used explicitly.

The gauge invariant Lagrangian density for QCD is 
\begin {eqnarray}
    {\cal L} & = & - {1\over2} 
    {\rm Tr} \bigl({\bf F}^{\mu\nu} {\bf F}_{\mu \nu} \bigr)
 + {1\over2} \bigl[ \overline \Psi \bigl(i\gamma^\mu {\bf D} _\mu 
   - {\bf m} \bigr)  \Psi   + \ {\rm h.c.} \bigr] 
,\nonumber \\
             & = & - {1\over4}  F^{\mu\nu}_a  F_{\mu \nu}^a
  + {1\over2} \bigl[ \overline \Psi \bigl(i\gamma^\mu  
   {\bf D} _\mu -  {\bf m} \bigr) \Psi   + \ {\rm h.c.} \bigr] 
,\nonumber %
\end {eqnarray}
The  color-electro-magnetic fields and the covariant derivative 
are now
\begin {eqnarray}
   {\bf F}^{\mu\nu} \equiv \partial^\mu  {\bf A}^\nu  
   - \partial^\nu  {\bf A}^\mu 
   + i g \bigl[ {\bf A}^\mu, {\bf A}^\nu\bigr] 
   &=& {\bf T}^a\left(
   \partial^\mu  A^\nu_a
   - \partial^\nu  A^\mu_a  
   - g  f^{ars}  A^\mu_r  A^\nu_s \right) 
,\nonumber\\
   {\bf D} ^\mu _{cc^\prime} = 
  \delta_{cc^\prime} \partial^\mu 
  + i g {\bf A}^\mu _{cc^\prime} &=& 
  \delta_{cc^\prime} \partial^\mu 
  + i g T^a_{cc^\prime} A^\mu _{a}
.\nonumber 
\end {eqnarray}
As compared to QED, each local gauge field $ A^\mu (x)$ is 
replaced by the $3\times 3$ matrix ${\bf A} ^\mu (x) $.
All such matrices can be parametrized  
$ {\bf A} ^\mu \equiv T ^a _{cc ^\prime} A ^\mu _a$.
More generally for SU(N), the vector potentials $ {\bf A}^\mu $
are hermitian and traceless $N \times N$ matrices.
The color index $c$ (or $c^\prime$) runs now from 1 to $n_c$, 
and correspondingly the gluon index $a$ (or $r,s,t$) 
from 1 to $n_c^2-1$. 
No distinction will be made between raising or lowering them.
In order to make sense of expressions like 
$ \overline \Psi {\bf A} ^\mu \Psi $ the quark fields 
$\Psi _{c, \alpha } (x)$ must carry a color index $c$. 
They, as well as the Dirac indices, are usually suppressed.
The color matrices $ T ^a_{cc ^\prime}$ obey 
\begin{equation}
  \Bigl[ T ^r, T ^s \Bigr] _{cc ^\prime} 
  = i f ^{rsa}  T _{cc ^\prime} ^a 
  \qquad\quad {\rm and } \qquad
  {\rm Tr}\ \bigl(T ^r  T ^s \bigr) = {1\over2}\delta _r ^s 
.\label{eq:2.26}\end{equation}
For SU(2) the color matrices are $T ^a = {1\over 2} \sigma ^a $,  
with $\sigma ^a$ being the Pauli matrices.
The structure constants $\displaystyle f ^{rst}$ 
are therefore the totally antisymmetric tensor
$\displaystyle \epsilon _{rst}$.
For SU(3), $ T ^a = {1\over 2} \lambda ^a$
with the Gell-Mann matrices $\lambda ^a$, with
the corresponding structure constants tabulated 
in the literature \cite{mut87}.
 Everything proceeds in analogy with QED. 
The energy-momentum vector, 
\begin {eqnarray} 
    P_\nu & = & \int_\Omega \! d\omega _0 
    \ \biggl(F^{0\kappa}_a F ^a_{\kappa\nu} +
    {1\over4} g^0_\nu
    F^{\kappa\lambda}_a F_ {\kappa\lambda}^a +
    {1\over2}\Bigl[i\overline\Psi\gamma^0 
    T^a D^a_\nu\Psi  
  + \ {\rm h.c.} \Bigr] \biggr) 
, \nonumber \\
    P_\nu & = & \int_\Omega \! d\omega _+ 
    \ \biggl(F^{+\kappa}_a F ^a_{\kappa\nu} 
    + {1\over4} g^+_\nu
    F^{\kappa\lambda}_a F_ {\kappa\lambda}^a
    + {1\over2}\Bigl[i\overline\Psi\gamma^+
    T^a D^a_\nu\Psi  
  + \ {\rm h.c.} \Bigr] \biggr) 
,\label {eq:2.60} 
\end {eqnarray}
is manifestly gauge-invariant both in the instant and the front form.

\subsection{The Poincar\'e symmetries in the front form}

The ten constants of motion $ P ^\mu $ and $ M ^{\mu \nu} $ 
are  observables with real eigenvalues.
It is advantageous to construct representations 
in which the constants of motion are diagonal. 
But one cannot diagonalize all ten constants of motion 
simultaneously because they do not commute. 
The algebra of the four-energy-momentum $ P^\mu= p ^\mu $ 
and four-angu\-lar-momen\-tum  
$ M^{\mu \nu} = x ^\mu p^\nu - x ^\nu p ^\mu $ 
for free particles 
with the basic commutator 
$ {1\over i\hbar}[x ^\mu, p _\nu ] = \delta ^\mu _\nu $ is
 \begin {eqnarray}
     {1 \over i \hbar}
     \left[ P ^\rho, M^{\mu \nu} \right]  
     &=& 
     g ^{\rho \mu}  P  ^\nu - g ^{\rho \nu}  P  ^\mu \ , \quad
             {1 \over i \hbar}
     \left[ P ^\rho,  P ^\mu \right]  = 0 \  ,  
\nonumber \\  
     \quad {\rm and} \ \quad {1 \over i \hbar}
     \left[ M ^{\rho \sigma} ,  M ^{\mu \nu} \right]  
     &=& 
     g  ^{\rho \nu} M ^{\sigma \mu}
     + g  ^{\sigma \mu} M ^{\rho \nu} 
     - g  ^{\rho \mu}  M ^{\sigma \nu}
     - g  ^{\sigma \nu}  M ^{\rho \mu}   
.\nonumber 
\end{eqnarray}
It is postulated that the generalized momentum 
operators satisfy the same commutator relations
as a single particle.
They form thus a group, the Poincar\'e group. 

It is convenient to discuss the structure of the Poincar\'e 
group in terms of the 
Pauli-Lubansky vector
$  V ^\kappa \equiv \epsilon  ^{\kappa \lambda \mu \nu} 
                 P _\lambda  M _{\mu \nu} $.
$ V  $ is orthogonal to the generalized momenta, 
$  P _\mu  V ^\mu = 0$.
The two group invariants are 
the operator for the invariant mass-squared 
$  M^2  =  P ^\mu  P _\mu $
and the operator for intrinsic spin-squared
$  V^2  =  V ^\mu  V _\mu $.
They are Lorentz scalars and commute with all 
generators $  P ^\mu $ and $  M ^{\mu \nu} $, 
as well as with all $ V ^\mu$.
A convenient choice of six mutually commuting 
operators is:
The invariant mass squared 
$M^2 =  P ^\mu  P _\mu $,  
the three space-like momenta $P ^+$ and $\vec P _{\!\bot}$, 
the total spin squared $ S^2 = V ^\mu   V _\mu $,  
and one component of $ V $, say $V ^+\equiv S_z$. 

Inspecting the definition of  boost-angular momentum 
$M_{\mu\nu}$ in Eq.(\ref{eq:2.17}) one identifies which 
components are dependent on the interaction and which are not.
Dirac \cite{dir49} calls them complicated and simple, or 
dynamic and kinematic, or  Hamiltonians and Momenta, respectively. 
In the instant form, the three components of the 
boost vector $K_i = M_{i0}$ are dynamic, and
the three components of angular momentum
$J_i=\epsilon_{ijk} M_{jk}$ are kinematic. 
As noted already by Dirac, the front form is special 
in having four kinematic components of $M _{\mu\nu}$
($ M _{+-}, M _{12}, M _{1-}, M _{2-}$)  and only
two dynamic ones ($ M _{+1}$ and $ M_{+2}$). 
In the front form one deals thus with seven mutually 
commuting operators 
\[
     (M _{+-}, M _{12}, M _{1-}, M _{2-}), \quad{\rm and\ all\ } P^\mu
,\nonumber %
\]
instead of the six in the instant form. 
These symmetries imply the very important aspect
of the front form that both the Hamiltonian 
and all amplitudes obtained in light-cone perturbation 
theory are manifestly invariant under 
a large class of Lorentz transformations:
\begin{eqnarray}
\begin{array}{llll}
   & p^+ \rightarrow C_{\parallel}\ p^+          \,,
   & \vec p_{\!\perp}\rightarrow \vec p_{\!\perp}\,,
   & p^- \rightarrow C_{\parallel}^{-1} p^-      \,,
\\
   & p^+ \rightarrow p^+                           \,,
   & \vec p_{\!\perp} \rightarrow \vec p_{\!\perp}
     + p^+ \vec C_{\!\perp}                        \,,
   & p^- \rightarrow p^- 
     + 2\vec p_{\!\perp}\cdot \vec C_{\!\perp}
     + p^+ \vec C_{\!\perp}^2                      \,,
\nonumber 
\\
   & p^+ \rightarrow p^+                 \,,
   & \vec p_{\!\perp}^{\;2} \rightarrow 
         \vec p_{\!\perp}^{\;2}          \,,
\end{array}
\nonumber 
\end{eqnarray}
{\it i.e.} for parallel boosts, transverse boosts,
and rotations, respectively.
All of these hold for every single particle momentum 
$p^\mu$, and for any set of dimensionless c-numbers
$ C_{\parallel}$ and $ \vec C_{\!\perp}$.

\subsection{The light-cone Hamiltonian operator}

The four-vector of energy-momentum for gauge theory  
in Eq.(\ref{eq:2.60}) contains time-derivatives and 
other constraint field components.
They will be eliminated in this section using 
the equations of motion with the 
goal to express $P^\nu$ in terms of the free fields
and to isolate the dependence on the coupling constant.

The color-Maxwell equations are four
equations for determining the four functions $A^\mu_a$. 
One of the equations of motion is identically fullfilled by 
choosing the light-cone gauge $A ^+ _a = 0$.
Two of the equations give expressions the time derivatives of the
two transversal components $\vec A ^a _{\!\perp} $. 
The fourth is the Gauss' law in the front form, 
$ \partial _\mu  F  _a^{\mu +} = g  J   _a^+$, 
or explicitly
\begin{equation} 
   - \partial^+\partial_-  A  _a^-  
   - \partial^+ \partial_i  A  _{\!\perp a}^i   = g J^+_a
.\label{eq:2.56}\end{equation} 
It contains only space-derivatives and is a constrained
equation for the components of $A^\mu_a$.
For the free case ($g=0$), $A^\mu_a$ reduces 
to $\widetilde A_a^\mu $ and therefore to 
\[  
   \widetilde A_a^\mu = 
   \left(\widetilde A^+_a,\vec A_{\!\perp a},\widetilde A^-_a \right) =
   \left(0,\vec A_{\!\perp a},
   -{1\over\partial_-}\left[\partial_i A^i_{\!\perp a}\right]
   \right) 
.\nonumber %
\] 
As a consequence, $\widetilde A ^\mu _a $ is purely transverse.
The formal inversion of Eq.(\ref{eq:2.56}) is therefore
\begin{equation} 
   A^-_a = \widetilde A^-_a + {g\over (i\partial_-)^2}\,J^+_a
,\label{eq:2.57}\end{equation} 
which must be substituted everywhere. 
The inverse space derivatives 
$\left(i\partial^+\right)^{-1}$ and 
$\left(i\partial^+\right)^{-2}$, used here and below, 
are actually Green's functions. 
Since they depend only on $x^-$, they are comparatively simple. 

The color-Dirac equations 
$(i \gamma^\mu{\bf D} _\mu -{\bf m})\Psi = 0$ can be used 
to express the time derivatives $\partial_+ \Psi $ 
as function of the other fields. 
After multiplication with $\gamma^0$ one gets first
\[    \bigl(
       i \gamma^0\gamma^+ T^a D^a _+ + 
       i \gamma^0\gamma^-  T^a D^a _-   +
       i \alpha^i_{\!\perp}   T^a D^a _{\!\perp i} \bigr) \Psi
       = m \beta\Psi
,\nonumber %
\] 
with the usual Dirac matrices 
$\beta=\gamma^0$ and $\alpha^k=\gamma^0\gamma^k$. 
In order to isolate the  time derivative 
one introduces the projectors  $\Lambda _\pm$
and projected spinors $ \Psi_\pm$ by
\[ 
    \Lambda _\pm 
   = {1\over2}(1\pm\alpha^3)     
   \qquad{\rm and}\qquad
   \Psi_\pm = \Lambda_\pm\Psi
\ . \nonumber 
\] 
Multiplying the color-Dirac equation
once with $\Lambda_+$ and once with $\Lambda_-$ gives
\begin{eqnarray} 
   2i\partial _+\Psi _+ &=&  \left( m \beta - 
   i\alpha_{\!\perp}^i T^a D^a _{\!\perp i} \right) \Psi _-  
   +  2g   A  _+^a T^a \Psi _+
,\nonumber \\ {\rm and} \quad 
   2i\partial _- \Psi _-  &=&  \left( m \beta - 
   i\alpha_{\!\perp}^i T^a D^a _{\!\perp i}\right)  \Psi_+
    + 2gA_-^a T^a \Psi _-
,\label{eq:2.61}\end{eqnarray} 
a coupled set of spinor equations. 
The first of them contains a time derivative.
The second contains a space derivative and is a constraint equation.
The component
\begin{equation} 
       \Psi _{-} =  {1\over 2i\partial_-}    \left( m\beta - 
       i\alpha_{\!\perp}^i T^a D^a _{\!\perp i} \right) \Psi_+
\label{PsiMinus}\end{equation} 
must therefore be substituted everywhere.
The time derivative becomes then
\[  
   2i\partial _+ \Psi _+  
   =  2g   A  _+^a T^a \Psi _+ +   \left(m\beta -
   i\alpha_{\!\perp}^j T^a D^a _{\!\perp j}\right) 
   {1\over 2i\partial _-} \left( m\beta -
   i\alpha_{\!\perp}^i T^a D^a _{\!\perp i} \right) \Psi _+
\,.\nonumber 
\]  
In analogy to the color-Maxwell case one defines free spinors by
\[  
   \widetilde\Psi = \Psi _+
   + \left(m\beta - i\alpha_{\!\perp}^i \partial_{\!\perp i}\right) 
   {1\over 2i\partial _-} \Psi _+
\ .\nonumber 
\] 
Contrary to the full spinor,
$\widetilde\Psi $ is independent of the interaction.

Inserting the above expressions into Eq.(\ref{eq:2.60}),
the space-like components of $P^\nu$ become
\begin{equation} 
   P_k = \int\!dx_+d^2x_{\!\perp} \Bigl(
   \overline{\widetilde\Psi}\ \gamma^+ i\partial _k  
   \widetilde \Psi  
   + \widetilde A ^\mu_a \ \partial^+\partial _k  
   \widetilde A_\mu^a \Bigr)  
   \ ,\qquad{\rm for}\ k=1,2,-
.\label{eq:total-momenta}\end{equation} 
Inserting them into $P_+$
gives rather lengthy expressions, 
which are conveniently written as a sum of five terms 
\begin{equation} 
   P_+ = T + V + W_1+W_2+W_3
.\label{eq:2.87}\end{equation} 
Only the first term survives the limit $g\rightarrow 0$, 
and therefore is called the free part of the Hamiltonian, 
or its `kinetic energy' 
\[ 
   T = {1\over2}\int\!dx_+d^2x_{\!\perp} 
   \biggl(\overline{\widetilde\Psi} \gamma^+
   {m^2 +(i\nabla_{\!\!\perp}) ^2 \over i\partial^+}
   \widetilde\Psi   +
   \widetilde A ^\mu_a (i\nabla_{\!\!\perp}) ^2 
   \widetilde A _\mu^a     \biggr)
.\nonumber %
\] 
The vertex interaction 
\begin{equation} 
    V = g  \int\!dx_+d^2x_{\!\perp} 
      \ \widetilde J ^\mu_a \widetilde A _\mu^a        
,\qquad{\rm with}\quad
   \widetilde J ^\nu_a (x) 
   = \overline{\widetilde\Psi}\gamma^\nu T^a \widetilde\Psi  
   + f^{a b c}\partial^\mu \widetilde A^\nu_b \widetilde A_\nu^c
,\label{eq:2.94}\end{equation} 
is linear in the coupling constant and is
the light-cone analogue of the conventional 
$J_\mu A^\mu$-structures in the instant form.
Note that the current $\widetilde J ^\mu_a $
has contributions from both quarks and gluons.
The  four-point gluon interaction 
\[ 
     W_1 = { g  ^2 \over 4} \int\!dx_+d^2x_{\!\perp} 
      \ \widetilde B ^{\mu\nu}_a \widetilde B _{\mu\nu}^a 
,\qquad{\rm with}\quad
   B ^{\mu\nu}_a =f^{a b c}\widetilde A^\mu  _b \widetilde A ^\nu _c
,\nonumber 
\]  
describes the four-point gluon-vertices which is
quadratic in $g$.
The remainders are 
the `instantaneous interactions'. 
They are characterized by the inverse derivatives.
The instantaneous gluon interaction 
arises from the Gauss equation,
\[ 
     W_2 = { g  ^2 \over 2} \int\!dx_+d^2x_{\!\perp} 
      \ \widetilde J ^+_a 
     {1\over \left(i \partial ^+ \right)^2} \widetilde J ^+_a 
,\nonumber 
\] 
and is the light-cone analogue of the Coulomb energy.
The instantaneous fermion interaction 
originates from the light-cone specific decomposition 
of Dirac's equation
\[ 
    W_3 = { g  ^2 \over2} \int\!dx_+d^2x_{\!\perp} 
    \ \overline{\widetilde\Psi}  \gamma ^\mu  T  ^a 
    \widetilde A ^a_\mu \ {\gamma ^+\over i\partial ^+}
    \left( \gamma ^\nu  T  ^b \widetilde A ^b_\nu 
    \widetilde\Psi  \right)
.\nonumber 
\] 
It has no analogue in the instant form.

Most remarkable, however, is that the relativistic
Hamiltonian is  additive 
in the `kinetic' and the `potential' energy, 
very much like a non-relativistic Hamiltonian
$H = T + U$.
In this respect the front form is distinctly different  from  
the conventional instant form. 

\subsection{The free field solutions}  

The free solutions of the Dirac and the Maxwell equations
are in the front form 
\begin{eqnarray} 
   \widetilde\Psi  _{\alpha cf} (x) &=&
   \sum _\lambda\!\int\!\!  
   {dp^+ d^2p_{\!\bot}\over \sqrt{2p^+(2\pi)^3}}
   \left( b (q) u_\alpha (p ,\lambda ) e^{-ipx} +
    d^\dagger (q)v_\alpha(p,\lambda) 
   e^{+ipx}\right)
,\nonumber \\
   \widetilde A _\mu^a (x) &=& 
   \sum _\lambda\!\int\!\!  
   {dp^+ d^2p_{\!\bot}\over \sqrt{2p^+(2\pi)^3}}
   \left( a(q)\epsilon_\mu(p,\lambda) e^{-ipx} +
   a^\dagger(q)\epsilon_\mu^\star(p,\lambda)
   e^{+ipx} \right) 
.\nonumber 
\end{eqnarray} 
The properties of the Dirac spinors $u_\alpha$ and $v_\alpha$, 
and of the polarization vectors $\epsilon_\mu$, 
are given for example in \cite{bpp97}.
The single particle state are specified by 
the quantum numbers 
$q = (p^+,\, p_{\!\perp\,x},\, p_{\!\perp\,y}, \,\lambda, \,c , \,f )$.
Their creation and destruction operators are subject to
the usual relations
\[  
   \left [a(q), a^\dagger(q^\prime)\right]  = 
   \left\{b(q), b^\dagger(q^\prime)\right\} = 
   \left\{d(q),\ d^\dagger(q^\prime)\right\} = 
   \delta (p^+-p^{+\,\prime}) 
   \delta ^{(2)}(\vec p_{\!\bot}-\vec p_{\!\bot} ^{\,\prime}) 
   \delta _\lambda ^{\lambda ^\prime} 
   \delta _c^{c^\prime} 
   \delta _f^{f^\prime} 
,\nonumber %
\] 
which carry the operator structure of the theory.
When inserting the free fields into $P_\mu$
one can integrate out the dependence on $x^\mu$, producing
essentially Dirac delta-functions in the single particle
momenta, which reflect momentum conservation:
\begin{eqnarray} 
   \int\!{dx_+\over2\pi} e^{i x_+\big(\sum_j p_j^+\big)} 
   = \delta\Big( \sum_j p _j^+\Big)
, \qquad 
   \int\!{d^2x_{\!\perp} \over(2\pi)^2}    
   e^{-i\vec x_{\!\perp} \big(\sum_j \vec p_{\!\perp j}\big)}  
   = \delta^{(2)}\Big(  \sum_j\vec p_{\!\perp j}\Big)
.\nonumber %
\end{eqnarray} 
In detail this can be quite laborious, as shown by the example 
with the fermionic contribution to the vertex interaction 
\begin{eqnarray} 
   V _f &=&        
   g  \int\!dx_+d^2x_{\!\perp}\left.
   \ \overline{\widetilde\Psi}(x)  \gamma ^\mu  T  ^a 
   \widetilde\Psi(x)  \widetilde A ^a_\mu (x) 
   \right\vert_{x^+=0}
\nonumber\\
   &=& {g\over\sqrt{(2\pi)^3}}
   \sum _{\lambda_1,\lambda_2,\lambda_3}
   \sum _{c_1,c_2,a_3}
   \int\! {dp^+_1 d^2p_{\!\bot 1}\over \sqrt{2p^+ _1}}
   \int\! {dp^+_2 d^2p_{\!\bot 2}\over \sqrt{2p^+ _2}}
   \int\! {dp^+_3 d^2p_{\!\bot 3}\over \sqrt{2p^+ _3}}
\nonumber\\
   &\times& \int\!{dx_+d^2x_{\!\perp} \over(2\pi)^3} 
   \left[\left( b ^\dagger (q_1) \overline u_\alpha 
   (p_1,\lambda_1) e^{+ip_1x}  +    
   d(q_1)\overline v_\alpha(p_1,\lambda_1)  
   e^{-ip_1x}\right)T^{a_3}_{c_1,c_2} \right.
\nonumber\\
   &\times& \phantom{T^{a_3}_{c_1,c_2}}
   \phantom{d^\dagger q}\gamma ^\mu _{\alpha\beta}
   \left.\left( 
   d^\dagger (q_2)v_\beta(p_2,\lambda_2) 
   e^{+ip_2x} +
   b (q_2) u_\beta (p_2 ,\lambda _2) 
   e^{-ip_2x} \right)\right]
\nonumber\\
   &\times& \phantom{\gamma ^\mu _{\alpha\beta}
  T^{a_3}_{c_1,c_2} d^\dagger q}
   \left.\left( a^\dagger(q_3)\epsilon_\mu^\star
   (p_3,\lambda_3) e^{+ip_3x} +
   a(q_3)\epsilon_\mu(p_3,\lambda_3) 
   e^{-ip_3x} \right) \right.
.\nonumber %
\end{eqnarray} 
Note that the sum of these single particle momenta is
essentially the sum of the particle momenta minus the sum
of the hole momenta. Consequently, if a particular term
has  only creation or only destruction operators as in
\[ 
   b^\dagger (q_1) d^\dagger (q_2) a^\dagger (q_3)
   \ \delta\Big( p _1^+ + p _2^+ + p _3^+\Big) \simeq 0
,\nonumber %
\] 
its contribution vanishes since the light-cone longitudinal 
momenta $p ^+$ are all positive and can not add to zero.
As a consequence,  all
energy diagrams which generate the vacuum fluctuations 
in the usual formulation of quantum field theory are absent 
in the front form.

\begin{table} [t]
\begin{center}
\begin{tabular}{|ll|} 
\hline
\begin{tabular}{l}  
 \epsfysize=32ex\epsfbox{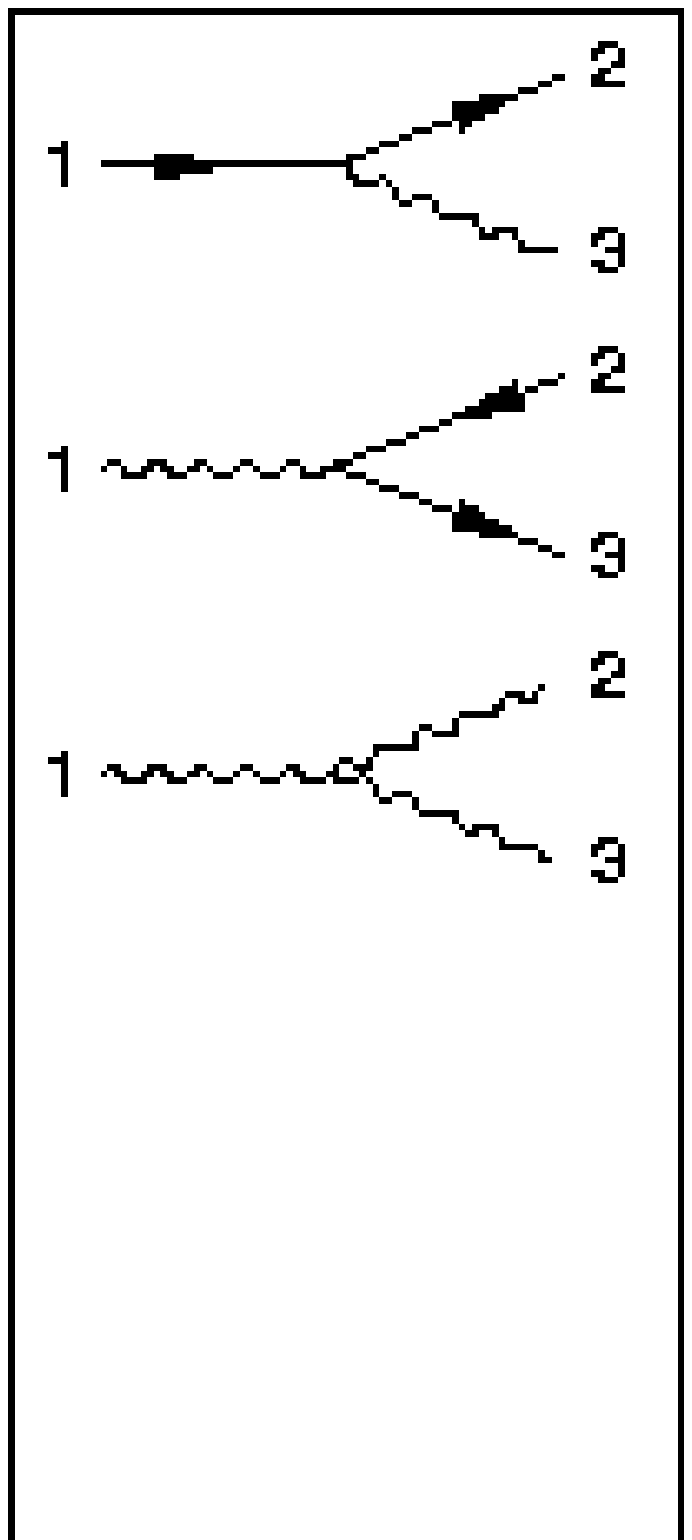}
\\ 
\end{tabular}
&
\begin{tabular}{l}                
  $\displaystyle  V_{1\phantom{,3}} +  
  {\Delta_V\over \sqrt{k^+_1k^+_2 k^+_3}}
  \ (\bar u _1 /\!\!\!\epsilon_3T^{a_3}u_2)$
\\ 
  $\displaystyle V_{3\phantom{,3}} +  
  {\Delta_V \over \sqrt{k^+_1k^+_2 k^+_3}}
  \ (\bar v_2 /\!\!\!\epsilon^\star_1T^{a_1}u_3)$
\\ 
  $\displaystyle V_4 =  
  {iC_{a_2a_3}^{a_1}\,\Delta_V \over \sqrt{k^+_1k^+_2 k^+_3}}
  \ (\epsilon^\star_1k_3)\ (\epsilon_2\epsilon_3) $
\\ 
  $\displaystyle \phantom{V_4} +
  {iC_{a_2a_3}^{a_1}\,\Delta_V \over \sqrt{k^+_1k^+_2 k^+_3}}
  \ (\epsilon_3k_1)\ (\epsilon^\star_1\epsilon_2) $
\\ 
  $\displaystyle \phantom{V_4} +
  {iC_{a_2a_3}^{a_1}\,\Delta_V \over \sqrt{k^+_1k^+_2 k^+_3}}
  \ (\epsilon_3k_2)\ (\epsilon^\star_1\epsilon_2) $
\\ 
\end{tabular}
\\ \hline
\end{tabular}
\caption {\label {tab:verspi} \sl
   The vertex interaction in terms of Dirac spinors.
   The matrix elements $V_{n}$ are displayed on the right, 
   the corresponding (energy) graphs on the left.
   All matrix elements are proportional to 
   $\Delta_V = \widehat g \delta (k^+_1| k^+_2 +_3) 
   \delta ^{(2)} (\vec k _{\!\perp,1} | \vec k _{\!\perp,2} +
   \vec k _{\!\perp,3} ) $, 
   with $\widehat  g = g/\sqrt{2(2\pi)^3} $.
   For the periodic boundary conditions 
   one uses $\widehat  g= g/\sqrt{\Omega} $. 
}\end{center}
\end{table}

\subsection{The Hamiltonian as a Fock-space operator}

The {\em kinetic energy} $T$ becomes a sum of 3   
diagonal operators 
\begin{eqnarray} 
     T &=& 
     \int dk_- d^2\vec k_{\!\perp} \sum_{\lambda,c,f} 
     \Big({m^2 + \vec k_{\!\bot} ^2 \over k_-}\Big)_q
     \left( b_q^\dagger b_q + d_q^\dagger d_q +  a_q^\dagger a_q \right)
\nonumber\\ &\equiv& 
     \sum_{q} \Big({m^2 + \vec k_{\!\bot} ^2 \over k_-}\Big)_q
     \left( b_q^\dagger b_q + d_q^\dagger d_q +  a_q^\dagger a_q \right)
.\nonumber 
\end{eqnarray} 
Here and in the sequel it is convenient to abbreviate
the integration and summation over the single particle
coordinates by the symbol $\sum$ to replace for instance
$b(q)$ with $b_q$.

The {\em vertex interaction} $V$ becomes a sum of  4 operators
\begin{eqnarray} 
   V  &=& V_1 + V_2 + V_3 + V_4   \nonumber\\ &=& 
   \sum_{1,2,3}\left[b^\dagger_1b_2a_3\,V_{1}(1;2,3)+{\rm h.c.}\right]+ 
   \sum_{1,2,3}\left[d^\dagger_1d_2a_3\,V_{2}(1;2,3)+{\rm h.c.}\right]+ 
\nonumber \\ &+& 
   \sum_{1,2,3}\left[a^\dagger_1d_2b_3\,V_{3}(1;2,3)+{\rm h.c.}\right]+ 
   \sum_{1,2,3}\left[a^\dagger_1a_2a_3\,V_{4}(1;2,3)+{\rm h.c.}\right] 
\,.\nonumber 
\end{eqnarray} 
It connects Fock states whose particle number differs by 1.
The {\em matrix elements} $V_{n}(1;2,3)=V_n(q_1;q_2,q_3)$ 
are simple functions of the three single-particle states $q_i$,
which are given in Table~\ref{tab:verspi}. 

\begin{table}[t]
\begin{center}
\begin{tabular}{|ll|} 
\hline 
\begin{tabular}{l}  
 \epsfysize=54ex\epsfbox{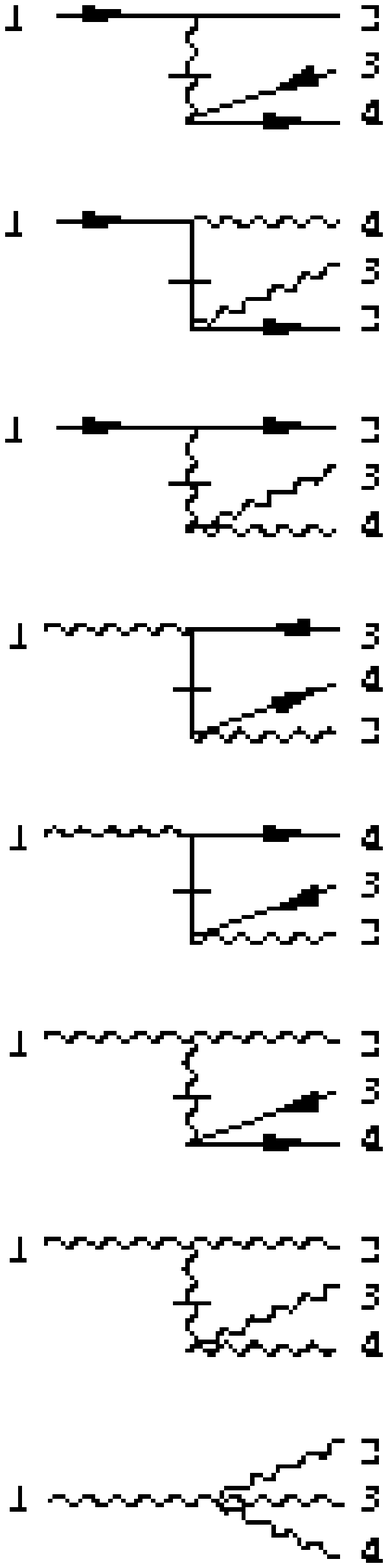}
\end{tabular}
&
\begin{tabular}{l}  
  $\displaystyle F_{1\phantom{,1}} = + 
  {2\Delta\over \sqrt{k^+_1k^+_2 k^+_3 k^+_4}}
  {(\bar u _1T^a\gamma^+u_2)\ (\bar v_3\gamma^+T^au_4)
   \over(k^+_1-k^+_2)^2}$
\\                                           
  $\displaystyle F_{3,1} = +
  {\Delta\over \sqrt{k^+_1k^+_2 k^+_3 k^+_4}}
  {(\bar u _1T^{a_4}/\!\!\!\epsilon_4\gamma^+
   /\!\!\!\epsilon_3T^{a_2}u_2)
   \over( k^+_1-k^+_4)}$
\\                                           
  $\displaystyle F_{3,2} = -
  {2k^+_3\Delta\over \sqrt{k^+_1k^+_2 k^+_3 k^+_4}}
  {(\bar u _1T^a\gamma^+u_2)
  \ (\epsilon_3iC^a\epsilon_4)\over( k^+_1-k^+_2)^2}$
\\                                           
  $\displaystyle F_{5,1} = +
  {\Delta\over \sqrt{k^+_1k^+_2 k^+_3 k^+_4}}
  {(\bar v _3T^{a_1}/\!\!\!\epsilon^\star_1\gamma^+
  /\!\!\!\epsilon_2T^{a_2}u_4)
  \over( k^+_1-k^+_3)}$
\\                                           
  $\displaystyle F_{5,2} = -
  {\Delta\over \sqrt{k^+_1k^+_2 k^+_3 k^+_4}}
  {(\bar v _3T^{a_2}/\!\!\!\epsilon_2\gamma^+
  /\!\!\!\epsilon^\star_1T^{a_1}u_4)
  \over( k^+_1-k^+_4)}$
\\                                           
  $\displaystyle F_{5,3} = +
  {2(k^+_1+k^+_2)\Delta\over \sqrt{k^+_1k^+_2 k^+_3 k^+_4}}
  {(\bar v _3T^a\,\gamma^+\,u_4)   
   \ (\epsilon^\star_1iC^a\epsilon_2)  \over( k^+_1-k^+_2)^2}$
\\                                           
  $\displaystyle F_{6,1} = +
  {2k^+_3(k^+_1+k^+_2)\Delta\over 
  \sqrt{k^+_1k^+_2 k^+_3 k^+_4}\hfill}
  \ {(\epsilon^\star_1C^a\epsilon_2)
  \   (\epsilon_3C^a\epsilon_4)  \over( k^+_1-k^+_2)^2}$
\\                                           
  $\displaystyle F_{6,2} = +
  {2\Delta\over \sqrt{k^+_1k^+_2 k^+_3 k^+_4}}
  \ (\epsilon^\star_1\epsilon_3)\ (\epsilon_2\epsilon_4) 
  \ C^a_{a_1a_2} C^a_{a_3a_4}$
\end{tabular}
\\  \hline
\end{tabular}
\caption{\label{tab:forspi} \sl
   The fork interaction in terms of Dirac spinors.
   The matrix elements $F_{n,j}$ are displayed on the right, 
   the corresponding (energy) graphs on the left.
   All matrix elements are proportional to 
   $\Delta = \widetilde g^2  \delta (k^+_1 | k^+_2 + k^+_3+k^+_4) 
   \delta ^{(2)} (\vec k _{\!\perp,1} | \vec k _{\!\perp,2} +
                  \vec k _{\!\perp,3} + \vec k _{\!\perp,4} ) $, 
   with $\widehat  g = g/\sqrt{2(2\pi)^3} $.
   For the periodic boundary conditions 
   one uses $\widehat  g= g/\sqrt{\Omega} $. 
}\end{center}
\end{table}
%

The four-point interactions are broken up conveniently 
into fork and seagull interactions, $F$ and $S$, 
depending on whether they have an odd or an even number of creation
operators, thus
\[ 
      P_+ = T + V + F + S 
.\nonumber %
\] 
The {\em fork interaction} $F$ becomes then a sum of 6 operators,
\begin{eqnarray} 
   F &=& F_1+F_2+F_3+F_4+F_5+F_6 
\nonumber \\ &=&
   \sum_{1,2,3,4}
   \left[b_1 ^\dagger b_2 d_3 b_4\,F_1(1;2,3,4)+{\rm h.c.}\right]+ 
   \left[d_1 ^\dagger d_2 b_3 d_4\,F_2(1;2,3,4)+{\rm h.c.}\right]   
\nonumber \\ &+&
   \sum_{1,2,3,4}
   \left[b_1 ^\dagger b_2 a_3 a_4\,F_3(1;2,3,4)+{\rm h.c.}\right]+
   \left[d_1 ^\dagger d_2 a_3 a_4\,F_4(1;2,3,4)+{\rm h.c.}\right]  
\nonumber \\ &+&
   \sum_{1,2,3,4}
   \left[a_1 ^\dagger a_2 d_3 b_4\,F_5(1;2,3,4)+{\rm h.c.}\right]+
  \left[a_1 ^\dagger a_2 a_3 a_4\,F_6(1;2,3,4)+{\rm h.c.}\right]
.\label{eq:fork_int}
\end{eqnarray} 
It changes the particle number by 2. 
The matrix elements are given in Table~\ref{tab:forspi}.

The {\em seagull interaction} $S$, finally, becomes a sum of 7 operators  
\begin{eqnarray} 
      S  &=& S_1+S_2+S_3+S_4+S_5+S_6+S_7 
\nonumber \\ &=&
      \sum_{1,2,3,4}\, b_1^\dagger b_2^\dagger b_3 b_4\,S_1(1,2;3,4) + 
      \sum_{1,2,3,4}\, d_1^\dagger d_2^\dagger d_3 d_4\,S_2(1,2;3,4) 
\nonumber \\ &+&
      \sum_{1,2,3,4}\, b_1^\dagger d_2^\dagger b_3 d_4\,S_3(1,2;3,4) + 
      \sum_{1,2,3,4}\, b_1^\dagger a_2^\dagger b_3 a_4\,S_4(1,2;3,4)
\nonumber \\ &+&
      \sum_{1,2,3,4}\, d_1^\dagger a_2^\dagger d_3 a_4\,S_5(1,2;3,4) +
      \sum_{1,2,3,4}  (b_1^\dagger d_2^\dagger a_3 a_4\,S_6(1,2;3,4) +{\rm h.c.})
\nonumber \\ &+&
      \sum_{1,2,3,4}\, a_1^\dagger a_2^\dagger a_3 a_4\,S_7(1,2;3,4) 
.\nonumber %
\end{eqnarray} 
Its matrix elements are given in Table~\ref{tab:seaspi}.
It can act only between Fock states with the same particle number.

\begin{table} 
\begin{center}
\begin{tabular}{|ll|} 
\hline 
\begin{tabular}{l}  
 \epsfysize=86ex\epsfbox{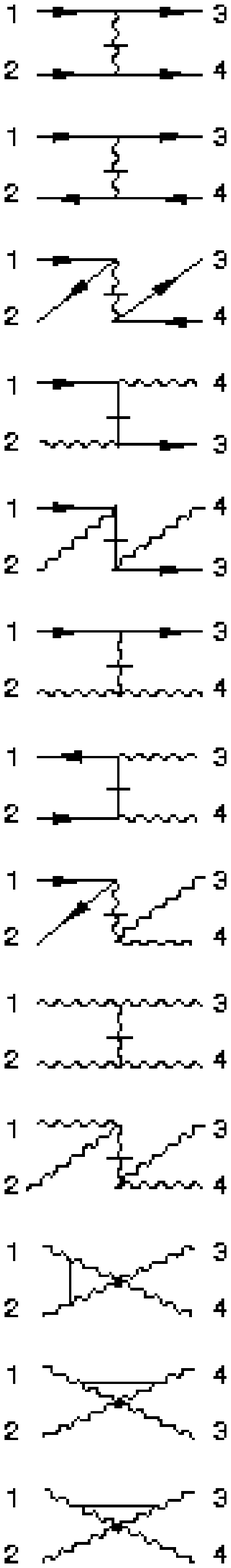}
\\ 
\end{tabular}
&
\begin{tabular}{l}  
  $\displaystyle S_{1\phantom{,1}} = - 
  {\Delta\over \sqrt{k^+_1k^+_2 k^+_3 k^+_4}}
  {(\bar u _1T^a\gamma^+u_3)\ (\bar u _2\gamma^+T^au_4)
   \over(k^+_1-k^+_3)^2}$
\\ 
  $\displaystyle S_{3,1} = +
  {2\Delta\over \sqrt{k^+_1k^+_2 k^+_3 k^+_4}}
  {(\bar u _1T^a\gamma^+u_3)\ (\bar v _2\gamma^+T^av_4)
  \over(k^+_1-k^+_3)^2}$
\\ 
  $\displaystyle S_{3,2} = -
  {2\Delta\over \sqrt{k^+_1k^+_2 k^+_3 k^+_4}}
  {(\bar v_2T^a\gamma^+u_1)\ (\bar v _4\gamma^+T^au_3)
  \over(k^+_1+k^+_2)^2}$
\\ 
  $\displaystyle S_{4,1} = +
  {\Delta\over \sqrt{k^+_1k^+_2 k^+_3 k^+_4}}
  {(\bar u _1T^{a_4}/\!\!\!\epsilon_4\gamma^+
   /\!\!\!\epsilon^\star _2T^{a_2}u_3)
   \over( k^+_1-k^+_4)}$
\\ 
  $\displaystyle S_{4,2} = +
  {\Delta\over \sqrt{k^+_1k^+_2 k^+_3 k^+_4}}
  {(\bar u _1T^{a_2}/\!\!\!\epsilon^\star _2\gamma^+
  /\!\!\!\epsilon_4T^{a_4}u_3)
  \over( k^+_1+k^+_2)}$
\\ 
  $\displaystyle S_{4,3} = +
  {2(k^+_2+k^+_4)\Delta\over \sqrt{k^+_1k^+_2 k^+_3 k^+_4}}
  {(\bar u _1T^a\gamma^+u_3)
  \ (\epsilon_2^\star iC^a\epsilon_4)\over( k^+_1-k^+_3)^2}$
\\ 
  $\displaystyle S_{6,1} = +
  {\Delta\over \sqrt{k^+_1k^+_2 k^+_3 k^+_4}}
  {(\bar u _1T^{a_3}/\!\!\!\epsilon_3\gamma^+/\!\!\!\epsilon_4
   T^{a_4}v_2) \over( k^+_1-k^+_3)}$
\\ 
  $\displaystyle S_{6,2} = -
  {(k^+_3-k^+_4)\Delta\over \sqrt{k^+_1k^+_2 k^+_3 k^+_4}}
  {(\bar u _1T^a\,\gamma^+\,v_2)   
   \ (\epsilon_3iC^a\epsilon_4)  \over( k^+_1+k^+_2)^2}$
\\ 
  $\displaystyle S_{7,1} = -
  {(k^+_1+k^+_3)(k^+_2+k^+_4)\Delta\over 
  \sqrt{k^+_1k^+_2 k^+_3 k^+_4}\hfill}
  \ {(\epsilon^\star_1C^a\epsilon_3)
   \ (\epsilon^\star_2C^a\epsilon_4)  \over( k^+_1-k^+_3)^2}$
\\ 
  $\displaystyle S_{7,2} = +
  {2k^+_3k^+_4\Delta\over \sqrt{k^+_1k^+_2 k^+_3 k^+_4}}
   \ {(\epsilon^\star_1C^a\epsilon^\star_2)
    \ (\epsilon_3C^a\epsilon_4)  \over( k^+_1+k^+_2)^2}$
\\ 
  $\displaystyle S_{7,3} = +
  {\Delta\over \sqrt{k^+_1k^+_2 k^+_3 k^+_4}}
  \ (\epsilon^\star_1\epsilon_3)\ (\epsilon^\star_2\epsilon_4) 
  \ C^a_{a_1a_2} C^a_{a_3a_4}$
\\ 
  $\displaystyle S_{7,4} = +
  {\Delta\over \sqrt{k^+_1k^+_2 k^+_3 k^+_4}}
  \ (\epsilon^\star_1\epsilon_3)\ (\epsilon^\star_2\epsilon_4) 
  \ C^a_{a_1a_4} C^a_{a_3a_2}$
\\ 
  $\displaystyle S_{7,5} = +
  {\Delta\over \sqrt{k^+_1k^+_2 k^+_3 k^+_4}}
  \ (\epsilon^\star_1\epsilon^\star_2)\ (\epsilon_3\epsilon_4) 
  \ C^a_{a_1a_3} C^a_{a_2a_4}$
\\ 
\end{tabular}
\\  \hline
\end{tabular}
\caption{\label{tab:seaspi} \sl
   The seagull interaction in terms of Dirac spinors.
   The matrix elements $S_{n,j}$ are displayed on the right, 
   the corresponding (energy) graphs on the left.
   All matrix elements are proportional to 
   $\Delta = \widetilde g^2  \delta (k^+_1 + k^+_2 | k^+_3+k^+_4) 
   \delta ^{(2)} (\vec k _{\!\perp,1} + \vec k _{\!\perp,2} |
                  \vec k _{\!\perp,3} + \vec k _{\!\perp,4} ) $, 
   with $\widehat  g = g/\sqrt{2(2\pi)^3} $.
   For  periodic boundary conditions 
   one uses $\widehat  g= g/\sqrt{\Omega} $. 
}\end{center}
\end{table}
%

The above results are are quite generally applicable: 
They hold for arbitrary non-abelian gauge theory $SU(N)$. 
They hold for abelian gauge theory (QED), 
formally  by replacing the color-matrices 
$T^a_{c,c^\prime}$ with the unit matrix and by setting to 
zero the structure constants $f^{abc}$, 
thus $B^{\mu\nu}=0$ and $\chi^{\mu}=0$.
They hold for 1 time dimension and arbitrary
$d+1$ space dimensions, with $i=1,\dots,d$.
All what has to be adjusted is the volume integral
$\int\!dx_+ dx_{\!\perp,1}\dots dx_{\!\perp,d}$.

\section{The hadronic bound-state problem}

One has to find a language in which one can represent 
hadrons in terms of relativistic confined quarks and
gluons. As reviewed in \cite{lsg91}, 
the Bethe-Salpeter formalism 
has been the central method  for analyzing hydrogenic atoms 
in QED and provides a completely 
covariant procedure for obtaining bound state solutions. 
However, calculations using this method are extremely 
complex and appear to be intractable much beyond the 
ladder approximation. It also appears impractical to extend 
this method to systems with more than a few constituent
particles. 

An intuitive approach  for solving relativistic bound-state 
problems would be to solve the instant form Hamiltonian eigenvalue 
problem
\[ 
   H\left|{\Psi}\right\rangle = \sqrt{M^2 + \vec P ^{\,2}}
   \left|{\Psi}\right\rangle
\] 
for the hadron's mass and wave function.
Here, one imagines that
$\left|\Psi\right\rangle$ is an expansion in multi-particle
occupation number Fock states, and that the operators $H$ 
and $\vec P $ are second-quantized Heisenberg 
operators. Unfortunately, this method is
complicated by its non-covariant structure and the 
necessity to first understand its complicated vacuum 
eigenstate over all space and time. The presence of the
square root operator presents severe mathematical 
difficulties. Even if these problems could be solved, 
the eigensolution is only determined in its rest system
($\vec P = 0$); determining the boosted wave function
is as complicated as diagonalizing $H$ itself.
This is why instant form wave function cannot be applied 
in practice to scattering problems.
Structure functions for example cannot be calculated.
 
In principle, the front form approach works in the
same way.  One aims at solving  the Hamiltonian 
eigenvalue problem
\begin{equation}
   H\left|{\Psi}\right\rangle = 
  { M^2 + \vec P _{\!\perp}^{\,2}\over P ^+}
   \left|{\Psi}\right\rangle
,\label{eq:4.2}\end{equation}
which for several reasons is easier: 
Contrary to $ P _z$  the operator $ P ^+$ is positive,
having only positive eigenvalues. 
The square-root operator is absent.
The boost operators are kinematic.
Having determined the wave function in a particular frame
with fixed total momenta $P ^+$ and $\vec P _{\!\perp}$
the kinematic boost operators allow to covariantly transcribe 
to any other frame. In fact, as discussed below,
one can formulate the theory frame-independently.

The ket $\left|\Psi\right\rangle $ can be
calculated in terms of a complete set of functions
$\left|\mu\right\rangle$ or $\left|\mu_n\right\rangle$, 
\[ 
   \int\!d[\mu]
   \ \left|\mu\right\rangle\left\langle\mu\right|
   = \sum\limits_{n}\int\!d[\mu_n]
   \ \left|\mu_{n}\right\rangle\left\langle\mu_{n}\right|
   = \mathbf {1}
.\] 
The transformation  between the complete set
of eigenstates $ \left|\Psi\right\rangle $ and the complete set
of basis states  $ \left| \mu_n \right\rangle $ are then 
$\left\langle\mu_n|\Psi\right\rangle $
and usually called the  {\em wavefunctions}  
$\Psi_{n/h)}(\mu) \equiv \left\langle\mu_n|\Psi\right\rangle $.
In addition to the quantum numbers of the Lorentz group,
the eigenfunction is labeled by quantum numbers 
like charge, parity, or baryon number which specify a particular 
hadron $ h $, thus
\[ 
   \left|\Psi\right\rangle 
   = \sum _n \int\! d[\mu_n] \ \left|\mu_n\right\rangle
    \Psi_{n/h} (\mu)
.\] 
One constructs the complete basis of Fock states 
$ \left| \mu_n \right\rangle $ in the usual way by
applying products of free field creation operators to the 
vacuum state $\left| 0 \right\rangle$:
\begin{eqnarray}
\begin{array}{lrllrl}
    n=0: \qquad
     & {}&\left| 0 \right\rangle\ ,   \nonumber \\
     n=1: \qquad\hfill
 & {}&\left|{q\bar q: k^+_i, \vec k _{\!\perp i},\lambda_i}
    \right\rangle  &=& 
    b^\dagger( q _1)\,
    d^\dagger( q _2)\,
&\left| 0 \right\rangle\ , \nonumber \\
   n=2: \qquad
   &{}&\left|{q\bar q g: k^+_i, \vec k _{\!\perp i},\lambda_i}
   \right\rangle &=& 
    b^\dagger( q _1)\,
    d^\dagger( q _2)\,
    a^\dagger( q _3)\, 
&\left| 0 \right\rangle\ , \nonumber \\
    n=3: \qquad
    & {}&\left|{g g: k^+_i, \vec k _{\!\perp i},\lambda_i}
    \right\rangle &=&  
    a^\dagger( q _1)\,
    a^\dagger( q _2)\,
&\left| 0 \right\rangle\ , \nonumber \\    
    \vdots & {} & \vdots & \vdots & \vdots & \left|0\right\rangle\ .
    \end{array}
\label{eq:4.7}\end{eqnarray}
The operators
$b^\dagger(q)$, $d^\dagger(q)$ and $a^\dagger(q)$ 
create bare leptons (electrons or quarks), bare anti-leptons
(positrons or antiquarks) and bare vector bosons (photons or
gluons).
All of these particles are `on-shell', 
$(k^\mu k_\mu)_i=m_i^2$.
The various Fock-space classes are conveniently labeled 
with a running index $n$. 
Each Fock state 
$\left|\mu_n \right\rangle 
= \big|{n: k^+_i, \vec k _{\!\perp i},\lambda_i}\big\rangle$ 
is an eigenstate of $P^+$ and $\vec P _{\!\perp}$
and the free part of the energy $P^-_0 $, 
with eigenvalues 
\[ 
   P^+ = \sum\limits_{i\in n} k^+_i 
   ,\quad
   \vec P _{\!\perp} = \sum\limits_{i\in n} \vec k _{\!\perp i}
   ,\quad
   P^-_0 = 
   \sum\limits_{i\in n} \frac{m_i^2 + k^2 _{\!\perp i}}{k^+_i} 
.
\]
The free invariant mass square of a Fock-state is 
$M_0^2 = (p_1+p_2+\dots+p_{n_{i}})^2 $, thus 
\begin{eqnarray} 
   M_0^2 = 
   P^\mu_0 P_{0,\mu} = P^+ P^-_0 - \vec P _{\!\perp} ^2 =
   P^+ \left(\sum\limits_{i\in n} 
   \frac{m_i^2 + k^2 _{\!\perp i}}{k^+_i} \right) - \vec P _{\!\perp} ^2
.\label{eq:freeMass}\end{eqnarray}
The Fock and the physical vacuum have eigenvalues $0$.

The restriction to $k^+ > 0$ is a key difference
between light-cone quantization and ordinary equal-time
quantization. In equal-time quantization, the state of a parton is
specified by its ordinary three-momentum $\vec k =
(k_x,k_y,k_z)$. Since each component of $\vec k$  can
be either positive or negative, there exist zero total momentum 
Fock states of arbitrary particle number, and these will mix with 
the zero-particle state to build up the ground state, the
physical vacuum. However, in light-cone quantization each of the 
particles forming a zero-momentum state must have vanishingly 
small $k^+$.  
The free or Fock space vacuum $ \left|0\right\rangle$ is then
an exact  eigenstate of the full front form Hamiltonian $H$, in
stark contrast to the quantization at equal usual-time.
However, the vacuum in QCD is
undoubtedly more complicated due to the possibility of 
color-singlet states with $P^+ = 0$ built on  
zero-mode massless gluon quanta, 
but the physical vacuum in the front form 
is {\em still far simpler} than in the usual instant form.

Since $k_i^+ > 0$ and $ P ^+ > 0$,  one can define 
boost-invariant longitudinal momentum fractions
\[ 
   x_i = {k_i^+ \over P^+ }
   \ , \qquad {\rm with}\quad
   0 < x_i  < 1
,\] 
and boost-invariant intrinsic transverse momenta 
$\vec k _{\!\perp i} $
Their values are constrained, 
\begin{equation} 
   \sum\limits_{i\in n} x_i = 1
   \quad{\rm and}\quad
   \sum\limits_{i\in n} \vec k _{\!\perp i} = \vec 0
,\label{eq:4.12}\end{equation} 
corresponding to the intrinsic frame $\vec P _{\!\perp} = \vec 0$.
All particles in a Fock state $\left|\mu_n \right\rangle 
 = \big|{n: x_i, \vec k _{\!\perp i},\lambda_i}\big\rangle$
have a boosted four-momentum
\[ 
   p^\mu_i \equiv (p^+,\vec p _{\!\perp},p^-) _i = 
   \left( x_i P^+, \vec k _{\!\perp i} + x_i \vec P _{\!\perp}  ,
   \frac{m_i^2 + (\vec k _{\!\perp i} + x_i \vec P _{\!\perp})^2 }
   {x_i P^+} \right)
.\] 
The free invariant mass square of the Fock state, Eq.(\ref{eq:freeMass}), 
is therefore
\[ 
      M_0^2 =  \sum\limits_{i\in n} 
      \left({m_i^2 + (\vec k _{\!\perp i} + x_i\vec P _{\!\perp})^{\,2}
      \over x_i} \right) - \vec P _{\!\perp}^{\,2}
      = \sum\limits_{i\in n} 
      \left({m^2+\vec k_{\!\perp} ^{\,2}\over x}\right)_i 
,\] 
as a direct consequence of the transverse boost properties.
 
The phase-space differential
$d[\mu_n]$ depends on how one normalizes 
the single particle states. 
In the convention where commutators are normalized to 
a Dirac-delta function, the phase space integration is 
\begin{eqnarray}
   \int\!d[\mu_n] \dots &=& \sum _{\lambda_i \in n} 
   \int \left[dx_i d^2 k_{\!\perp i} \right] \dots
\ , \qquad{\rm with} \nonumber \\ 
   \left[dx_i d^2 k_{\!\perp i} \right] 
   &=& \delta \Big(1-\sum_{j\in n}  x_j\Big)
   \delta^{(2)} \Big(\sum _{j\in n}  \vec k_{\perp j}\Big) 
   \ dx_1 \dots  dx_{N_n} 
   \ d^2 k _{\!\perp 1} \dots d^2 k _{\!\perp N_n}
,\nonumber\end{eqnarray}
where $ N_n$ is the number of particles in Fock state $\mu_n$.
The additional Dirac $\delta$-functions account for the 
constraints (\ref{eq:4.12}).
The eigenvalue equation (\ref{eq:4.2}) stands then
for an infinite set of  coupled integral equations
\begin{eqnarray}
   &&\sum_{n^\prime} 
   \int [d\mu^\prime_{n^\prime}]
   \ \langle n: x_i, \vec k _{\!\perp i}, \lambda_i \vert H
   \vert n^\prime: x_i^\prime, \vec k _{\!\perp i}^{\,\prime},
   \lambda_i^\prime \rangle \,\Psi_{n^\prime/h}
   (x_i^\prime, \vec k _{\!\perp i}^{\,\prime},\lambda_i^\prime)
\nonumber\\ 
   &=&{ M^2 + \vec P _{\!\perp}^{\,2}\over P ^+}
   \Psi _{n/h}(x_i, \vec k _{\!\perp}, \lambda_i )
,\qquad{\rm for}\quad
    n=1,\dots,\infty
.\end{eqnarray}
Since $P ^+$ and 
$\vec P _{\!\perp}$ are diagonal operators
one can rewrite this equation as 
\begin{eqnarray}
   &&\sum_{n^\prime} 
   \int [d\mu^\prime_{n^\prime}]
   \ \langle n: x_i, \vec k _{\!\perp i}, \lambda_i \vert 
   HP ^+ - \vec P _{\!\perp}^{\,2}
   \vert n^\prime: x_i^\prime, \vec k _{\!\perp i}^{\,\prime},
   \lambda_i^\prime \rangle \,\Psi_{n^\prime/h}
   (x_i^\prime, \vec k _{\!\perp i}^{\,\prime},\lambda_i^\prime)
\nonumber\\
   &=& M^2 
   \Psi _{n/h}(x_i, \vec k _{\!\perp}, \lambda_i )
.\label{eq:4.17}\end{eqnarray}
It is therefore possible to define a `light-cone Hamiltonian'
as the operator
\begin{equation} 
   H_{LC} = H P^+ - \vec P^2_{\!\perp} = P^\mu P_\mu 
,\label{eq:LC-hamiltonianI}\end{equation}
so that its eigenvalues correspond to the 
invariant mass spectrum $M_i$ of the theory.
Eq.(\ref{eq:4.17}) thus stands for 
\begin{equation} 
   H_{LC} \left| \Psi \right\rangle = 
   M^2    \left| \Psi \right\rangle
,\label{eq:LC-hamiltonianII}\end{equation}
in analogy to Eq.(\ref{eq:4.2}).

The Lorentz invariance of $H_{LC}$ and the boost invariance 
of the wave functions reflects the fact that the boost 
operators are kinematical. 
In fact one can boost the system to an `intrinsic frame'
in which the transversal momentum vanishes
$\vec P_{\!\perp} = \vec 0$, thus 
$H_{LC} = P^- P^+$.
The transformation to an arbitrary
frame with finite values of $\vec P_{\!\perp}$ 
is then trivially performed.
Consider a pion in QCD with momentum 
$ P = (P^+,\vec P _{\!\perp})$ as an example. 
It is described by
\[
   \left|{\pi: P}\right\rangle 
  = \sum_{n=1} ^{\infty} \int\!d[\mu_n] 
  \left|{n: x_i P^+, \vec k _{\!\perp i}
   + x_i \vec P _{\!\perp} , \lambda_i}\right\rangle\,
   \Psi_{n/\pi} (x_i,\vec k _{\!\perp i},\lambda_i)
,\]
where the sum is over all Fock space sectors of 
Eq.(\ref{eq:4.7}).
The ability to specify wavefunctions simultaneously
in any frame is a special feature of light-cone quantization. 
The light-cone wavefunctions $\Psi_{n/\pi}$ 
do not depend on the total momentum, 
since $x_i$ is the longitudinal momentum 
fraction carried by the $i^{\rm th}$ parton  and 
$\vec k _{\!\perp i} $ is its momentum ``transverse'' to the 
direction of the meson; both of these are 
frame-independent quantities.  They are the probability
amplitudes to find a Fock state of  bare particles in the
physical pion. But 
given these light-cone wavefunctions $\Psi_{n/h} (x_i,
\vec k _{\!\perp i} , \lambda_i)$, one can compute  any  
hadronic quantity by convolution with the appropriate quark 
and gluon matrix elements, 
see for example \cite{bpp97}. 

\begin{figure} [t]
 \epsfxsize=148mm\epsfbox{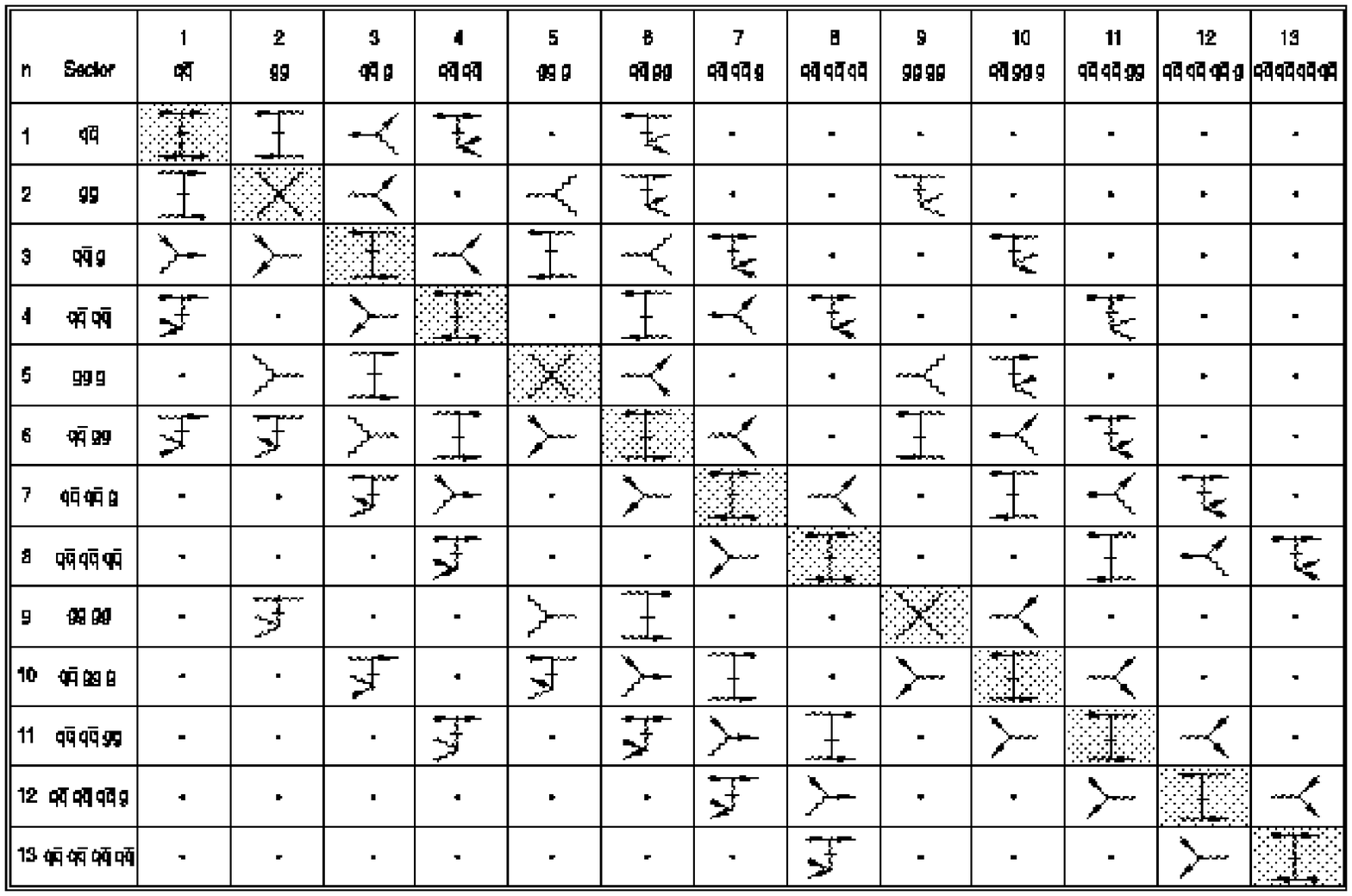}
\caption{\label{fig:holy-1} \sl
     The Hamiltonian matrix for a meson. 
     The matrix elements are represented by energy 
     diagrams. 
     Within each block they are all of the same type: 
     either vertex, fork or seagull diagrams.
     Zero matrices are denoted by a dot ($\cdot$).
     The single gluon is absent since it cannot be 
     color neutral.
}\end{figure}

In addressing to solve Eq.(\ref{eq:LC-hamiltonianII})
one faces several major difficulties, among them
that the above equations are 
ill-defined for very large values of the transversal momenta 
(`ultraviolet singularities') and for values of the 
longitudinal momenta close to the endpoints
$x\sim0$ or $x\sim1$ (`endpoint singularities').
One has  to introduce cut-offs $\Lambda$ to regulate 
the theory in some convenient way. 
Subsequently one has to remove the cut-off dependence
by renormalization group analysis.
Renormalization theory is known however only
for perturbation theory, for example when calculating 
Feynman scattering amplitudes in a certain order.
Renormalization theory is not available 
for the bound-state problem.~-- 
But even if one has found a convenient regularization scheme, 
one faces the large (infinite) 
number of coupled integral equations. 
Their nature resides in the complicated structure of the kernel
\[ 
   \langle n \vert H \vert n^\prime \rangle \equiv 
   \langle n: x_i, \vec k _{\!\perp i}, \lambda_i \vert H
   \vert n^\prime: x_i^\prime, \vec k _{\!\perp i}^{\,\prime},
   \lambda_i^\prime \rangle 
.\] 
In analyzing the structure of this very complicated many-body
problem, as done in Figure~\ref{fig:holy-1},
one realizes that most of its matrix elements are zero
by nature of the operator structure of the Hamiltonian.
The Hamiltonian is zero for all sectors whose 
particle number difference is larger than 2.
As an example, consider the fork interaction given in
Eq.(\ref{eq:fork_int}) particularly its matrix element $F_3$.
It scatters a quark into an other momentum state 
and distroys two gluons. In the block matrix element 
$\langle 1 \vert H \vert 6 \rangle =
 \langle q\bar q \vert F_3 \vert q\bar q \,gg\rangle $
one has many ways to realize that, without that the 
anti-quark $\bar q$ changes its momentum.
In Fig.~\ref{fig:holy-1},
all matrix elements in the block $\langle 1 \vert H \vert 6 \rangle $
are therefore represented 
diagrammatically by the same energy diagram as 
in Table~\ref{tab:forspi}.
Usually, the typical seagull matrix elements occur
for the diagonal blocs of this matrix, with one exception:
In the graph $S_6$ of Table~\ref{tab:seaspi}
a $q\bar q $-pair is annihilated and scatteredg instantaneously
into a $gg$-pair. Correspondingly, not all entries of
the block matrix element
$\langle 5 \vert H \vert 5 \rangle =
 \langle q\bar q\,g \vert S_6 \vert gg\,g\rangle $
can vanish.
In order to cope with the formidable many-body problem
imposed by Eq.(\ref{eq:LC-hamiltonianII}), 
the method of discretized light-cone quantization is
very useful.

\subsection{Perturbation theory in the front form}

Let us devote a section to the
perturbative treatment of front-form gauge theory. 
Light-cone perturbation
theory is really Hamiltonian perturbation theory, and we give
the complete set of rules which are the analogues of  the
Feynman rules. We  shall demonstrate in
a selected example, that one gets  the same covariant and
gauge-invariant scattering amplitude as in Feynman theory, 
see also \cite{bpp97}.

The Green's functions $\widehat G_{fi}(x^+)$ are
the probability amplitudes  that a state starting in Fock 
state $\vert i\rangle$ ends up in Fock state 
$\vert f\rangle$ at a later time $x^+$ 
\begin{eqnarray}
   \Big\langle f \mid \widehat G(x^+) \mid i \Big\rangle  
   = \langle f\vert e^{-i P_+ x^+} \vert i \rangle 
   = i \int {d\epsilon \over 2\pi} \, e^{-i\epsilon x^+}
   \langle f \vert G(\epsilon)\vert i \rangle 
.\nonumber\end{eqnarray}
Its Fourier transform 
$\left\langle f \mid G(\epsilon) \mid i \right\rangle $  
is called the resolvent of the Hamiltonian $H$, {\it i.e.}
\[ 
\left\langle f \mid G(\epsilon) \mid i \right\rangle 
   = \Bigl\langle f \left| {1\over \epsilon - H + i0_+}
   \right| i \Bigr\rangle 
   = \Bigl\langle f \left| {1\over \epsilon - H_0  - U + i0_+}
   \right| i \Bigr\rangle 
.\] 
Separating the Hamiltonian $H=H_0 + U$ into a free part  
$ H_0 $ and an interaction 
$ U $, one can expand the resolvent into the series
\[ 
   \langle f \vert G(\epsilon) \vert i \rangle 
   =\langle f \vert \widetilde G (\epsilon) + 
   \widetilde G (\epsilon) U \widetilde G (\epsilon) + 
   \widetilde G (\epsilon) U \widetilde G (\epsilon) U 
   \widetilde G (\epsilon) + 
   \ldots \vert i \rangle 
.\] 
The rules for $x^+$-ordered perturbation theory follow 
when the resolvent of the free Hamiltonian
$\widetilde G (\epsilon)= 1/(\epsilon - H_0 + i0_+) $ 
is replaced by its spectral decomposition 
\begin{equation} 
   \widetilde G (\epsilon)=
   \sum_{n=1} ^{\infty} \widetilde G _n(\epsilon)
   ,\quad \widetilde G _n(\epsilon) = 
   \int d[\mu_n]
   \left| n: k^+_i,\vec k_{\!\perp i},\lambda_i\right\rangle \,
   {1\over \Delta _n} 
   \left\langle n:k^+_i,\vec k_{\!\perp i},\lambda_i\right| 
,\label{eq:spec-decom}\end{equation} 
with the energy denominator 
$\Delta _n = 
   \epsilon - \sum_{i \in n} 
   \Big( ( k_{\!\perp}^{\,2} +m^2)/ k^+ \Big)_i + i0_+$. 
The sum runs over all Fock states $n$ 
intermediate between two interactions $U$.
 
To calculate then $\langle f|(\epsilon)|i\rangle$ 
perturbatively, all $x^+$-ordered diagrams must be 
considered, the contribution from each graph computed 
according to the rules of old-fashioned Hamiltonian 
perturbation theory \cite{kos70,leb80}:
\begin{enumerate}
\item
   Draw all topologically distinct $x^+$-ordered diagrams.
\item
   Assign to each line a single particle momentum $k^\mu$, a helicity
   $\lambda$, as well as color and flavor. 
   With fermions (electrons or quark)
   associate a spinor $u_\alpha(k,\lambda)$,
   with antifermions $v_\alpha(k,\lambda)$,
   and with vector bosons (photons or gluons)  a
   polarization vector $\epsilon_\mu(k,\lambda)$.
\item
   For each vertex include the matrix element 
   $\langle n\vert V\vert n'\rangle$ between Fock state $n$
   and $n'$ as given in Table~\ref{tab:verspi}.
\item
   For each intermediate state there is an 
   energy denominator $1/\Delta_n$ in which
   $\epsilon = P^-_{0,\rm in}$ is the incident free
   light-cone energy.
\item 
   To account for three-momentum conservation include 
   for each intermediate state the delta-functions  
   $\delta \Big( P ^+ - \sum_i k ^+_i \Big)$ 
   and $\delta ^{(2)}\Big( \vec P _{\!\perp} 
         - \sum_i \vec k _{\!\perp i} \Big)$. 
\item
   Sum over all internal helicities (and colors for gauge theories)
   and integrate over each internal $k$ with the weight
   $\int d^2k_{\!\perp}dk^+ \theta(k^+) (2\pi)^{-3/2}$.    
\item
   Include a factor $-1$ for each closed fermion loop, 
   for each fermion
   line that both begins and ends in the initial state, 
   and for each diagram in which fermion
   lines are interchanged in either of the initial or final states.
\item
   Imagine that every internal line is a sum
   of a `dynamic' and an `instantaneous' line, and draw
   all diagrams with $1,2,3,\dots$ instantaneous lines. 
\item
   Two consecutive instantaneous interactions give a 
   vanishing contribution.
\item
   For each instantaneous line insert a factor 
   $\langle n\vert W \vert n'\rangle$,
   with the matrix element given in Table~\ref{tab:forspi} 
   and~\ref{tab:seaspi}.
\end{enumerate}
The light-cone Fock state representation can thus be used
advantageously in perturbation theory. 
The sum over intermediate Fock states is equivalent to 
summing all $x^+-$ordered diagrams and integrating over 
the transverse momentum and light-cone fractions
$x$. Because of the restriction to positive $x,$ diagrams
corresponding to vacuum fluctuations or those containing
backward-moving lines are eliminated. 

\subsection{The $q\bar q$-scattering amplitude}

The simplest application of the above rules is  
the calculation of the electron-muon scattering amplitude 
to lowest non-trivial order. But the quark-antiquark 
scattering is only marginally more difficult.
We thus imagine an initial $(q ,\bar q)$-pair with different 
flavors $f \neq \bar f$ to be scattered off each other by
exchanging a gluon.

Let us treat this problem as a pedagogical example
to demonstrate the rules.
Rule 1: There are two time-ordered diagrams associated with
this process. In the first one the gluon is emitted by
the quark and absorbed by the antiquark, and in the second
it is emitted by the antiquark and absorbed by the quark.
For the first diagram, we assign the momenta required 
in rule 2 by giving explicitly the initial Fock state 
$      \vert q ,\bar q \rangle 
       ={1\over \sqrt{n_c}} \sum _{c=1} ^{n_c}
       b^\dagger_{cf} (k_q,\lambda_q) 
       d^\dagger_{c\bar f}  (k_{\bar q},\lambda_{\bar q})
       \vert 0 \rangle $.
Note that it is invariant under $SU(n_c)$.
The final Fock state is
$      \vert q^\prime, \bar q^\prime \rangle 
       ={1\over \sqrt{n_c}} \sum _{c=1} ^{n_c}
       b^\dagger_{cf} (k_q^\prime,\lambda_q^\prime) 
       d^\dagger_{c\bar f}(k_{\bar q}^\prime,\lambda_{\bar q}^\prime)
       \vert 0 \rangle$.
The intermediate state
\begin  {eqnarray}
      \vert q^\prime, \bar q,g  \rangle 
&=& \sqrt{{2\over n_c^2-1}} 
       \sum _{c=1} ^{n_c} 
       \sum _{c^\prime=1} ^{n_c} 
       \sum _{a=1} ^{n_c^2-1} 
       T^a_{c,c^\prime}
       b^\dagger_{c\bar f} (k_q^\prime,\lambda_q^\prime) 
       d^\dagger_{c'\bar f}  (k_{\bar q},\lambda_{\bar q})
       a^\dagger_a  (k_{g},\lambda_{g})
       \vert 0 \rangle 
\  , \end {eqnarray} 
has `a gluon in flight'. Since the gluons
longitudinal momentum is positive, the diagram allows
only for  $k_q^{\prime+} < k_q^+ $.
Rule 3 requires at each vertex the factors
\begin {eqnarray} 
       \langle q ,\bar q \vert \,V\,
       \vert q^\prime, \bar q , g  \rangle 
       &=& {g\over (2\pi)^{{3\over2}}}
       \sqrt{{n_c^2-1\over 2n_c}}
       \ {\left[ \overline u (k_q,\lambda_q) 
       \,\gamma^\mu\epsilon_\mu (k_g,\lambda_g)\,
        u(k_q^\prime,\lambda_q^\prime)\right] 
        \over \sqrt{2k^+_{q}} \sqrt{2 k^+_{ g}} \sqrt{2k'^+_{q}}  }
\ ,\\
       \langle q^\prime, \bar q ,g  \vert \,V\,
       \vert q ^\prime ,\bar q ^\prime \rangle 
       &=& {g\over (2\pi)^{{3\over2}}}
       \sqrt{{n_c^2-1\over 2n_c}}
        \ {\left[ \overline 
        v(k_{\bar q}^\prime,\lambda_{\bar q}^\prime)
        \,\gamma^\nu\epsilon^\star _\nu(k_g,\lambda_g)\,
	v (k_{\bar q},\lambda_{\bar q}) 
	\right] 
        \over \sqrt{2k^+_{\bar q}} \sqrt{2k^+_{g}}
        \sqrt{2k'^+_{\bar q}}} 
,\end {eqnarray} 
respectively, wich are found by means of Table~\ref{tab:verspi}.
Working with  color neutral Fock states,
all color structure reduces to
an overall factor $C$, with $C^2=(n_c^2-1)/2n_c$. 
Rule 4 requires the energy denominator $1/\Delta_3$.
It is useful to work the four-momentum
transfers of the quark 
$Q^2 = -(k_q-k_q')^2 = k_g^+(k_g+k_q^\prime-k_q)^-$.
The anti-quark has
$\overline Q^2 = (k_{\bar q}-k_{\bar q}')^2 
 = k_g^+(k_g+k_{\bar q}- k_{\bar q} ^\prime)^-$. 
With the initial energy $\epsilon= \widetilde P _+
   = (k_q+k_{\bar q})_+ = {1\over2}(k_q+k_{\bar q})^-$,  
the energy denominator becomes then 
\[  
   \Delta _3 
   = (k_q+k_{\bar q})^- - (k_g+k'_q+k_{\bar q})^-
   = -{Q^2 \over k^+_g} 
\]  
Rule 5 requires two Dirac-delta  functions, one at each 
vertex, to account for conservation of three-momentum. 
One of them is removed by the requirement of  rule 6, 
namely to integrate over all intermediate internal momenta and
the other remains in the final equation
(\ref{eq:4s.15}).  The gluon momentum is thus 
fixed by the external legs of the graph. 
The  polarization sum over the gluon helicity gives
\begin {eqnarray} 
        d_{\mu\nu}(k_g) \equiv \sum_{\lambda_g}
        \epsilon_\mu (k_g,\lambda_g)\,
        \epsilon^\star_\nu (k_g,\lambda_g)
        = -g_{\mu\nu} + 
        {k_{g,\mu}\eta_\nu + k_{g,\nu}\eta_\mu 
        \over k_g^\kappa \eta_\kappa } 
.\nonumber\end {eqnarray} 
The null vector has the components 
$\eta^\mu = (\eta^+,\vec\eta_{\!\perp},\eta^-)
        = (0,\vec 0_{\!\perp},2)$
and thus the properties $\eta^2\equiv\eta^\mu\eta_\mu=0$ 
and $k\eta=k^+$. As shown explicitly in \cite{bpp97} one gets
for the second order diagram 
$\langle q ,\bar q \vert V \widetilde G_3 V\vert q' ,\bar q' \rangle$
after some non-trivial steps
\begin{eqnarray} 
       \langle q ,\bar q \vert V \widetilde G_3 V \vert q' ,\bar q' \rangle
       =  {g^2 C^2\over (2\pi)^3} 
       \ {\left[ \overline u (k_q,\lambda_q) 
       \gamma^\mu u(k'_q,\lambda'_q)\right]
        \over \sqrt{4k^+_q k'^+_q} }
       {1\over  Q^2}
       {\left[ \overline u (k_{\bar q},\lambda_{\bar q}) 
       \gamma_\mu u(k'_{\bar q},\lambda'_{\bar q})
       \right]\over \sqrt{4k^+_{\bar q} k'^+_{\bar q}}}
\nonumber\\
       -{g^2 C^2\over (2\pi)^3} 
       \ {\left[ \overline u (k_q,\lambda_q) 
       \gamma^+ u(k'_q,\lambda'_q)\right]
        \over \sqrt{4k^+_q k'^+_q} }
       {1\over  \big( k^+_g \big) ^2}
       {\left[ \overline u (k_{\bar q},\lambda_{\bar q}) 
       \gamma^+ u(k'_{\bar q},\lambda'_{\bar q})
       \right]\over \sqrt{4k^+_{\bar q} k'^+_{\bar q}}}
.\end{eqnarray} 
The delta-functions and a step function
$\Theta(k'^+_q\leq k^+_q)$ are omitted for simplicity.
One proceeds with rule 8, by including 
the instantaneous lines. 
Table~\ref{tab:seaspi} gives
\begin {eqnarray} 
       \langle q ,\bar q \vert \,S\,
       \vert q', \bar q' \rangle 
       =  {g^2 C^2\over (2\pi)^3} 
       \ {\left[ \overline u (k,\lambda) 
       \gamma^+ u(k',\lambda')\right]_q
        \over \sqrt{4k^+_q k'^+_q} }
       {1\over  \big( k^+_{q} - k'^+_{q}\big) ^2}
       {\left[ \overline u (k,\lambda) 
       \gamma^+ u(k',\lambda')
       \right]_{\bar q}
       \over \sqrt{4k^+_{\bar q} k'^+_{\bar q}}}
.\nonumber\end {eqnarray} 
The $q\bar q$-scattering amplitude, the sum  
$\langle q ,\bar q \vert \,S + V \widetilde G_3 V \,
        \vert q', \bar q' \rangle $
has then the correct gauge-invariant result 
known from Feynman theory up to second order
 \begin{eqnarray} 
        \langle q ,\bar q \vert S + V \widetilde G_3 V 
	\vert q', \bar q' \rangle 
        &=& {(-1)\over (k_q-k'_{\bar q}) ^2}
        {1\over \sqrt{k^+_q k^+_{\bar q}  k'^+_q k'^+_{\bar q}}}    
         \delta(P ^+ - P '^+)
        \delta^{(2)}(\vec P_{\!\perp} -\vec P'_{\!\perp})
\nonumber\\  
        &\times&       
        {(gC)^2\over (2\pi)^3}
        \left[ \overline u (k,\lambda) 
        \gamma^\mu 
        u(k^\prime,\lambda^\prime)\right]_{q} 
        \left[ \overline u (k,\lambda) 
        \gamma_\mu
        u(k^\prime,\lambda^\prime)\right]_{\bar q} 
.\label{eq:4s.15}\end{eqnarray}
The instantaneous diagram is cancelled exactly.
\section{Discretized Light-Cone Quantization}

The infinitely many coupled integral equations 
in the preceeding section a very difficult 
to cope with in practice.
Because of the many integrations and summations
it is very difficult to even write them down.
But field theory becomes much more transparent when one 
works with periodic boundary conditions.
The coupled integral equations become then coupled
matrix equations. 
Since rows and colums of a matrix can be denumerated,
one can keep track of the necessary manipulations much easier.
In fact, working with periodic boundary conditions, or with 
`Discretized Light-Cone Quantization' (DLCQ), 
one has obtained the first total solutions to non-trivial 
quantum field theories in 1+1 dimensions. 
In 3+1 dimensions the method has the 
ambitious goal to calculate the spectra and wavefunctions of 
physical hadrons from a covariant gauge field theory.
The ingredients of the method shall be reviewed in short
in this section.

\subsection {The non-relativistic A-body problem in one dimension}

Let us first briefly review the difficulties for a
conventional non-relativistic many-body theory.
One starts with a many-body
Hamiltonian $ H  =  T + U $.
The kinetic energy $ T$
is usually a one-body operator and thus simple.
The potential energy $U $ is at least a two-body operator
and thus complicated.
One has solved the problem if one has found
the eigenvalues and eigenfunctions
of the Hamiltonian equation, $ H \Psi  =  E \Psi  $.
One always can expand the eigenstates
in terms of products of single particle states
$\langle \vec x \vert m \rangle $ belonging to a
complete set of ortho-normal functions.
When antisymmetrized, one refers to them as `Slater-determinants'.
All Slater-determinants with a fixed particle number
form a complete set.

One can proceed as follows. In the first step one chooses
a complete set of single particle wave functions.
These single particle wave functions are solutions of an arbitrary
single particle Hamiltonian.
In a second step, one selects one Slater determinat as a reference state.
All Slater determinants can be classified relative to this
reference state as 1-particle-1-hole (1-ph) states, 
2-particle-2-hole (2-ph) states, and so on.
The Hilbert space is truncated at some level.
In a third step, one calculates the Hamiltonian matrix
within this Hilbert space.

\begin{figure}
\centering 
 \epsfxsize=148mm\epsfbox{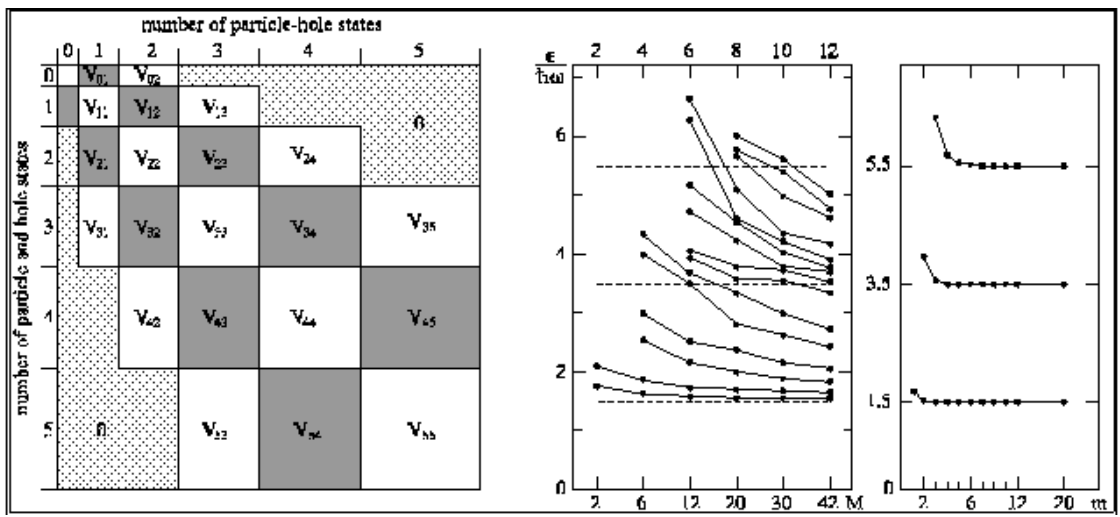} 
\caption{\label{fig-kyf-1} \sl
    Non-relativistic many-body theory. 
}\end{figure}

In Figure~\ref{fig-kyf-1},
the Hamiltonian matrix for a two-body interaction
is displayed schematically. Most of the matrix-elements vanish,
since a 2-body Hamiltonian changes the state by up to 2 particles.
Therefore the structure of the Hamiltonian is a 
finite penta\--diagonal block matrix. 
The dimension within a block is made finite by an artificial cut-off
on the kinetic energy, {\it i.e.}
on the single particle quantum numbers $m$.
A finite matrix can be diagonalized on a computer.
At the end one must verify that the physical results
are reasonably insensitive to the cut-off(s) and other
formal parameters.

This procedure was actually carried out in one space dimension
\cite {pau84} with two different sets of single-particle functions,
\begin {equation}
   \langle x \vert m \rangle = N_m H_m \Big( {x\over L} \Big)
   \exp {\left\{-{1\over 2} \Big( {x\over L} \Big) ^2 \right\}}
   \quad\quad{\rm and} \quad
   \langle x \vert m \rangle
 = N_m \exp {\left\{ i m{x\over L}\pi\right\}}
\ . \label{eold} \end {equation}
The two sets are the eigenfunctions of the harmonic oscillator
($ L \equiv \hbar / m \omega  $) with its Hermite polynomials 
$H_m$, and the eigenfunctions of the momentum of a 
free particle with periodic boundary conditions.
Both are suitably normalized ($N_m $), and both depend
parametrically on a characteristic length parameter $L$.
The calculations are particularly easy for particle number $2$,
and for a harmonic two-body interaction.
The results are displayed  in Figure~\ref{fig-kyf-1},
and surprisingly different.
For the plane waves, the results converge rapidly
to the exact eigenvalues
$E= {3\over2}, {7\over2}, {11\over2}, \dots$,
as shown in the right part of the figure.
Opposed to this, the results with the oscillator states
converge extremely slowly. Obviously, the larger part
of the Slater determinants is wasted on building
up the plane wave states of center of mass motion from the
Slater determinants of oscillator wave functions.
It is obvious, that the plane waves are superior, since they
account for the symmetry of the problem, namely Galilean covariance.
The approach was successfully applied for getting the exact
eigenvalues and eigenfunctions for up to 30 particles.

From these calculations, one should conclude that
discretized plane waves are a useful tool for many-body problems,
and that the generate good wavefunctions
even for a `confining' potential like the harmonic oscillator.

\subsection{QED in 1+1 dimension} 

DLCQ had been applied first to Yukawa theory \cite {pab85a} 
in 1-space and 1-time dimensions followed by 
an application to QED \cite{epb87} and to QCD \cite{hbp90}, 
but the advantages of working with periodic boundary conditions
in the front form,
particularly when discussing the `zero modes' in the $\phi^4$-theory, 
had been noted also by Maskawa and Yamawaki 
\cite{may76} in 1976. 

In one space dimension are no rotations --- hence no spin.
The Dirac spinor has two components 
and the Dirac matrices are 2 by 2 matrices.
The gauge field $A^\mu$ field has two components. 
One of them is eliminated by fixing the gauge, 
and the other determined by Gauss' law. 
The gauge field carries no dynamical degree of freedom. 
The theory confines quarks (or electrons) 
because the Poisson equation in 1 space dimension 
gives rise to a linearly rising potential. 

Quantum electrodynamics in 1+1 dimension has played an important
role in field theory because its massless version, the 
Schwinger model, is analytically soluable. The fundamental 
solution  corresponds to a composite $e\bar e$-state 
-- the Schwinger-boson -- with
invariant mass $m_B\equiv g/\sqrt{\pi}$.
Note that the coupling constant $g$ 
in 1+1 dimension has the dimension of a mass.

\def\d{D} \def\v{V} \def\b{$\cdot$} \def\s{S} \def\f{F} 
\def\g{g\phantom{\bar q}}
\begin {table} 
\begin{minipage}[t] {92mm} \makebox[0mm]{}
\begin {tabular}  {||lc||cccccccc||}
\hline \hline 
\rule[-1ex]{0ex}{4ex} 
 Sector & n & 
     1 & 2 & 3 & 4 & 5 & 6 & 7 & 8  
\\ \hline \hline   
 $ q\bar q\, $ &  1 & 
    \d &\f &\b &\b &\b &\b &\b &\b  
\\
 $ q\bar q\, q\bar q\, $ &  2 & 
    \f &\d &\f &\b &\b &\b &\b &\b  
\\
 $ q\bar q\, q\bar q\, q\bar q\, $ &  2 & 
    \b &\f &\d &\f &\b &\b &\b &\b  
\\ 
 $ q\bar q\, q\bar q\, q\bar q\, q\bar q\, $ & 4 & 
    \b &\b &\f &\d &\f &\b &\b &\b  
\\ 
 5 $ q\bar q$-pairs & 5 & 
    \b &\b &\b &\f &\d &\f &\b &\b  
\\ 
 6 $ q\bar q$-pairs & 6 & 
    \b &\b &\b &\b &\f &\d &\f &\b  
\\
 7 $ q\bar q$-pairs & 7 & 
    \b &\b &\b &\b &\b &\f &\d &\f  
\\ 
 8 $ q\bar q$-pairs & 8 & 
    \b &\b &\b &\b &\b &\b &\f &\d  
\\ \hline\hline
\end {tabular}
\end{minipage}
\hfill
\begin{minipage}  {50mm} \makebox[0mm]{}
\caption [masses] {\label {tab:qed1+1} \sl
   Fock-space sectors and block
   matrix structure for QED in 1+1 dimension.
   Diagonal blocs (D=T+S) refer to seagull, 
   off-diagonal (F) to fork matrix elements.
   A zero block matrix is marked by ($\cdot$).
}\end{minipage} 
\end {table}
Consider first the massive Schwinger model.
The Lagrangian for the theory takes the same
form as  in Eq.(\ref{eq:2.7}).
Again one works in the light-cone gauge $A^+=0$, and
uses the same projection operators $\Lambda_\pm$. 
The Dirac equation decomposes again into a time-derivative
$2i \partial^+ \Psi_+ = m\beta \Psi_- + g A^- \Psi_+ $
and a constraint $2i \partial^- \Psi_- = m \beta \Psi_+ $.
One is left with only one independent field,
$\Psi_+$, which is canonically quantized at $x^+=0$, 
\begin{equation}
   \bigl\{\Psi_+(x^-,x^+),\Psi_+^\dagger(y^-,y^+)\bigr\}
   _{x^+=y^+=0} 
   = \Lambda_+\delta(x^--y^-) 
.\label{ceq:19a}\end{equation}
The formalism gets somewhat simplified if one uses
the chiral representation \cite{hbp90}
\[
   \gamma^0=
   \pmatrix{\phantom{-}0 \phantom{-}\phantom{-}1\cr 
            \phantom{-}1 \phantom{-}\phantom{-}0\cr}
   ,\quad
   \gamma^1=
   \pmatrix{\phantom{-}0 \phantom{-}\phantom{-}1\cr 
                    {-}1 \phantom{-}\phantom{-}0\cr}
   ,\quad
   \gamma^5=\gamma^0\gamma^1= 
   \pmatrix{        {-}1 \phantom{-}\phantom{-}0\cr 
            \phantom{-}0 \phantom{-}\phantom{-}1\cr}
.\]
The $\Lambda_\pm=(1\pm \gamma^0\gamma^1)/2$ 
are then diagonal and project  
on the chiral components 
\[
   \Psi = \pmatrix{\Psi_L \cr\Psi_R}
   ,\quad
   \Psi_+ = \pmatrix{0 \cr \Psi_R}
   ,\quad
   \Psi_- = \pmatrix{\Psi_L \cr 0}
.\]
The (light-cone) momentum and energy operators become
\begin {eqnarray}
   P^+ &=& \int_{-L}^{+L} dx^- \Psi_R^\dagger \partial_-\Psi_R
,\nonumber 
\\ 
   P^- &=&
   m^2 \int_{-L}^{+L}dx^- 
   \Psi_R^\dagger 
   \frac{1}{i\partial_-}
   \Psi_R 
   \ + {g^2\over 2} \int_{-L}^{+L} dx^- 
   \Psi_R^\dagger \Psi_R 
   \frac{1}{(i\partial_-)^2}
    \Psi_R^\dagger \Psi_R
.\nonumber 
\end {eqnarray}
One can expand $\Psi_R$ (or $\Psi_+$) 
with periodic boundary conditions \cite{epb87}, 
but anti-periodic boundary conditions \cite{hbp90} 
avoid the zero mode: 
\[
   \Psi_R(x^-) = {1\over \sqrt{2L}}
   \sum_{n={1\over 2}, {3\over 2},\dots}^{\infty}
    \left( b_{n} e^{-i{n\pi \over L}x^-}
     + d_{n}^\dagger  e^{i{n\pi \over L}x^-} \right)
.\]
The creation and destruction operators obey
$\left\{ b_{n}^\dagger, b_{m} \right\} =
 \left\{ d_{n}^\dagger, d_{m} \right\} =
 \delta_{n,m}$,
consistent with Eq.(\ref{ceq:19a}). One redefines units by
\[
   P^+ = {2\pi \over L } K
   ,\quad
   P^-_0 = {L \over 2\pi } \left(m^2\,H_0 + {g^2 \over \pi} \,U \right)
.\]  
The length $L$ drops out in  $H_{\rm LC}= P^+P^-$. 
Inserting the above fields gives 
\begin{eqnarray}
   K &=& 
   \sum_{n={1\over 2}, {3\over 2},...}^{\infty}
   n \left( b_{n}^\dagger b_{n} + d_{n}^\dagger d_{n} \right)
   ,\quad
   H_0 = 
   \sum_{n={1\over 2}, {3\over 2},...}^{\infty} {1\over n}
   \left( b_{n}^\dagger b_{n} + d_{n}^\dagger d_{n} \right)
,\nonumber\\
   U &=& :U: \ +\  
   \sum_{n={1\over 2}, {3\over 2},...}^{\infty}
   { I_n \over n} 
   \left( b_{n}^\dagger b_{n} + d_{n}^\dagger d_{n} \right)
   ,\quad{\rm with}\quad
   I_n = - {1\over 2n} + \sum_{m=1}^{n+{1\over2}} {1\over m^2}   
,\nonumber\\
   :U: &=& \sum_{n_1,n_2,n_3,n_4}
   \frac {\delta(n_1+n_2|n_3+n_4)} {2(n_1-n_3)^2}
   \left( b_{1}^\dagger b_{2}^\dagger b_{3} b_{4} +
          d_{1}^\dagger d_{2}^\dagger d_{3} d_{4} \right) 
\nonumber\\
   &+& \sum_{n_1,n_2,n_3,n_4}
   \delta(n_1+n_2|n_3+n_4) \left(
   \frac {1} {(n_1-n_3)^2} - \frac {1} {(n_1+n_2)^2} \right)
   \ b_{1}^\dagger d_{2}^\dagger b_{3} d_{4}  
\nonumber\\
   &+& \sum_{n_1,n_2,n_3,n_4}
   \frac {\delta(n_1|n_2+n_3+n_4)} {(n_1-n_3)^2}
   \left( b_{1}^\dagger b_{2} b_{3} d_{4} +
          d_{1}^\dagger d_{2} d_{3} b_{4} + h.c.\right) 
.\nonumber\end{eqnarray} 
The symbols $\delta(n|m)=\delta_{n,m}$ 
are Kronecker delta's and $b_1\equiv b_{n_{1}}$.
In 1+1 dimensions it is very important to keep the 
`self-induced inertias' $I_n$ from the normal ordering. 
They are needed to cancel the infrared singularity 
in the interaction term in the continuum limit.

\begin {table}[t]
\begin{minipage}[t] {70mm}
\centering
\begin {tabular}  {|l|l||}
\hline \hline 
  1 & $\frac{1}{2};\frac{7}{2}$  
\rule[-1ex]{0ex}{4ex} 
\\
  2 & $\frac{3}{2};\frac{5}{2}$  
\\
  2 & $\frac{5}{2};\frac{3}{2}$  
\\
  4 & $\frac{7}{2};\frac{1}{2}$  
\\
  5 & $\frac{1}{2},\frac{3}{2};\frac{1}{2},\frac{3}{2}$  
\rule[-2ex]{0ex}{4ex} 
\\ \hline\hline
\end {tabular}
\caption{\label{tab:five_states} \sl
   The 5 Fock states for $K=4$.
}\end{minipage}
\hfill
\begin{minipage} [t] {75mm}
\centering
\begin {tabular}  {||c||llll||l||}
\hline \hline 
\rule[-1ex]{0ex}{4ex} 
  K  &1 $q\bar q$&2 $q\bar q$ &3 $q\bar q$ &4 $q\bar q$& Total 
\\ \hline \hline  
  1  & 1  & -   & -  & - & 1
\\
  4  & 4  & 1   & -  & - & 5
\\
  9  & 9  & 20  &  1 & - & 30
\\
 16  & 16 & 140 & 74 & 1 & 231
\\ \hline\hline
\end {tabular}
\caption{\label{tab:number_states} \sl
   Number of Fock states versus harmonic resolution.
}
\end{minipage} 
\end{table}
The next step is to actually solve the equations of motions in the
discretized space. Typically one proceeds as follows: 
One constructs the Fock space
\[
   \vert \mu_n \rangle = 
   \vert q_1,\dots,q_n; \bar q_1,\dots,\bar q_n\rangle = 
   b_{q_1}^\dagger,\dots,b_{q_n}^\dagger 
   d_{\bar q_1}^\dagger,\dots,d_{\bar q_n}^\dagger 
   \vert 0 \rangle 
,\]
in the same way as above in Eq.(\ref{eq:4.7}), and arranges 
it in denumerated Fock space sectors, 
as illustrated in Table~\ref{tab:qed1+1}. 
Each Fock state must be an eigenstate to $P^+$ and thus of the 
{\em harmonic resolution} $K$ \cite {epb87},
with eigenvalue $K=\sum _{i\in\mu_n} n_i$.
One selects now one value for $K$ and constructs all Fock 
states. Both the number of Fock space sectors 
and the number of Fock states within each sector are finite
for every finite $K$, 
due to the positivity condition on the light-cone momenta.
For $K=1$ one has only one Fock-space class with one
Fock state 
$\vert \frac{1}{2}; \frac {\bar 1}{2}\rangle = 
   b_{\frac{1}{2}}^\dagger d_{\frac{1}{2}}^\dagger \vert 0 \rangle $,
for $K=4$ one has the 5 Fock states given in Table~\ref{tab:five_states}.
The numbers of Fock states increase with given $K$, 
as shown in Table~\ref{tab:number_states},
but less than exponentially due to the exclusion principle.
Next, one calculates the matrix elements of $H=P^-$.
In the last step one diagonalizes $H$.
Any of its eigenvalues $E(K)$ depends on $K$ and 
corresponds to an invariant mass 
$ M^2(K) \equiv P ^+ P ^- = K E (K)$.  
Notice that one gets a spectrum of invariant mass-squares
for any value of $K$.

\begin{figure}
\unitlength1cm   \centering 
\begin{minipage}[t]{88mm}
 \epsfxsize=88mm\epsfbox{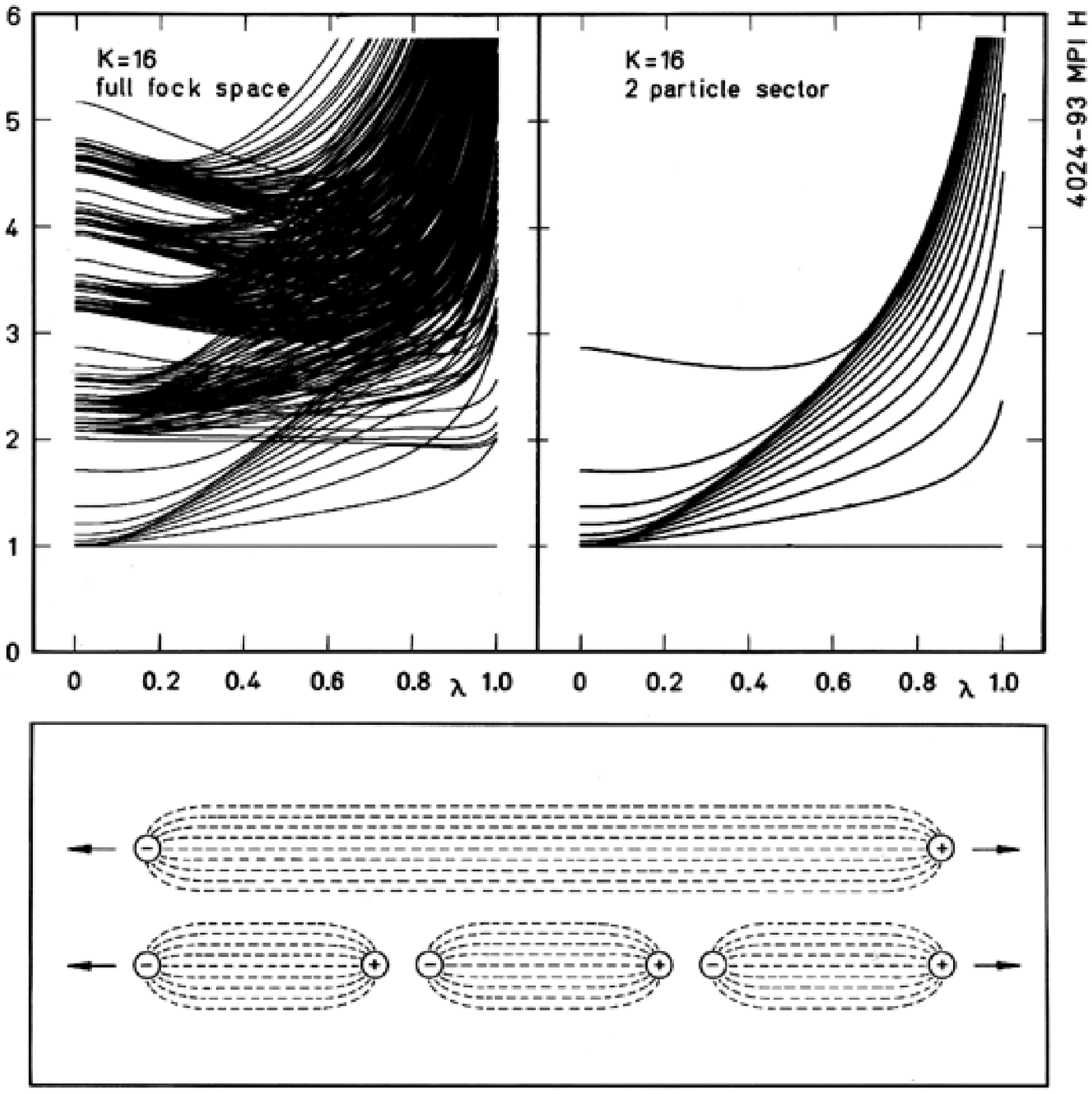}
\end{minipage}
\hfill
\begin{minipage}[t]{58mm}
 \epsfxsize=58mm\epsfbox{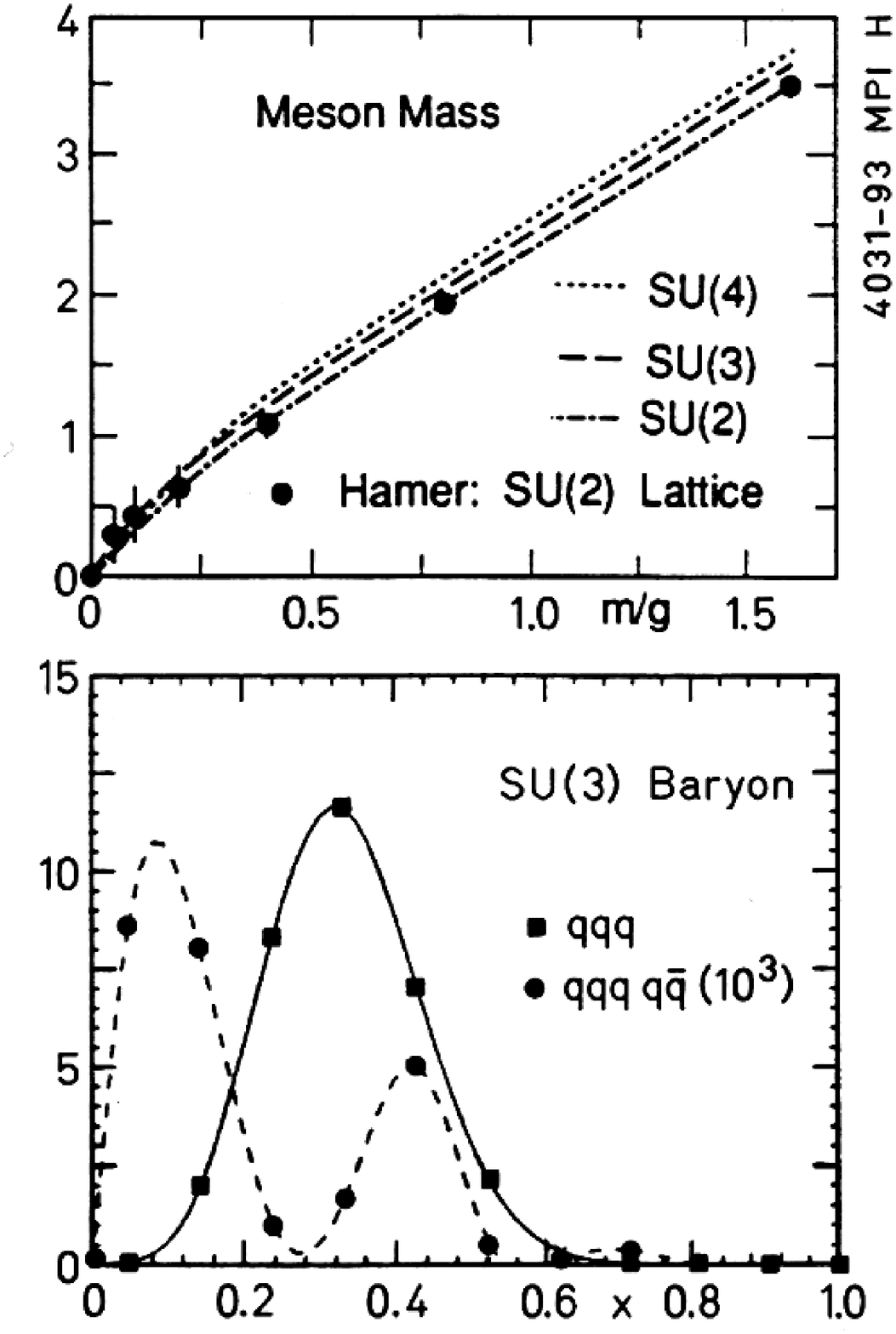}
\end{minipage}
\caption{\label{fig-kyf-4} \sl
    Spectra and wavefunctions in 1+1 dimensions, taken from 
    \protect{\cite{epb87,hbp90}}. 
}\end{figure}

The eigenvalue spectrum of QED$1+1$ was given first
by Eller \cite{epb87}, 
for periodic boundary conditions on the fermion fields.
The plot likeon the left side in Figure~\ref{fig-kyf-4} was 
calculated \cite {els94} with anti-periodic boundary conditions. 
It shows the full mass spectrum  of QED in the charge zero sector 
for all values of the 
coupling constant and the fermion mass, parametrized by  
$\lambda = (1+ \pi (m/g)^2) ^{-{1\over2}}$. 
The eigenvalues $M_i$ are plotted in units where the mass 
of the lowest `positronium' state has the numerical value 1. 
All states with $M>2$ are unbound. 
The plot includes the free case $\lambda=0$ ($g=0$) and the 
the Schwinger limit $M=1$ for $\lambda=1$ ($m=0$),
where DLCQ generates the exact eigenvalue.~--
The lower left part of the figure illustrates the following point.
The rich complexity of the spectrum allows for multi-particle 
Fock states {\em at the same invariant mass} as the 
`simple $q\bar q$-states' shown in the figure as the
`2 particle sector'.  The spectrum includes not only
the simple bound state spectrum, but also the associated
discretized continuum of the same particles in relative motion.  
One can identify the simple bound states as two quarks 
connected by a confining string as displayed in the figure. 
The smallest residual interaction mixes the simple 
configuration with the large number of
`continuum states' at the same mass. 
The few simple states have a much smaller statistical weight, 
and it looks as if the long string
`breaks' into several pieces of smaller strings.
Loosely speaking one can interpret such a process as the
decay of an excited pion into multi-pion configurations
$\pi^\star\rightarrow\pi\pi\pi$.

In $1+1$ dimensions quantum electro dynamics \cite {epb87} 
and quantum chromo dynamics  \cite {hbp90} 
show many similarities, both from the technical and from the 
phenomenological point of view. 
In the right part of Figure~\ref{fig-kyf-4} some of the results of
Hornbostel \cite {hbp90} on the spectrum and the 
wavefunctions for QCD are displayed. 
Fock states in non-abelian gauge theory SU(N) 
can be made color singlets for any order of the gauge group 
and thus one can calculate mass spectra for mesons and
baryons for almost arbitrary values of N. In the upper right
part of the figure the lowest mass eigenvalue of a meson
is given for $N=2,3,4$. 
Lattice gauge calculations are available only for 
$N=2$  and for the lowest two eigenstates
\cite{crh80}. In general the agreement is very 
good. In the left lower part of the figure the
structure function of a baryon is plotted versus (Bj\o rken-)$x$ for
$m/g=1.6$. With DLCQ it is possible to calculate also higher 
Fock space components. As an example, the figure includes the
probability distribution to find a quark in a $qqq\,q\bar q\,$-state.

\subsection {Fermion condensates and the small mass limit}

Based on the low energy theorems from times prior to QCD,
it is believed that the square of the pion mass is linear
in the quark mass $m$ for sufficiently small $m$.
The proportionality constant has to have a dimension of mass,
and since there is no other scale in
the problem except the {\em quark condensates in the vacuum}
$\langle 0 \vert \overline \Psi \Psi \vert 0\rangle$,
one believes that the square of the pion mass
is approximatively given by
$m^2_\pi \sim 2 \langle 0 \vert \overline \Psi \Psi \vert 0\rangle m $,
a theorem which was succesful in many phenomenological
applications. 

The Schwinger model has played an important role 
as a paradigm for our understanding of hadronic physics.
Among other aspects it has the  desired feature that 
the invariant mass square of the $e\bar e$-boson is linear  
in the electron mass $m$
\begin{equation}
   M^2 = m_B ^2 + 2m \,
   \langle 0 \vert \overline \Psi \Psi \vert 0\rangle
,\quad{\rm with} \quad 
   \langle 0 \vert \overline \Psi \Psi \vert 0\rangle =
   e^\gamma\ m_B 
,\label{eq:schwinger-mass}\end{equation}
where $m_B\equiv g/\sqrt{\pi}$ is the invariant mass in the
Schwinger limit. 
Euler's constant $\gamma$
was understood as a signal for non-perturbative physics.
In his analysis of the Schwinger model, 
Bergknoff \cite{ber77} showed that the take-off from the 
Schwinger limit obeys
\begin{equation}
   \langle 0 \vert \overline \Psi \Psi \vert 0\rangle =
   \frac {\pi} {\sqrt{3}}\ m_B 
.\label{eq:schwinger-bergknoff-mass}\end{equation}
The result of this `chiral perturbation theory' 
to first order, $\frac {\pi} {\sqrt{3}}\sim 1.81 $, 
is numerically very close to $e^\gamma\sim 1.78 $.

\begin{figure} [t]
\begin{minipage}[t]{71mm}
 \epsfxsize=71mm\epsfbox{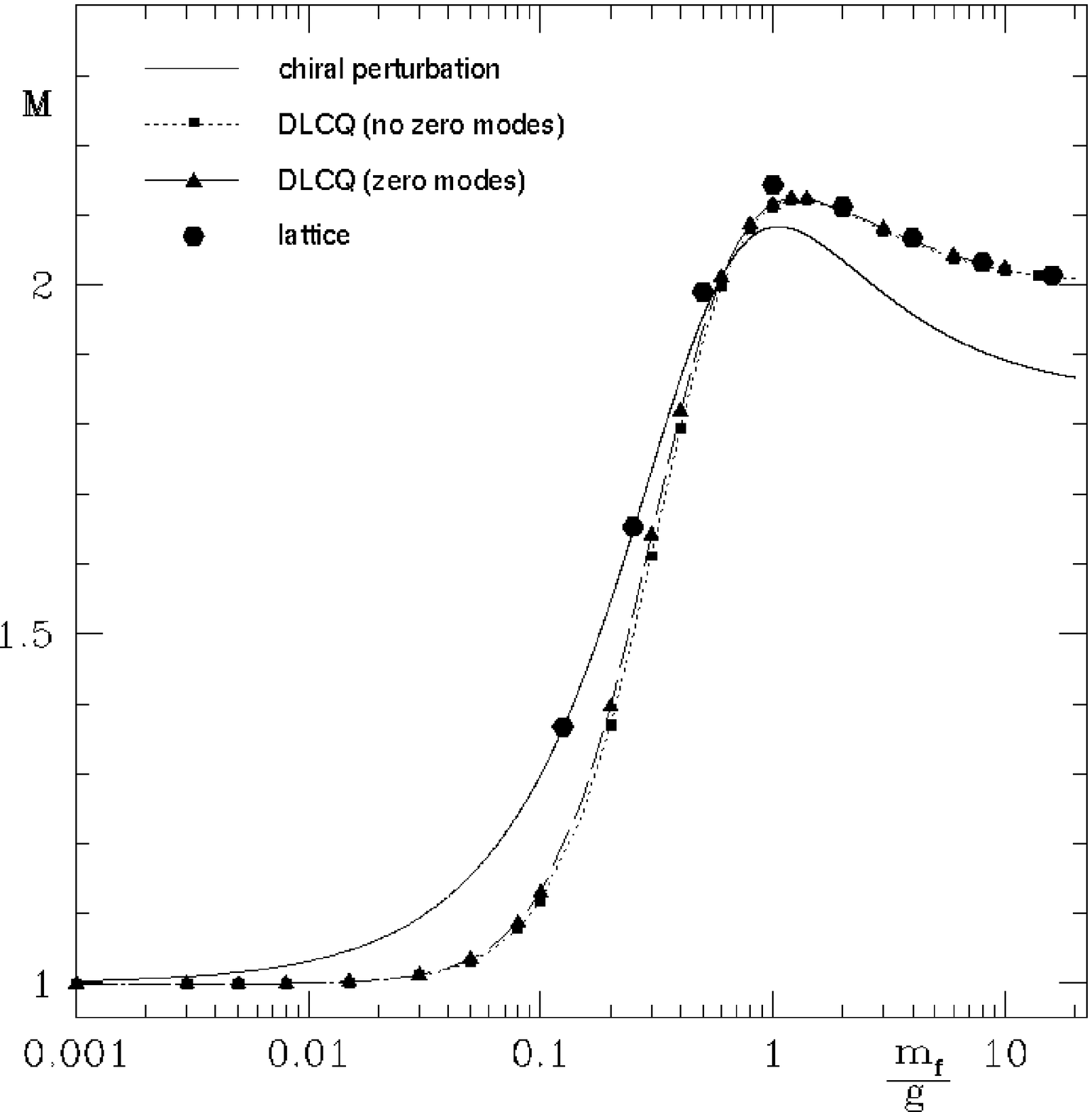} 
\caption{\label{fig:vol-1} \sl
   The lowest mass eigenvalue of QED$_{1+1}$ in units of $m_B$
   is plotted versus the fermion mass $m_f$ in units of 
   the coupling constant $g$,
   as calculated with `conventional DLCQ' ($K=16$).
   Taken from Ref.~\protect{\cite{voe96}}.
   See also discussion in the text.
}\end{minipage}
\hfill
\begin{minipage}[t]{71mm}
 \epsfxsize=71mm\epsfbox{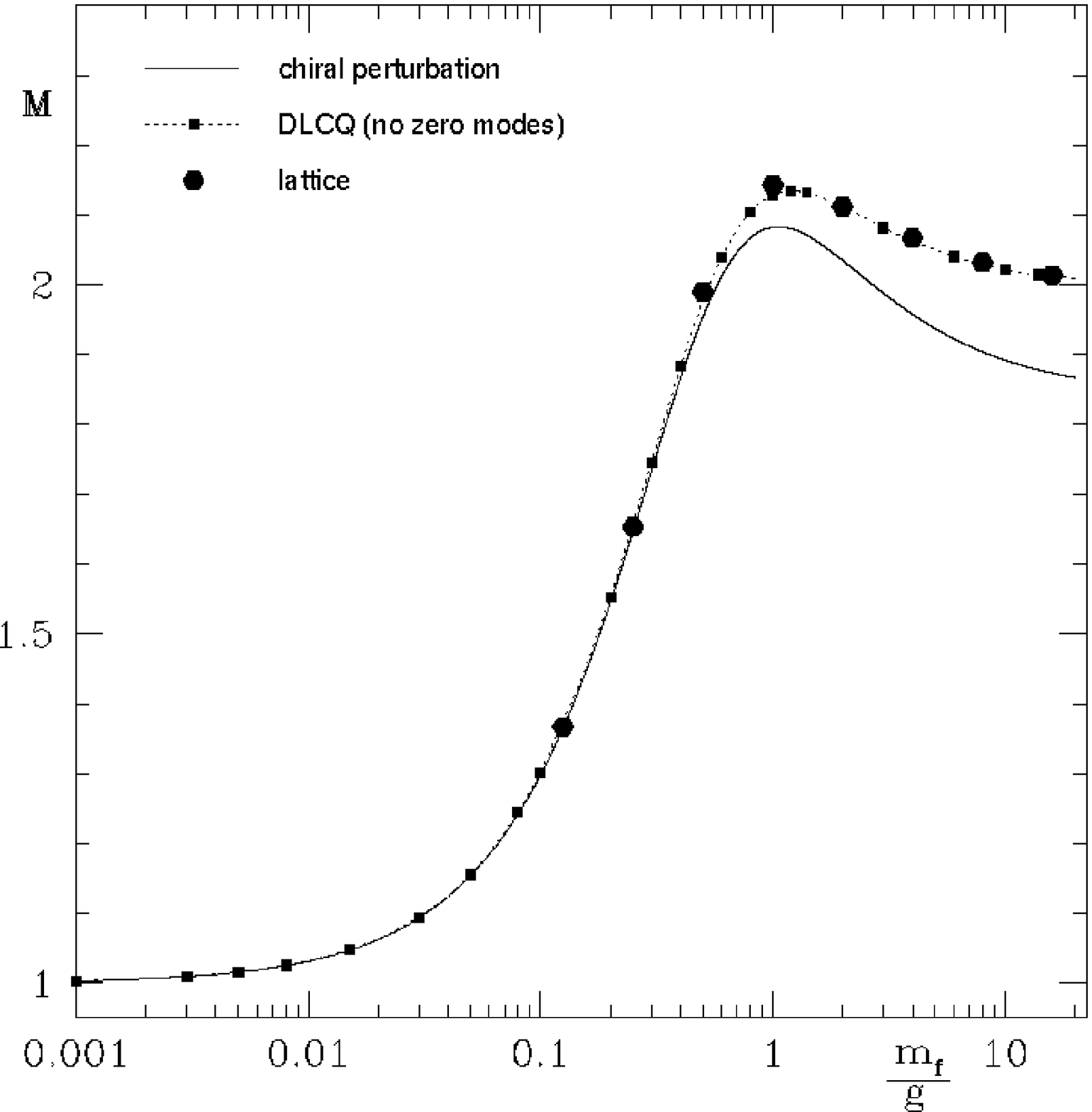} 
\caption{\label{fig:vol-2} \sl
   The lowest mass eigenvalue of QED$_{1+1}$ in units of $m_B$
   is plotted versus the fermion mass $m_f$ in units of 
   the coupling constant $g$,
   as calculated with `improved DLCQ' ($K=20$),
   as proposed in Ref.~\protect{\cite{van96}}.
   Taken from Ref.~\protect{\cite{voe96}}.
   See also discussion in the text.
}\end{minipage}
\end{figure}

Its is actually quite easy to derive the 
Bergknoff equation from DLCQ \cite{elp89}.
For a fixed resolution $K$, a $q\bar q$-state is fixed
by the quantum number $n$ of the electron.
The eigenvalue equation is then given by
\[
   M^2\langle n\vert \Psi \rangle = \sum _{n'=\frac{1}{2}} ^{[K]}
   \langle n \vert H_{LC} \vert n' \rangle\, 
   \langle n' \vert \Psi \rangle 
,\quad{\rm for}\quad 
   n=\frac{1}{2},\frac{3}{2}, \dots,\ [K]   
,\] 
with $[K] \equiv K-\frac{1}{2}$.
Using the Hamiltonian matrix elements are given above, one
introduces $x=p^+/P^+=n/K$ and goes to the continuum limit.
After a few steps \cite{elp89} one gets 
the integral equation of Bergknoff \cite {ber77},
\[ 
   M^2 \langle x \vert \Psi \rangle = 
   \frac {m^2} {x(1-x)} \langle x \vert \Psi \rangle + 
   m_B^2 \int\limits_0^1 dx' \langle x' \vert \Psi \rangle +  
   m_B^2 \int\limits_0^1 \!\!\!\!\!\!-\ dx' 
   \frac {\langle x\vert \Psi \rangle -
   \langle x'\vert \Psi \rangle }{(x-x')^2} 
.\] 
For $m=0$, the solution $\langle x\vert \Psi \rangle =1$ 
has the eigenvalue $M^2=m_B^2$, the Schwinger boson.

Eqs.(\ref{eq:schwinger-mass}) and (\ref{eq:schwinger-bergknoff-mass})
state that the mass-squared of the Schwinger boson
is {\em linear in the fermion mass} $m$, 
in the limit when $m$ goes to zero.
For a long time, this result was in conflict with the explicit 
DLCQ-calulations \cite{epb87,els94,voe96}. 
The shortcoming was taken as a hint \cite{vfp96}
that light-cone quantization was in failure because 
of the trivial vacuum structure which does not allow for
condensates. 

The most recent results by V\"ollinger \cite{voe96} 
for DLCQ are displayed in Figure~\ref{fig:vol-1}.
They are compared with the lattice calculations
of Crewther and Hamer \cite{crh80} and with chiral perturbation
theory up to second order \cite{vfp96}. 
The lattice results and DLCQ show some discrepancy
for very small $m$ which however fades away for
larger $m$. On the other hand, the lattice results and chiral
perturbation theory agree perfectly at small 
and deviate for large $m$, where chiral perturbation
theory is not suposed to work.
The figure show aslo that a correct inclusion of
the (gauge field) zero modes has no significant impact.
The re-solution of this puzzle came by van de Sande \cite{van96}.
He realized that conventional DLCQ has difficulties to reproduce
the wave function $\langle x \vert \Psi \rangle$
near the endpoints $x\rightarrow0$ and $x\rightarrow1$
very close to the Schwinger limit. 
His `improved DLCQ' accounts for that, and indeed
when properly included as shown in Figure~\ref{fig:vol-2} 
all discrepancies fade away.

What should be learned from this exercise is that not
everything what is called a `vacuum condensate'
in the literature deserves this name in a physical sense:
In naive light-cone quantization particularly DLCQ
the vacuum is trivial and can not have condensates.

\subsection {$\Phi^4$ in 1+1 dim's: Zero modes and phase transitions}
 
The naive front-form vacuum is simple.
However, one commonly associates important long range properties
of a field theory with the vacuum like spontaneous symmetry breaking, 
the Goldstone pion, or color confinement.  
If one cannot associate long range phenomena with the vacuum state 
itself, then the only alternative is the zero momentum components 
of the field, the `zero modes'.  
In some cases, the zero mode operator is not an independent degree 
of freedom but obeys a constraint equation. 
Consequently, it is a complicated operator-valued function of all 
the other modes of the field. 
Zero modes of this type have been investigated first by Maskawa 
and Yamawaki as early as in 1976 \cite{may76}.
An analysis of the zero mode constraint equation for (1+1)--dimensional
$\phi^4$ field theory, by van de Sande and Pinsky \cite{piv94}, 
shows how spontaneous symmetry breaking occurs 
within the context of this model. 

\begin{figure} [t]
\begin{minipage}[t]{60mm} \makebox[0mm]{}
 \epsfxsize=60mm\epsfbox{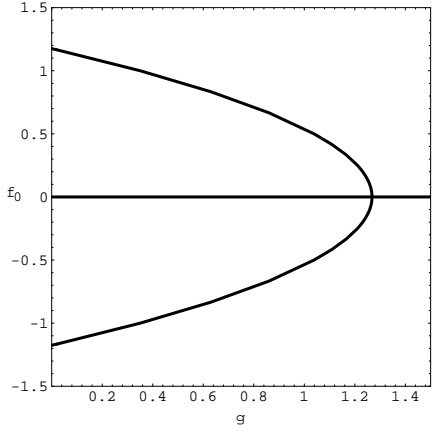} 
\end{minipage}
\hfill
\begin{minipage}[t]{80mm} \makebox[0mm]{}
\caption{\label{ff1} \sl
The vacuum expectation value of $\phi$,
$f_0=\protect\sqrt{4\pi} \langle 0|\phi| 0\rangle$ 
is plotted versus $g=24\pi\mu^2/\lambda$,
which is the inverse of the coupling constant $\lambda$.
Taken from \protect{\cite{piv94}}.~---
In the front form, the vacuum state $| 0\rangle$ is simple,
but the operator $\phi$, or $a_0$, is complicated.
In the conventional instant form, the vacuum $| 0\rangle$ is complicated
but the operator $\phi$ is simple.
}\end{minipage}
\end{figure}

The model represents a new paradigm for spontaneous 
symmetry breaking and shall be reviewed shortly. 
The Lagrangian in one space and one time dimension is 
${\cal L} = \partial_+\phi\partial_-\phi 
- {{\mu^2}\over 2} \phi^2
-  {\lambda
\over 4!} \phi^4
$. 
Imposing periodic boundary conditions with a length parameter 
$d=2L$ gives
$ 
   \phi\!\left(x\right) 
   = {1\over{\sqrt d}} 
   \sum_{n=-\infty}^\infty q_n(x^+) e^{i \frac{\pi}{L} n x^-} 
$. 
The field integral 
$\Sigma_n = \int dx^- \, \phi(x)^n-(zero\ modes)$
is convenient to discuss the problem and becomes
\[  
\Sigma_n = {1\over{n!}} \sum_{i_1, i_2, \ldots, i_n \neq 0} q_{i_1}  q_{i_2}
\ldots q_{i_n}\, \delta_{i_1 + i_2 + \ldots + i_n, 0}. 
\]  
Also the canonical Hamiltonian can be disentangled 
into $\Sigma_n$ and the zero mode $q_0$
\[  
   P^- = 
   {{\mu^2 }\over 2}q_0^2 + \mu^2 \Sigma_2 + 
   {{\lambda }\over{4! d}} q_0^4 + 
   {{\lambda }\over{2! d}} q_0^2 \Sigma_2+ 
   {{\lambda }\over  d} q_0 \Sigma_3+
   {{\lambda }\over d} \Sigma_4
.\]  
One can apply canonical quantization.
Following the Dirac-Bergman prescription, described in \cite{bpp97},
one identifies  first-class constraints which define the conjugate momenta
$ 
   0 = p_n - i k_n^+ q_{-n}\;, 
$,  
where $\left[q_m,p_n\right] = {{\delta_{n, m}}/ 2}$
and $m ,n \neq 0$.
The secondary constraint  
\begin{equation} 
  0 = \mu^2 q_0 
  + {{\lambda }\over{3! d}} q_0^3
  + {{\lambda }\over d}     q_0 \Sigma_2
  + {{\lambda }\over d}     \Sigma_3
\label{eq:eom_phi}\end{equation} 
determines the zero mode $q_0$. 
This result can also be obtained by integrating 
the equations of motion, 
$\partial_+\partial_-\phi + \mu^2 \phi = {\lambda \over 3!} \phi^3$.
To quantize the system one replaces the Dirac bracket by  
a commutator. One must choose
a regularization and an operator-ordering prescription 
in order to make the system well-defined.
One begins by defining creation and annihilation operators 
$a_k^\dagger$ and $a_k$,
\[ 
   q_k = \sqrt {d\over{4 \pi \left| k \right|}} \: a_k\;,
   \quad a_k = a_{-k}^\dagger\;,\quad  k\neq 0\; , 
\] 
which satisfy the usual commutation relations
$  
\left[a_k,a_l^\dagger\right] =\delta_{k, l}
$.  
Likewise, one defines the zero mode operator $q_0 = a_0 \sqrt{d/4\pi}$.
In the quantum case, one  normal orders the operator $\Sigma_n$. 
One redefines this the fields in terms of operators
\begin{equation}
   \phi\!\left(x\right) 
  = {1\over \sqrt{d} } \left(a_0 + \sum_{n=1}^\infty 
   \frac {1} {n}\left(
   a_{n} e^{-i \frac{\pi}{L} n x^-} +
   a_{n} ^\dagger e^{i \frac{\pi}{L} n x^-}
   \right)\right)
,\end{equation} 
and notes that they obey the canonical commutation relations
\begin{equation}
   \Big[\phi(x^+,x^-),\partial_-\phi(y^+,y^-)\Big] _{x^+=y^+=0} = 
   \delta(x^--y^-)
,\end{equation} 
for a boson field in the front form, see also Eq.(\ref{ceq:19a}).

The solution of the constraint Eq.(\ref{eq:eom_phi}) is very difficult.
Van de Sande and Pinsky \cite{piv94} have shown that
the zero mode $a_0$ could acquire finite values, 
\[
   \langle 0\vert \phi \vert 0\rangle = a_0 \neq 0
,\]
depending on the coupling constant.

One finds the following general behavior:  
for small coupling 
(large $g$, where  $g \propto 1/{\rm coupling}$) 
the constraint equation has a single solution and the field has no
vacuum expectation value (VEV). As one increases the coupling 
(decreases $g$) to  the ``critical coupling'' $g_{\rm critical}$, 
two  additional solutions which give the
field a nonzero VEV appear.   
These solutions differ only infinitesimally from the
first solution near the  critical coupling, 
indicating  the presence of a second order phase transition. 
Above the critical coupling ($g < g_{\rm critical}$), there
are three solutions:  one with zero VEV, the ``unbroken phase,'' 
and two with nonzero VEV, the ``broken phase''.   
The ``critical curves'' shown in Figure~\ref{ff1}, is a
plot of the VEV as a function of $g$.  

\section {DLCQ in 3+1 dimensions}

Periodic boundary conditions on ${\cal L} $ can be realized by
periodic boundary conditions on the vector potentials $A _\mu$
and anti-periodic boundary conditions on the spinor fields,
since ${\cal L} $ is bilinear in the $\Psi _\alpha$.
In momentum representation one expands these fields into plane
wave states $ e^{-ip _\mu x ^\mu} $,
and satisfies the boundary conditions by {\it discretized momenta}
\begin {eqnarray}
    p _- 
 &=& \cases{ {\pi\over \scriptstyle L} n ,
             &with $n =
              {\scriptstyle1\over\scriptstyle2},
              {\scriptstyle3\over\scriptstyle2},\dots,
              \infty\ $ for fermions, \cr
              {\pi\over L} n  ,
             &with $n = 1,2,\dots,\infty\ $ for bosons, \cr}
\nonumber \\
   \quad{\rm and}\ \vec p _{\!\bot}
 &=& {\pi\over L_{\!\bot}} \vec n _{\!\bot},
  \ \quad{\rm with} \ n_x,n_y = 0,\pm1,\pm2,\dots,\pm\infty
  \ \quad{\rm for}\ {\rm both} 
.\nonumber \end {eqnarray} 
As an expense, one has to introduce two artificial length 
parameters, $ L$ and $ L_{\!\bot}$. They also define
the normalization volume $ \Omega \equiv 2L(2L_{\!\bot})^2$.
More explicitly, the free fields are expanded as
\begin {eqnarray}
  \widetilde \Psi _\alpha (x)   = { 1\over \sqrt{\Omega } }
  \sum_q { 1\over \sqrt{ p^+} }
  \left(   b_q u_\alpha (p,\lambda) e^{ -ipx}
         + d^\dagger _q v_\alpha (p,\lambda) e^{ipx} \right) ,
\nonumber \\
  \quad{\rm and}\quad
  \widetilde A  _\mu (x) = { 1\over \sqrt{\Omega } }
  \sum_q { 1\over \sqrt{ p^+} }
  \left(   a _q \epsilon_\mu       (p,\lambda) e^{ -ipx}
         + a^\dagger _q \epsilon_\mu^\star (p,\lambda) e^{ipx} \right) 
,\label{enexdi}\end {eqnarray} %
particularly for the two transverse vector potentials
$ \widetilde A  ^i \equiv   \widetilde A  ^i _{\!\bot}$,
($i=1,2$).
As above, each single particle state `$q$' is specified by 
six quantum numbers, the three discrete momenta
$n, n _x, n _y$, helicity, color and flavor. 
The creation and destruction operators like $a^\dagger _q$ and 
$a _q$ create and destroy single particle states $q$, and obey
(anti-) commutation relations like
\[
  \big[  a _q, a^\dagger _{q^\prime} \big]  =
  \big\{ b _q, b^\dagger _{q^\prime} \big\} =
  \big\{ d _q, d^\dagger _{q^\prime} \big\} =
  \delta_{q,q^\prime}
.\]
The Kronecker symbol is unity only if all six quantum numbers
coincide. 

Inserting the free fields of Eq.(\ref{enexdi}) into the
Hamiltonian, one performs the space-like integrations
and ends up with the light-cone energy-momenta
$ P  ^\nu = P  ^\nu \big( a _q, a^\dagger _q,
        b _q, b^\dagger _q, d _q, d^\dagger _q \big) $
as operators acting in Fock space. 
The spatial components of $P^k$ are simple  and  diagonal, 
the temporal component $P^-$ is complicated and off-diagonal. 
The integrals over the coordinates $x^-$
are conveiently expressed in  Kronecker delta functions 
\[
    \delta (k^+|p^+)
= {1\over 2L}\int\limits_{-L}^{+L}\!dx^- e^{+i(k_--p_-)x^-}
= {1\over 2L}\int\limits_{-L}^{+L}\!dx^- e^{+i(n-m){\pi x^-\over L}}
= \delta _{n,m}
,\] 
and correspondingly for the transversal integrations.
In the tables given above they appear typically in the
overall factor 
\[ 
    \Delta(q_1;q_2,q_3,q_4) = 
    {g^2 \over 2\Omega }\ \delta (k^+_1| k^+_2 + k^+_3+k^+_4) \,
    \delta ^{(2)} (\vec k _{\!\perp 1} | \vec k _{\!\perp 2} +
                  \vec k _{\!\perp 3} + \vec k _{\!\perp 4} ) 
.\] 

\subsection{Retrieving the continuum formulation}
\label{sec:Retrieving}

The continuum formulation of the Hamiltonian problem in gauge 
field theory with its endless multiple integrals is usually 
cumbersome and untransparent. In DLCQ, the continuum
limit corresponds to harmonic resolution $K\rightarrow\infty$.
The compactified formulation with its simple multiple sums is 
straightforward. 
The key relation is the connection between sums and integrals
\begin{eqnarray}
      \int\! dk^+ f(k^+,\vec k _{\!\perp}) \Longleftrightarrow 
      {\pi\over 2L}\sum_{n} f(k^+,\vec k _{\!\perp})
,\nonumber \\
      \int\! d^2\vec k _{\!\perp}\  f(k^+,\vec k _{\!\perp}) 
      \Longleftrightarrow
      {\pi^2\over L^2_{\!\perp}}\sum_{n_{\!\perp}} f(k^+,\vec k _{\!\perp})
.\nonumber \end{eqnarray}
Combined they yield
\[ 
      \int\! dk^+ d^2\vec k _{\!\perp}\ f(k^+,\vec k _{\!\perp}) 
      \Longleftrightarrow
      {2(2\pi)^3\over\Omega}\sum_{n,n_{\!\perp}} f(k^+,\vec k _{\!\perp})
.\] 
Similarly, Dirac delta and Kronecker delta functions are
related by 
\[ 
      \delta(k^+)\ \delta^{(2)}(\vec k _{\!\perp}) 
      \Longleftrightarrow 
      {\Omega\over 2(2\pi)^3}\ \delta(k^+|0)\ \delta^{(2)}
      (\vec k _{\!\perp}|\vec0) 
.\] 
Because of  that, in order to satisfy the respective commutation 
relations, one must modify also the creation and destruction operators.
Denoting the single boson operators in the continuum by
$\widetilde a $ and in the discretized case by $a$, they 
must be related by
\[ 
      \widetilde a (q) \Longleftrightarrow 
      \sqrt{{\Omega\over 2(2\pi)^3}}\ a _q 
.\] 
and correspondingly for  fermion operators.
Of course, one has formally to replace sums by integrals, 
Kronecker delta by Dirac delta functions, and single particle
operators by their tilded versions.
In practice, it suffices to replace the tilded coupling constant
\[ 
      \widetilde g^2 = {g^2 \over 2\Omega} 
      \qquad{\rm by}\qquad
      \widetilde g^2 = {g^2 \over 4(2\pi)^3} 
\] 
in order to convert the discretized expressions in 
Tables~\ref{tab:verspi}-\ref{tab:seaspi} to the continuum 
formulation. 

The DLCQ method can be considered a general framework
for solving problems such as relativistic many-body theories
or approximate models. The general procedure is:
(1) Phrase the physics problem in DLCQ;
(2) Apply approximation and simplifications;
(3) Derive the final result;
(4) At the end convert th so obtained expressions to the continuum.

\subsection{Fock-space and vertex regularization} 

The finite number of Fock-space sectors is a consequence 
of the positivity of the longitudinal light-cone momentum $p^+$.
The transversal momenta $\vec p_{\!\bot}$ can take
either sign, and the number of Fock states within each sector
can be arbitrarily large.
In order to face a finite dimensional Hamiltonian matrix
one must have a finite number of Fock states, and this is 
achieved by {\em Fock space regularization}:
Following Lepage and Brodsky \cite{leb80}, a Fock state 
with $n$ particles is included only if its free invariant mass 
does not exceed a certain threshold 
\[ 
         (p_1+p_2+\dots p_n)^2 -        
         ( m_1+m_2+\dots m_n)^2
         \leq \Lambda_0 ^2 
.\] 
The sum extends over all $n$ particles in a Fock state.
The lowest possible value of  $M_0^2$ is taken when 
all particles are at rest relative to each other, {\it i.e.}
$\left(M_0^2\right)_{min}=\left(m_1+m_2+\dots m_n\right)^2$. 
This frozen invariant mass should be removed from the cut-off 
The mass scale $\Lambda_0 $ is a Lorentz scalar 
and one of the parameters of the theory.

However, it was not realized in the past \cite{brp91}, that 
Fock-space regularization is almost irrelevant in the 
continuum theory. {\em Vertex regularization} seem to be
a better alternative. 
At each vertex,  a particle with four-momentum 
$p^\mu$ is scattered into two particles with respective 
four-momentum $p_1^\mu$ and $p_2^\mu$. 
In order to avoid potential singularities 
one can {\em regulate the interaction}
by setting the matrix element to zero if the off-shell mass 
$(p_1+p_2)^2$ exceeds a certain scale  $\Lambda$.
The condition
\begin{eqnarray} 
      R(\Lambda) = \Theta \left( (p_1+p_2)^2-( m_1+m_2)^2 - \Lambda^2 \right)
\label{aeq:i1} 
\end{eqnarray} 
will be referred to as the sharp cut-off for {\em vertex regularization}. 

The dependence on the regularization parameter $\Lambda$
must be removed by the renormalization group.
Renormalization looks like a terrible problem
in the context of non-perturbative theory,
but with the above regularization scheme it could be simple 
in principle:
The eigenvalues may not depend on the regulator
scale(s) $\Lambda$.
To require this is easier than to find a practical realization.

\section{The many-body problem in gauge theory}

In principle one proceeds in 3+1 like in 1+1 dimensions: 
One selects a particular value of the harmonic
resolution $K$  and the cut-off $\Lambda$ and diagonalizes
the finite dimensional Hamiltonian matrix by numerical methods.
But the bottle neck of any Hamiltonian approach is that the 
dimension of the Hamiltonian matrix  increases exponentially 
fast with the cut-off.
As a concrete example consider the matrix structure 
as given for $K=4$ in Fig.~\ref{fig:holy-1}.
Suppose the regularization procedure allows for 10 discrete
momentum states in each direction. For every single particle 
one has about $10^3$ possibilities to define a momentum state.
A Fock-space sector with $n$ particles and fixed total momentum
has then roughly  $10^{n-1}$ different Fock states.
Sector 13 in Fig.~\ref{fig:holy-1} alone, with its 8 particles, 
has thus about  $10^{21}$ Fock states. 
Chemists are able to handle matrices with some $10^{7}$ 
dimensions, but $10^{21}$ dimensions exceeds the 
calculational capacity of any computer in the foreseeable future. 

For 3+1 dimensions one is thus confronted with a
a similar problem as in conventional many-body
physics, displayed in Fig.~\ref{fig-kyf-1}. 
One has to diagonalize finite matrices with exponentially 
large dimensions (typically $>10^{6}$).
In fact, the problem in quantum field theory is even more 
difficult since the particle number is unlimited. 
One needs an effective interaction which acts in
smaller matrix spaces and which has a well defined
relation to the full interaction.
One needs it also for the physical understanding. 
The effective interaction between 
two electrons, for example, is the Coulomb interaction, at least
to lowest order of approximation.

The goal is therefore to develop  
an exact effective interaction between a quark and an anti-quark
in a meson. 
To achieve this goal, the Hamiltonian DLCQ-matrix is discussed 
in terms of block matrices, since the Fock-space sectors 
appear quite naturally in a gauge theory, 
see also Eq.(\ref{eq:4.17}).
Each Fock-space sector has a finite number of Fock states,
which are kept track of collectively.
Eq.(\ref{eq:LC-hamiltonianII}) is therefore rewritten  
as a block matrix equation, 
with $H\equiv H_{LC}$ and $E\equiv M^2$,   
\begin {equation} 
      \sum _{j=1} ^{N} 
      \ \langle i \vert H \vert j \rangle 
      \ \langle i \vert \Psi\rangle 
      = E\ \langle i \vert \Psi\rangle 
\, \qquad {\rm for\ all\ } i = 1,2,\dots,N 
.\label {aeq:319}\end {equation} 
Rows and columns are denumerated in the same convention 
as in Table~\ref{tab:qcd_blocks}.
As to be shown, it can be mapped 
identically on a matrix equation which acts only 
in sector $\vert 1 \rangle$.
Once this is achieved, one can go to the continuum limit.
 
\subsection{The approach of Tamm and Dancoff}

Effective interactions are a well known tool in  many-body 
physics \cite{mof50}. In field theory
the method is known as the Tamm-Dancoff-approach, 
applied first by Tamm \cite{tam45} and by Dancoff \cite{dan50}, 
which shall be reviewed it in short.

The rows and columns of a matrix are split
into the $ P $- and the $Q$-space. 
In terms of the sector numbers of Eq.(\ref{aeq:319}), 
$P = \sum _{j=1} ^n  \vert j \rangle\langle j \vert $ and 
$Q = \sum _{j=n+1} ^N  \vert j \rangle\langle j \vert $,
where $1\leq n<N$. 
Eq.(\ref{aeq:319}) can thus be written as a 2 by 2 block matrix equation
\begin {eqnarray} 
   \langle P \vert H \vert P \rangle\ \langle P \vert\Psi\rangle 
 + \langle P \vert H \vert Q \rangle\ \langle Q \vert\Psi\rangle 
 &=& E \:\langle P \vert \Psi \rangle 
 ,  \label{aeq:321} \\ 
   \langle Q \vert H \vert P \rangle\ \langle P \vert\Psi\rangle 
 + \langle Q \vert H \vert Q \rangle\ \langle Q \vert\Psi\rangle 
 &=& E \:\langle Q \vert \Psi \rangle 
. \label{aeq:322}\end {eqnarray}
If one can invert the quadratic matrix 
$ \langle Q\vert E -  H \vert Q \rangle $ 
one could express the Q-space  
in terms of the $ P $-space wavefunction. 
But here is a problem:   
The eigenvalue $ E $ is unknown at this point. 
But one can replace it by another number, 
{\em the starting point energy} $\omega$,
which is at first a free parameter.
The matrix inverse to 
$ \langle Q\vert \omega -  H \vert Q \rangle $ 
is called the resolvent of the Hamiltonian
in the $Q$-space and is denoted by $G _ Q (\omega)$.
The $Q$-space wave function becomes then
\begin{eqnarray} 
   \langle Q \vert \Psi (\omega)\rangle    = 
   G _ Q (\omega) \langle Q \vert H \vert P \rangle 
   \,\langle P \vert\Psi\rangle 
,\qquad 
   G _ Q (\omega) =  
   {1\over\langle Q \vert\omega- H \vert Q \rangle} 
.\label{aeq:332} \end{eqnarray} %
Substituting it in Eq.(\ref{aeq:321}) produces an 
eigenvalue equation in the $P$-space,
\begin{equation} 
      H _{\rm eff} (\omega ) 
      \vert P\rangle\,\langle P \vert\Psi_k(\omega)\rangle =
      E _k (\omega )\,\vert\Psi _k (\omega ) \rangle 
,\label{aeq:345}\end{equation} 
and defines the effective $P$-space Hamiltonian
\[ 
      H _{\rm eff} (\omega) =  H +
      H \vert Q \rangle\,G_Q(\omega)\,\langle Q\vert H 
.\] 
Varying $\omega$ one generates a set of  
{\em energy functions} $ E _k(\omega) $. 
Every solution of the {\em fix-point equation} 
\[ 
E _k (\omega ) = \omega 
,\] 
generates one of the eigenvalues $H$, 
in fact, it generates all of them.

\def\d{D} \def\v{V} \def\b{$\cdot$} \def\s{S} \def\f{F} 
\def\g{g\phantom{\bar q}}
\begin {table}[t]
\begin{center}
\begin {tabular}  {||lc||ccccccccccccc||}
\hline \hline 
\rule[-1ex]{0ex}{4ex} 
 Sector & n & 
     1 & 2 & 3 & 4 & 5 & 6 & 7 & 8 & 9 &10 &11 &12 &13 
\\ \hline \hline   
 $ q\bar q\, $ &  1 & 
    \d &\s &\v &\f &\b &\f &\b &\b &\b &\b &\b &\b &\b 
\\ 
 $ \g\,\g$ &  2 & 
    \s &\d &\v &\b &\v &\f &\b &\b &\f &\b &\b &\b &\b 
\\
 $ q\bar q\, \g $ &  3 & 
    \v &\v &\d &\v &\s &\v &\f &\b &\b &\f &\b &\b &\b 
\\
 $ q\bar q\, q\bar q\, $ &  4 & 
    \f &\b &\v &\d &\b &\s &\v &\f &\b &\b &\f &\b &\b 
\\ 
 $ \g\,\g\,\g $ &  5 & 
    \b &\v &\s &\b &\d &\v &\b &\b &\v &\f &\b &\b &\b 
\\
 $ q\bar q\,\g\,\g $ &  6 & 
    \f &\f &\v &\s &\v &\d &\v &\b &\s &\v &\f &\b &\b 
\\
 $ q\bar q\, q\bar q\,\g $ &  7 & 
    \b &\b &\f &\v &\b &\v &\d &\v &\b &\s &\v &\f &\b 
\\
 $ q\bar q\, q\bar q\, q\bar q\, $ &  8 & 
    \b &\b &\b &\f &\b &\b &\v &\d &\b &\b &\s &\v &\f 
\\ 
 $ \g\,\g\,\g\,\g $ &  9 & 
    \b &\f &\b &\b &\v &\s &\b &\b &\d &\v &\b &\b &\b 
\\
 $ q\bar q\,\g\,\g\,\g $ & 10 & 
    \b &\b &\f &\b &\f &\v &\s &\b &\v &\d &\v &\b &\b 
\\
 $ q\bar q\, q\bar q\,\g\,\g $ & 11 & 
    \b &\b &\b &\f &\b &\f &\v &\s &\b &\v &\d &\v &\b 
\\
 $ q\bar q\, q\bar q\, q\bar q\,\g $ & 12 & 
    \b &\b &\b &\b &\b &\b &\f &\v &\b &\b &\v &\d &\v 
\\
 $ q\bar q\, q\bar q\, q\bar q\, q\bar q\, $ & 13 & 
    \b &\b &\b &\b &\b &\b &\b &\f &\b &\b &\b &\v &\d 
\\ \hline\hline
\end {tabular}
\caption [masses] {\label {tab:qcd_blocks} \sl
   The Fock-space sectors and the Hamiltonian block
   matrix structure for QCD.
   Diagonal blocs are marked by  $D$.
   Off-diagonal blocks are labeled by $V$, $F$ and $S_6$, 
   corresponding to vertex, fork and seagull interactions, 
   respectively.  
   Zero-matrices are denoted by dots. 
   {\it Taken from} \cite{pau96}.
    See also Fig.~\protect{\ref{fig:holy-1}}.} 
\end{center}
\end {table}

If one identifies the $P$- with the $q\bar q$-space
one seems to have found the effective 
interaction which acts in the Fock space of a single quark
and a single anti-quark. 
It looks as if one has mapped a difficult problem, the 
diagonalization of a big matrix onto a simpler problem,  
the diagonalization of a small matrix. 
But the price to pay is to invert a matrix. 
Matrix inversion takes about the 
same numerical effort as its diagonalization. 
In view of having to vary $\omega$, the numerical work 
is therefore rather larger than smaller as compared 
to a direct diagonalization. 
The advantage of working with a resolvent is 
of analytical nature to the extent that resolvents can be
approximated systematically.  The two resolvents
\begin{eqnarray} 
     G _Q (\omega) =  {1\over \langle Q \vert 
           \omega - T - U  \vert Q \rangle} 
,\quad{\rm and}\quad 
     \widetilde G   (\omega) = {1\over \langle Q \vert 
           \omega - T \vert Q \rangle} 
,\label{aeq:352}\end{eqnarray} 
defined once with and once without the off-diagonal 
interaction $U$, are identically related by 
$G_Q=\widetilde G  +\widetilde G  U G_Q $, or by
the infinite series of (Tamm-Dancoff) perturbation theory 
\begin{equation}  
    G _Q =
    \widetilde G  + \widetilde G  U \widetilde G  +
    \widetilde G  U \widetilde G  U \widetilde G  + \dots 
,\label{aeq:m24}\end{equation} 
see also Eq.(\ref{eq:spec-decom}). 
The free resolvent $\widetilde G $ can be obtained trivially since 
the kinetic energy $T$ is diagonal.
Conceptually, it is the same object as the one in 
Eq.(\ref{eq:spec-decom}), except for two aspects:
The present $\widetilde G $ acts only in the $Q$-space,
and the starting point energy $\omega$ is a {\em constant} 
(one of the eigenvalues); the free energy $\epsilon$ 
in (\ref{eq:spec-decom}), however, is a 
{\em function of the incoming momenta}.
The starting point energy not being a kinetic
energy creates problems allover the place, 
since non-integrable singularities 
appear in every order of (Tamm-Dancoff) perturbation theory. 
Essentially two conclusions are possible:
Either gauge theory has no bound-state solution, or
the series in Eq.(\ref{aeq:m24}) has to be resumed 
to all orders of perturbation
theory before the singularities begin cancel eachother.

In practice, Tamm and Dancoff \cite{tam45,dan50} have restricted 
themselves to the first non-trivial order.
In order to make things work, 
they have replaced the energy
denominator with the eigenvalue $\omega$ by the energy
denominator with the function $\epsilon$, 
as it appears in the perturbative scattering amplitudes.
The same trick was applied in the later work with the 
front form \cite{kpw92,trp96}.

Perturbation theory within a bound state problem is known
to be a very difficult question. It has  motivated 
the formal work in the next few sections.
What we are after is the impossible, some kind of 
non-perturbative perturbation theory!

\subsection {The method of iterated resolvents}

The Tamm-Dancoff approach can be interpreted as the reduction 
of a block matrix dimension from 2 to 1. 
But having a matrix with block matrix dimension $N=13$
as in Table~\ref{tab:qcd_blocks}, it could be interpreted
as the  reduction from $N\rightarrow N-1$, simply by choosing the
$Q$-space appropriately. But then the procedure can be iterated,
one can reduce the block matrix dimension
from $N-1\rightarrow N-2$, and so on until one arrives at
$2\rightarrow 1$. 
This method of `iterated of resolvents' 
has certain advantages, which will be discussed  
as the formalism develops.

First, one needs a reasonable and compact notation.
One of them is to denumerate the Fock space sectors as 
in Table~\ref{tab:qcd_blocks}.
The full Hamiltonian will be denoted by $H\equiv H_{N}$, 
since by the definition in Eq.(\ref{aeq:319}) it has $N$ blocks.
Suppose, during the reduction 
one has arrived at block matrix dimension $n$, 
with $1 < n\leq N$.  
The eigenvalue problem corresponding to  
Eq.(\ref{aeq:345}) reads then
\[ 
   \sum _{j=1} ^{n} \langle i \vert H _n (\omega)\vert j \rangle 
                    \langle j \vert\Psi  (\omega)\rangle 
   =  E (\omega)\ \langle i \vert\Psi (\omega)\rangle 
,\qquad{\rm for}\ i=1,2,\dots,n
.\] 
In analogy to Eq.(\ref{aeq:332}), define the $n$-space resolvent, and get
\begin{eqnarray} 
   \langle n \vert \Psi (\omega)\rangle   
   = G _ n (\omega) 
   \sum _{j=1} ^{n-1} \langle n \vert H _n (\omega)\vert j \rangle 
   \ \langle j \vert \Psi (\omega) \rangle 
,\quad 
    G _ n (\omega)   
    =   {1\over \langle n \vert\omega- H_n (\omega)\vert n \rangle} 
.\label{aeq:410}\end{eqnarray}
The effective interaction 
in the  ($n -1$)-space becomes then 
\begin {equation}  
       H _{n -1} (\omega) =  H _n (\omega)
  +  H _n(\omega) G _ n  (\omega) H _n (\omega)
\label {aeq:414} \end {equation}
for every block matrix  element. 
It is unpleasant but unavoidable, that the symbol $n$ denotes both 
the block matrix dimension and the number of the last sector.
Else one proceeds like for Tamm-Dancoff, 
including the fixed point equation  $ E  (\omega ) = \omega $.
But one has achieved much more: Eq.(\ref{aeq:414}) is a 
{\em recursion relation}!

A few comments seem to be in order.
The method of iterated resolvents \cite{pau96,pau98}
is particularly suited for gauge theory with 
its many zero block matrices, see Table~\ref{tab:3}.
The zero matrices remove
many of the multiplications in Eq.(\ref{aeq:410}).
The Tamm-Dancoff procedure cannot make use of them.~--
Both the Tamm-Dancoff procedure and the iterated resolvents
can be put on a computer in model studies.
The iterated resolvents
require to invert several smaller matrices insted of one big one.
Since matrix diagonalization (and inversion) grows with
power 3 in the dimension, the technique of iterated resolvents
might thus even be faster.~-- 
The iterated and the Tamm-Dancoff resolvents are distinctly 
different in the following aspect:
$G _n$ conserves particle number, but $G _Q$ does not.
Both however conserve the incoming momentum.~--
The notation in Eq.(\ref{aeq:414}) is very compact and will
be explained further below.~--
The method applies also equally well to conventional 
many-body problems since a pair-interaction generates for example 
the bloc matrix structure shown in Figure~\ref{fig-kyf-1}.~--
Finally, it should be emphasized that the higher sector wavefunctions
can be retrieved by matrix multiplications from the eigenfunction 
in the lowest sector, from $\langle 1 \vert \Psi\rangle$.
No additional matrix diagonalizations or inversions are required.
To show this, consider Eq.(\ref{aeq:410}) for $n=1$, {\it i.e.}
\[ 
   \langle 2 \vert \Psi \rangle =
   \langle 2 \vert G_2 H_2 \vert 1 \rangle\,\langle 1 \vert \Psi\rangle
.\] 
Both $G_2$ and $H_2$ were calculated as a matrix for getting down to 
the effective interaction in the $1$-space. Required is thus
one additional matrix multiplication. Next get
$  \langle 3 \vert \Psi \rangle =
   \langle 3 \vert G_3 H_3 \vert 1 \rangle\,\langle 1 \vert \Psi\rangle +
   \langle 3 \vert G_3 H_3 \vert 2 \rangle\,\langle 2 \vert \Psi\rangle 
$. 
Substituting $\langle 2 \vert \Psi \rangle$ gives
\[ 
   \langle 3 \vert \Psi \rangle =
   \langle 3 \vert G_3 H_3 (1+G_2 H_2 ) 
   \vert 1 \rangle\,\langle 1 \vert \Psi\rangle
.\] 
The general case,
\begin{equation}
   \langle n \vert \Psi \rangle =
   \langle n \vert G_n H_n (1+G_{n-1} H_{n-1}) \dots (1+G_2 H_2 ) 
   \vert 1 \rangle\,\langle 1 \vert \Psi\rangle
,\label{eq:sector-wave}\end{equation}
can be proven by induction.

\subsection{A simple numerical example}

It might be interesting to study a Hamiltonian with 
a tridigonal band structure, such as was given for example in 
Table~\ref{tab:qed1+1}:
\begin{eqnarray}
   \pmatrix{
                \langle 1\vert 1+S \vert 1\rangle 
              & \langle 1\vert F \vert 2\rangle  
              & . 
              & . 
\cr
                \langle 2\vert F \vert 1\rangle 
              & \langle 2\vert 1+S \vert 2\rangle  
              & \langle 2\vert F \vert 3\rangle 
              &  . 
\cr
                . 
              & \langle 3\vert F \vert 2\rangle  
              & \langle 3\vert 1+S \vert 3\rangle 
              & \langle 3\vert F \vert 4\rangle  
\cr
                . 
              & . 
              & \langle 4\vert F \vert 3\rangle 
              & \langle 4\vert 1+S \vert 4\rangle  
\cr}
.\label{eq:example-matrix}\end{eqnarray}
According to the rules it develops
the structure a of continued fraction, since with
$G_n(\omega) = 1/(\omega - H_n)$ one has explicitly
\begin{equation}
\begin{array} {lllr}
    H_4 &= \langle 4\vert 1+S \vert 4\rangle , 
   &\langle 4\vert\Psi\rangle =  &\langle 4\vert G_4 F G_3 F G_2 F
    \vert 1\rangle\,\langle 1\vert\Psi\rangle , 
\\
    H_3 &= \langle 3 \vert 1+S +  F G_4 F \vert 3\rangle ,
   &\langle 3\vert\Psi\rangle =  &\langle 3\vert G_3 F G_2 F
    \vert 1\rangle\,\langle 1\vert\Psi\rangle , 
\\
    H_2 &= \langle 2 \vert 1+S +  F G_3 F \vert 2\rangle ,
   &\langle 2\vert\Psi\rangle =  &\langle 2\vert G_2 F
    \vert 1\rangle\,\langle 1\vert\Psi\rangle , 
\\
    H_1 &= \langle 1\vert 1+S + R G_2 R \vert 1\rangle , 
   &\langle 1\vert\Psi\rangle.  & 
\end{array}
\label{eq:cont-fraction}\end{equation}
Here and above a very compact notation is used,
which is explained by the example
\[ 
   \langle 3\vert D + V  G_4  V \vert 3\rangle
   =
   \langle 3\vert D   \vert 3\rangle + 
   \langle 3\vert F   \vert 4\rangle\,
   \langle 4\vert G_4 \vert 4\rangle\,
   \langle 4\vert F   \vert 3\rangle
.\] 
By reasons of space, the abbrevation 
$\langle i\vert D \vert i\rangle = 
 \langle i\vert T + S \vert i\rangle$
is used sometimes in the diagonal blocks.
The kinetic energies $T$ are diagonal matrix elements,
and the seagulls do not change particle number by definition.
Even that the above is very compact. Here is what it means in terms of
matrix operations: 
\begin{eqnarray}
   \langle 3,i\vert D + F  G_4  F \vert 3,i'\rangle &=&
   \langle 3,i\vert D   \vert 3,i'\rangle 
\nonumber\\ &+& \sum_{j,k}
   \langle 3,i\vert F   \vert 4,j \rangle\,
   \langle 4,j\vert G_4 \vert 4,k \rangle\,
   \langle 4,k\vert F   \vert 3,i'\rangle
.\nonumber\end{eqnarray}
Within each sector $\vert i\rangle$, the number of basis or Fock states is
finite, {\it i.e.} $\vert i\rangle \equiv \vert i;j\rangle$ with
$j=1,2,\,\dots\, ,N_i$.
The steps to be taken are then:
Take the reactangular block matrix 
$\langle 4,k\vert F   \vert 1,i'\rangle$;
matrix-multiply it from the left with the inverse matrix of 
$\langle 4\vert H \vert 4 \rangle$ which is
$\langle 4,j\vert G_4 \vert 4,k \rangle$; 
matrix-multiply the result with the transpose (and complex conjugated) 
matrix $\langle 4,j\vert F^+\vert 1,i\rangle$;
the result is the quadratic matrix 
$\langle 3,i\vert F  G_4  F \vert 3,i'\rangle$;
add this to the quadratic matrix 
$\langle 3,i\vert D\vert 3,i'\rangle$.
As a net result, the old matrix 
$\langle 3,i\vert D\vert 3,i'\rangle$
in Eq.(\ref{eq:4x4_matrix}) is
replaced the effective (and $\omega$-dependent) matrix 
$\langle 1,i\vert \overline D (\omega)\vert 1,i'\rangle$.

\begin{figure} [t]
\begin{minipage}[t]{60mm} \makebox[0mm]{}
\centering
 \epsfxsize=59mm\epsfbox{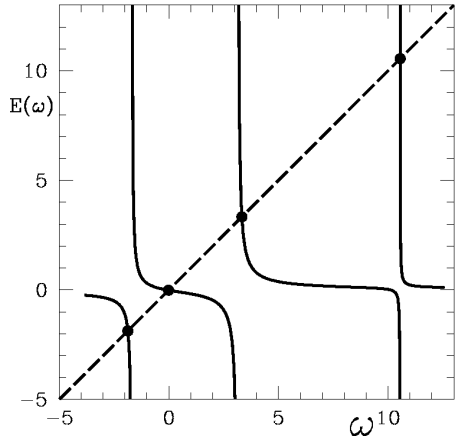}
\caption{\label{fig:4_1}  \sl
    The the energy function 
    $E(\omega) $ is plotted versus $\omega$. 
    The dotted line represents $E=\omega$. 
}\end{minipage}
\hfill
\begin{minipage}[t]{82mm} \makebox[0mm]{}
    {\sl 
    The $4\times4$ matrix
    $$\pmatrix{ 0 & 1 & . & . \cr 
                1 & 2 & 3 & . \cr
                . & 3 & 4 & 5 \cr 
                . & . & 5 & 6 \cr } $$
    has eigenvalues $E_i=$ 
    -1.872, -0,01518, 3.333, and 10.55.
    Its energy function is
    $$E(\omega) =  
    {\displaystyle 1\cdot 1  \over \displaystyle
     \omega - 2 - 
    {\displaystyle 3 \cdot 3 \over \displaystyle
     \omega - 4 - 
    {\displaystyle 5\cdot 5 \over \displaystyle
     \omega - 6 } } }.$$
    The solutions of  $ E(\omega) =  \omega$ agree
    with $E_i$ to within computer accuracy.
    The are marked with $\bullet$ in the figure. 
}\end{minipage}
\end{figure}

To play the game a little further, one can reduce
the number of states within a block to 1.
Eq.(\ref{eq:example-matrix}) stands then for a tri-diagonal matrix
and Eq.(\ref{eq:cont-fraction}) stands literally for a 
continued fraction. 
Such one is analyzed in Figure~\ref{fig:4_1},
with a perfect agreement between the
method of iterated resolvents and a direct numerical 
diagonalization.

\subsection{The 4 by 4 block matrix for gauge theory}

In the front form, the Fock-space for harmonic resolution
$K=2$ has only 4 Fock-space sectors, see Table~\ref{tab:qcd_blocks}.
$K=2$ provides thus a good example for all 4 by 4 hermitean bloc matrices,
in which the block matrix element $\langle 2\vert H \vert 4\rangle$ vanishes.
The case is considered as an exercise and a non-trivial example.

Since $N=4$ one starts out with the block matrix
\begin{eqnarray}
   \bordermatrix{
    & 1=(q\bar q) 
    & 2=(gg) 
    & 3=(q\bar q\,g) 
    & 4=(q\bar q\,q\bar q) 
\cr
  1=(q\bar q) & \langle 1\vert D \vert 1\rangle 
              & \langle 1\vert S \vert 2\rangle  
              & \langle 1\vert V \vert 3\rangle 
              & \langle 1\vert F \vert 4\rangle  
\cr
  2=(gg)          & \langle 2\vert S \vert 1\rangle 
                  & \langle 2\vert D \vert 2\rangle  
                  & \langle 2\vert V \vert 3\rangle 
                  &  0 
\cr
  3=(q\bar q\,g)  & \langle 3\vert V \vert 1\rangle 
                  & \langle 3\vert V \vert 2\rangle  
                  & \langle 3\vert D \vert 3\rangle 
                  & \langle 3\vert V \vert 4\rangle  
\cr
  4=(q\bar q\,q\bar q) & \langle 4\vert F \vert 1\rangle 
                       & 0 
                       & \langle 4\vert V \vert 3\rangle 
                       & \langle 4\vert D \vert 4\rangle  
\cr}
.\label{eq:4x4_matrix}\end{eqnarray}
The block matrix element $\langle 2\vert H_{\rm LC} \vert 4\rangle$ 
is a rectangular zero-matrix, 
since the light-cone operator has no matrix elements between
two gluons $gg$ and two $q\bar q$-pairs.
The notation keeps track of the field theoretic property, that 
each block matrix has only one type of interactions, namely
instaneneous  seagull (S) and fork (F), 
or dynamic vertex (V) interactions.

The Hamiltonian in the $4$-sector is quadratic and has the resolvent 
\begin{eqnarray}
    G_4(\omega) = \frac {1} {\langle 4\vert \omega - H_4 \vert 4\rangle}
    ,\qquad
    \langle 4\vert H_4 \vert 4\rangle = \langle 4\vert T + S \vert 4\rangle
.\label{eq:H_4}\end{eqnarray}
Using Eq.(\ref{aeq:414}) to reduce the block
matrix dimension from 4 to 3 gives
\begin{eqnarray}
   \bordermatrix{
    & 1=(q\bar q) 
    & 2=(gg) 
    & 3=(q\bar q\,g) 
\cr
      1=(q\bar q) 
    & \langle 1\vert D + F G_4 F \vert 1\rangle
    & \langle 1\vert S           \vert 2\rangle  
    & \langle 1\vert (1+ F G_4) V \vert 3\rangle
\cr
      2=(gg) 
    & \langle 2\vert S \vert 1\rangle 
    & \langle 2\vert D \vert 2\rangle  
    & \langle 2\vert V \vert 3\rangle 
\cr
      3=(q\bar q\,g) 
    & \langle 3\vert V (1+G_4 F) \vert 1\rangle
    & \langle 3\vert V           \vert 2\rangle  
    & \langle 3\vert D + V G_4 V \vert 3\rangle
\cr}
\nonumber\end{eqnarray}
Almost every block matrix element is replaced 
by an `effective' element.

Now continue to reduce from 3 to 2:
The resolvent of $\langle 3 \vert H\vert 3\rangle $ is now
\[ 
    G_3(\omega) = \frac {1} {\langle 3\vert \omega - H_3 \vert 3\rangle}
    ,\quad
    \langle 3 \vert H_3 \vert 3\rangle = 
    \langle 3 \vert T + S +  V G_4 V \vert 3\rangle
.\] 
With 
$\langle 1\vert A\vert 1\rangle \equiv 
 \langle 1\vert T + S + F G_4 F + (1+F G_4) V G_3 V(1+G_4 F) \vert 1\rangle$
one gets
\begin{eqnarray}
   \bordermatrix{
    & 1=(q\bar q) 
    & 2=(gg) 
\cr
      1=(q\bar q) 
    & \langle 1\vert A \vert 1\rangle 
    & \langle 1\vert S + (1 + F G_4) V G_3  V  \vert 2\rangle  
\cr
      2=(gg) 
    & \langle 2\vert S +  V G_3 V (1 + G_4 F) \vert 1\rangle 
    & \langle 2\vert T + S + V G_3 V  \vert 2\rangle  
\cr}
\nonumber
\end{eqnarray}
Finally, evaluate the last resolvent   
\[ 
    G_2(\omega) = \frac {1} {\langle 2\vert \omega - H_2 \vert 2\rangle}
    ,\quad
    H_2 = \langle 2 \vert T + S + V G_3 V \vert 2\rangle
,\] 
to end up with the 1 by 1 block matrix 
\begin{eqnarray}
    \langle 1\vert H_1 \vert 1\rangle &=& \langle 1\vert 
    D + F G_4 F + (1+ F G_4) V G_3 V (1+G_4 F) \vert 1\rangle
\\ 
    &+& \langle 1\vert 
    (S + V G_3 V  + F G_4 V G_3 V )\,G_2\,(S + V G_3 V + V G_3 V G_4 F )
    \vert 1\rangle
.\nonumber\end{eqnarray}
Here the procedure stops.
For gauge theory one can order the result 
with the power of the coupling constant
\begin{eqnarray}
    \langle 1\vert  H_1 \vert 1\rangle &=& \langle 1\vert T +
    \overline {V G_3 V} + 
    \overline {V G_3 V G_2 V G_3 V} 
    \vert 1\rangle 
\nonumber\\ &+&
    \langle 1\vert \,
    S G_2 V G_3 V G_4 F +
    F G_4 V G_3 V G_2 S +
\nonumber\\ &\phantom{+}& 
    \phantom{\langle 1\vert}
    V G_3 V G_2 V G_3 V G_4 F +
    F G_4 V G_3 V G_2 V G_3 V +
\nonumber\\ &\phantom{+}& 
    \phantom{\langle 1\vert}
    F G_4 V G_3 V G_4 F +
    F G_4 V G_3 V G_2 V G_3 V G_4 F
    \,\vert 1\rangle  
.\label{eq:4-interaction}\end{eqnarray}
with the abbreviations 
\begin{eqnarray}
   \overline {V G_3 V} &=& V G_3 V + S
,\nonumber\\  
   \overline {V G_3 V G_2 V G_3 V} &=& 
   V G_3 V G_2 V G_3 V + S G_2 V G_3 V + V G_3 V G_2 S 
\nonumber\\ &+&
   F G_4 V G_3 V  + V G_3 V G_4 F + S G_2 S + F G_4 F 
.\label{eq:overline-v}\end{eqnarray}
Finally, for completeness, the higher sector wave function
in Eq.(\ref{eq:sector-wave}) 
are written out explicitely 
\begin {eqnarray} 
     \langle 2\vert\Psi\rangle &=& 
     G_2 \left[ \overline {V G_3 V} + V G_3 V G_4 F \right]
    \vert 1 \rangle\,\langle 1 \vert \Psi\rangle
,\\ 
    \langle 3\vert\Psi\rangle  &=& 
     G_3 \left[V + \overline { V G_2 V G_3 V} + V G_2 V G_3 V G_4 F 
         \right] 
    \vert 1 \rangle\,\langle 1 \vert \Psi\rangle
,\\ 
    \langle 4\vert\Psi\rangle  &=&
    G_4 \left[ \overline{\overline {V G_3 V}} + 
    \overline{\overline {V G_3 V G_2 V G_3 V}} +
    V G_3 V G_2 V G_3 V G_4 F 
    \right]
    \vert 1 \rangle\,\langle 1 \vert \Psi\rangle
,\label{eq:wavefunction}\end{eqnarray}
with the additional abbreviations
\begin{eqnarray}
   \overline{\overline {V G_3 V}} &=& V G_3 V + F
\nonumber\\  
   \overline { V G_2 V G_3 V} &=& 
   V G_2 V G_3 V + V G_4 F + V G_2 S 
\nonumber\\ 
   \overline{\overline {V G_3 V G_2 V G_3 V}} &=& 
   V G_3 V G_2 V G_3 V + V G_3 V G_4 F + V G_3 V G_2 S
.\label{eq:overline-w}\end{eqnarray}
These abbreviations are not very relevant for the general case.
But for gauge theory they have a deeper physical meaning 
as to be discussed further below.

\begin{figure} [t]
\begin{minipage} {45mm} \makebox[0mm]{}
 \epsfxsize=40mm\epsfbox{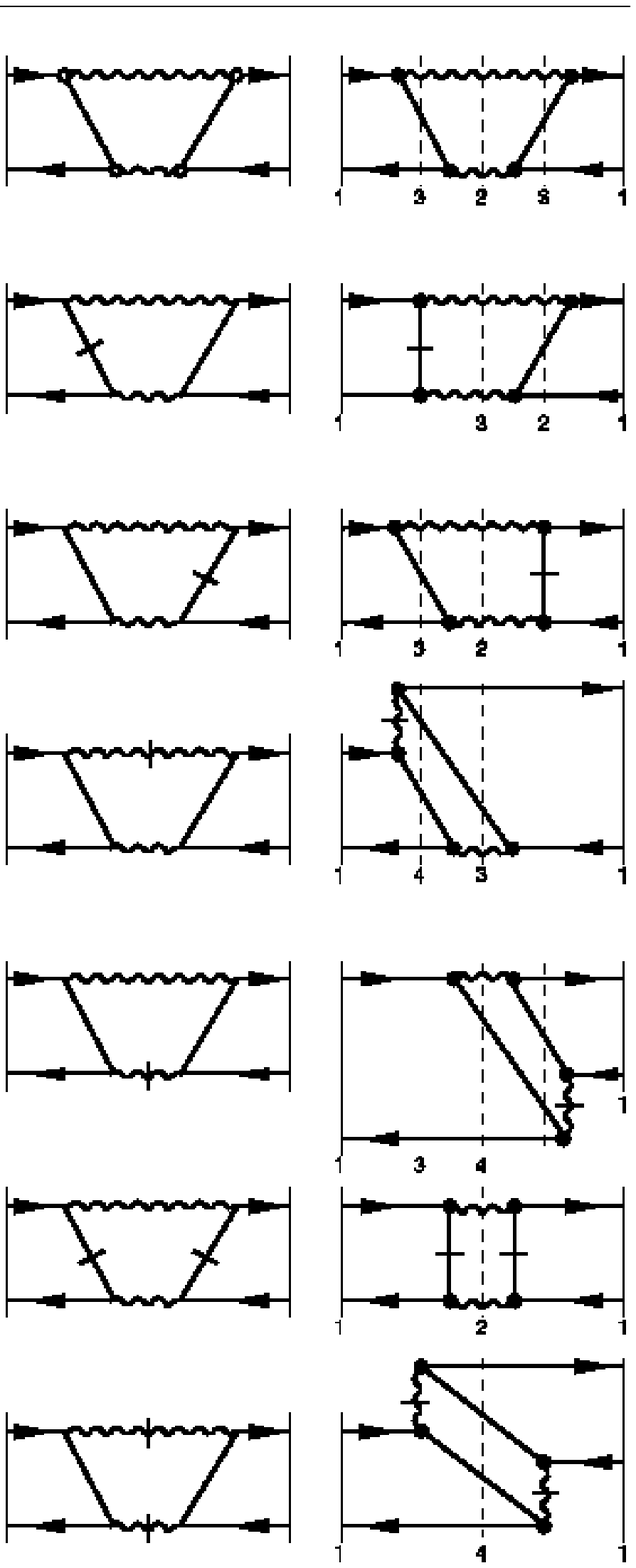} 
\makebox[0mm]{}
\end{minipage} 
\hfill
\begin{minipage} {100mm} \makebox[0mm]{}
\caption{\label{fig:6_6} \sl
A particular time-ordering of the two-gluon-annihilation 
interaction is drawn in the upper left of the figure.
If all intrinsic lines are understood as 
a sum of a `dynamical' and an 
` instantaneous' line, one gets the seven diagrams
in the figure, drawn in two different conventions.
One observes that the very same strings appear 
in $\overline{VG_3 VG_2 VG_3 V}$ as defined by
Eq.(\protect{\ref{eq:overline-v})}.
}
\hskip10em
\centering 
 \epsfxsize=60mm\epsfbox{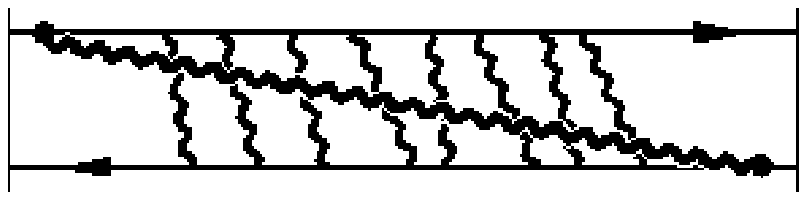} 
\caption{ \label{fig:gluons} \sl
Each propagator $G_n$
respresents an infinite resummation of Tamm-Dancoff
perturbative diagrams. The graph 
$\langle 1\vert VG_3V\vert 1\rangle$ respresents 
a resummation of all Tamm-Dancoff graphs 
with at least one $q\bar q$ pair, 
and where a first gluons is emitted by the quark 
and a (perhaps other) gluon is arbsorbed by the anti-quark. 
}\end{minipage}
\end{figure}

\subsection {Discussion and gauge invariant interactions}

All effective interactions have a \underline{finite}
number of \underline{finite} strings like 
$V\,GV\,\dots\,GV$.
This is a colossal simplification as compared 
infinite number of terms of Tamm-Dancoff perturbation theory.
Iterated resolvents resume them in a systematic way.
Both kinds of series can be identified term by term
if one expands the iterated resolvents also
about the kinetic energies.
Denoting the free resolvent by $\widetilde G_n$
it is related to the Tamm-Dancoff resolvent $\widetilde G$ by
\[ 
   \widetilde G_n = \sum_{i=1}^{N_n}\vert n,i \rangle 
   \frac {1}{\omega-T_{n,i}}\langle n,i\vert
   \quad ,\qquad
   \widetilde G_n = \sum_{n=1}^{N}\widetilde G_n  
\]
If one then expands $G_n$ using $H_n= T_n + U_n$
one gets 
\[
   G_n = \widetilde G_n + 
   \widetilde G_n U_n \widetilde G_n + 
   \widetilde G_n U_n \widetilde G_n U_n \widetilde G_n + 
   \dots
,\]
thus infinite series of infinite series, which after reordering
become identical with the Tamm-Dancoff series.

A look back to Eq.(\ref{eq:4-interaction}) is helpful.
The various term have been arranged there by order of
the coupling constant.
In the first line one finds the combination
$\overline {V G_3 V G_2 V G_3 V}$, 
defined in Eq.(\ref{eq:overline-v}).
In the caption to Figure~\ref{fig:6_6} it is explained
how one can simplify the formal procedures,
if one associates with each intrisic line a dynamic
plus an instantaneous line.
Doing that replaces the above sum of six terms
into a single one namely $V G_3 V G_2 V G_3 V$,
and similarily one replaces the term in the first line
by $V G_3 V + S\rightarrow V G_3 V$.

There is a physical reason behind that:
When calculating scattering diagrams to a fixed order
in perturbation theory, one obtains the gauge invariant
Feynman amplitudes only, if all instantaneous graphs
of the same order are included. An example for that was given above,
when calculating the scattering graph up to second order.
Obviously, the method of iterated resolvents accounts for this 
aspect automatically, in a formal way.

Since the remaining terms in Eq.(\ref{eq:4-interaction})
are of order 6 or 8 in the coupling constant,
one concludes that the restriction to the phase space
for $K=2$ satisfies gauge invariance only up to terms
of order 6 in the coupling constant. 
One can verify, that the order of violating gauge invariance
is pushed up if one includes more Fock-space sectors.
This rule was checked explicitly in many rather laborious 
calculations in \cite{pau96}. 

\begin{theorem}
An easy trick to achieve gauge invariance is the following:
Set to zero formally all instantaneous
interactions, perform the steps
required by the resolvents.
At the end, when having calculated
the effective interaction in terms of the vertex
interaction $V$, one replaces each internal line
by the sum of a dynamical and a instantaneous line,
according to the general rules of light-cone perturbation theory.
\end{theorem}

If one follows thes rules for the 4 by 4 matrix 
given in Eq.(\ref{eq:4x4_matrix})
one obtains for the sector Hamiltonians and wavefunctions simply
\begin {eqnarray} 
     H _4 &=&  T_4 , \quad
    \langle 4\vert\Psi\rangle =  \langle 4\vert
     G_4 V G_3 V +  G_4 V G_3 V G_2 V G_3 V 
     \vert 1\rangle\,\langle 1\vert\Psi\rangle
,\nonumber\\  
     H _3 &=&  T_3 + V G _4 V , \quad\hfill
    \langle 3\vert\Psi\rangle =  \langle 3\vert  
     G_3 V + G_3 V G_2 V G_3 V\vert 1\rangle\,\langle 1\vert\Psi\rangle
,\nonumber\\  
     H _2 &=&  T_2 + V G _3 V  , \quad
     \langle 2\vert\Psi\rangle =  \langle 2\vert 
     V G_3 V\vert 1\rangle\,\langle 1\vert\Psi\rangle
     ,\nonumber\\  
     H _1 &=&  T_1 + V G _3 V + V G _3 V  G _2 V G _3 V 
.\label{aeq:m36}\end {eqnarray}
These expressions are more transparent than
the full equations; they are also gauge invariant.

\subsection {The iterated resolvents for arbitrary $K$}

When one substitutes formally all instantaneous interactions
by block matrices with only zeros, keeping only
the vertex interactions $V$, one gets a block
matrix structure as displayed in Table~\ref{tab:3}.
Because of the many zero matrices,
the construction of the the sector Hamiltonians 
is now rather straightforward and simple.
Here they are for the first 12 sectors:
\begin{eqnarray}
    H_{12} 
&=& T_{12} +V G _{17} V + V G _{17} V  G _{16} V G _{12}V + V G _{13} V 
,\label{eq:U12}\\
    H_{11} 
&=& T_{11} +V G _{16} V + V G _{16} V  G _{15} V G _{16}V  + V G _{12} V 
,\\
     H_{10} 
&=&  T_{10} +V G _{15} V + V G _{15} V  G _{14} V G _{15}V +V G _{11} V  
,\label{aeq:622}\\ 
    H_9 
&=& T_{9} +V G _{10} V + V G _{14} V  
,\label{aeq:645}\\
    H_8 
&=& T_{8} + V G _{12} V + V G _{12} V  G _{11} V G _{12}V 
,\\
    H_7 
&=& T_{7} + V G _ {11} V+ V G _{11} V  G _{10} V G _ {11} V + V G _8 V 
,\\
     H_6 
&=& T_{6} +  V G _ {10} V+ V G _{10} V  G _9 V G _ {10} V + V G _7 V 
,\label{aeq:621}\\
    H_5 
&=& T_{5} + V G _ {6} V + V G _{9} V 
,\\
    H_4 
&=& T_{4} + V G _7 V + V G _7 V  G _6 V G _7 V 
,\\  
     H_3 
&=& T_{3} + V G _6 V + V G _6 V  G _5 V G _6 V + V G _4 V 
,\label{aeq:620}\\  
     H_2 
&=& T_{2} + V G _3 V + V G _5 V 
,\label{aeq:643}\\  
     H_1 
&=& T_{1} +  V G _3 V + V G _3 V  G _2 V G _3 V 
,\label{aeq:610} 
\end{eqnarray}
Actually, for $K=4$, the exact expressions 
corresponding to Table~\ref{tab:3} are obtained
by setting $G_{13}=1/(\omega - T_{13})$ and
$G _{n}=0 $ for $n\ge {14}$.
But it is easy to write down the expressions in
Eqs.(\ref{eq:U12})-(\ref{aeq:622}) valid for a higher value
of $K$. 
One also notes that they give 
all sector Hamiltonians for $K=2$ in Eqs.(\ref{aeq:m36}),
by setting formally $G _{n}=0 $ for $n\ge 5$.
\begin{theorem}
For sufficiently large harmonic resolution $K$
the formal expressions for effective Hamiltonian 
in sufficiently low sectors become 
independent of $K$.
\end{theorem}
One therefore can go to the limit $K\rightarrow\infty$
and thus to the continuum limit. 

\def\d{D} \def\v{V} \def\b{$\cdot$} \def\s{$\cdot$} \def\f{$\cdot$} 
\def\g{g\phantom{\bar q}}
\begin {table}[t]
\begin{center}
\begin {tabular}  {||lc||ccccccccccccc||}
\hline \hline 
\rule[-1ex]{0ex}{4ex} 
 Sector & n & 
     1 & 2 & 3 & 4 & 5 & 6 & 7 & 8 & 9 &10 &11 &12 &13 
\\ \hline \hline   
 $ q\bar q\, $ &  1 & 
    \d &\s &\v &\f &\b &\f &\b &\b &\b &\b &\b &\b &\b 
\\ 
 $ \g\,\g$ &  2 & 
    \s &\d &\v &\b &\v &\f &\b &\b &\f &\b &\b &\b &\b 
\\
 $ q\bar q\, \g $ &  3 & 
    \v &\v &\d &\v &\s &\v &\f &\b &\b &\f &\b &\b &\b 
\\
 $ q\bar q\, q\bar q\, $ &  4 & 
    \f &\b &\v &\d &\b &\s &\v &\f &\b &\b &\f &\b &\b 
\\ 
 $ \g\,\g\,\g $ &  5 & 
    \b &\v &\s &\b &\d &\v &\b &\b &\v &\f &\b &\b &\b 
\\
 $ q\bar q\,\g\,\g $ &  6 & 
    \f &\f &\v &\s &\v &\d &\v &\b &\s &\v &\f &\b &\b 
\\
 $ q\bar q\, q\bar q\,\g $ &  7 & 
    \b &\b &\f &\v &\b &\v &\d &\v &\b &\s &\v &\f &\b 
\\
 $ q\bar q\, q\bar q\, q\bar q\, $ &  8 & 
    \b &\b &\b &\f &\b &\b &\v &\d &\b &\b &\s &\v &\f 
\\ 
 $ \g\,\g\,\g\,\g $ &  9 & 
    \b &\f &\b &\b &\v &\s &\b &\b &\d &\v &\b &\b &\b 
\\
 $ q\bar q\,\g\,\g\,\g $ & 10 & 
    \b &\b &\f &\b &\f &\v &\s &\b &\v &\d &\v &\b &\b 
\\
 $ q\bar q\, q\bar q\,\g\,\g $ & 11 & 
    \b &\b &\b &\f &\b &\f &\v &\s &\b &\v &\d &\v &\b 
\\
 $ q\bar q\, q\bar q\, q\bar q\,\g $ & 12 & 
    \b &\b &\b &\b &\b &\b &\f &\v &\b &\b &\v &\d &\v 
\\
 $ q\bar q\, q\bar q\, q\bar q\, q\bar q\, $ & 13 & 
    \b &\b &\b &\b &\b &\b &\b &\f &\b &\b &\b &\v &\d 
\\ \hline\hline
\end {tabular}
\end {center}
\caption {\label{tab:3} \sl
The Fock space and the Hamiltonian matrix $H^\prime =T+V$ 
for a meson at fixed value of $K=4$.~--- 
See discussion in the text.
The diagonal blocks are denoted by $T$. 
Most of the block matrices 
are zero matrices, marked by a dot ($\cdot$).
The block matrices marked by $V$ are potentiall non-zero 
due to the vertex interaction. 
}\vspace{1em}
\end {table}

\subsection{Propagation in medium}

Here seems to be a problem:
For calculating $G_3$ one needs 
$G_6$, $G_5$ and $G_4$, 
for calculating $G_6$ one needs 
$G_{10}$, $G_9$ and $G_7$, and so on. 
In the next section will be shown how the hierarchy can
be broken in a rather effective way.
That final step will be comparatively simple if one has analyzed 
the propagators for the sectors with one $q\bar q$-pair 
and arbitrarily many gluons, as follows next.

\begin{figure}
\begin{minipage}[t]{71mm} \makebox[0mm]{}
 \epsfxsize=70mm\epsfbox{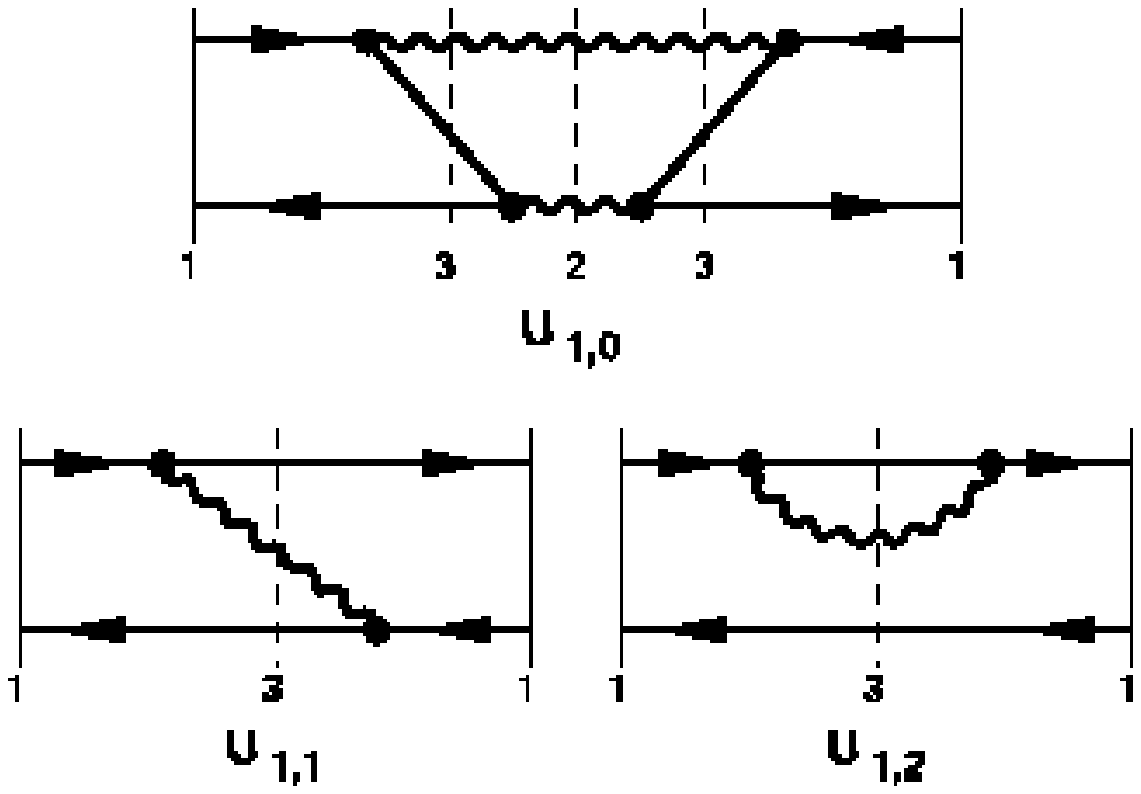}
 \caption{\label{fig:6_1} \sl
  The three graphs of the effective interaction in the 
  $q\bar q$-space. The vertical lines denote the 
  propagators $G_n$.
  On the right, the vertical lines denote the propagators
  $\overline G_n$. 
  The coupling function at the vertices is symbolized
  by graphs as they would appear in a perturbative analysis.
}\end{minipage}
\hfill
\begin{minipage} {76mm} \makebox[0mm]{}
 \epsfxsize=75mm\epsfbox{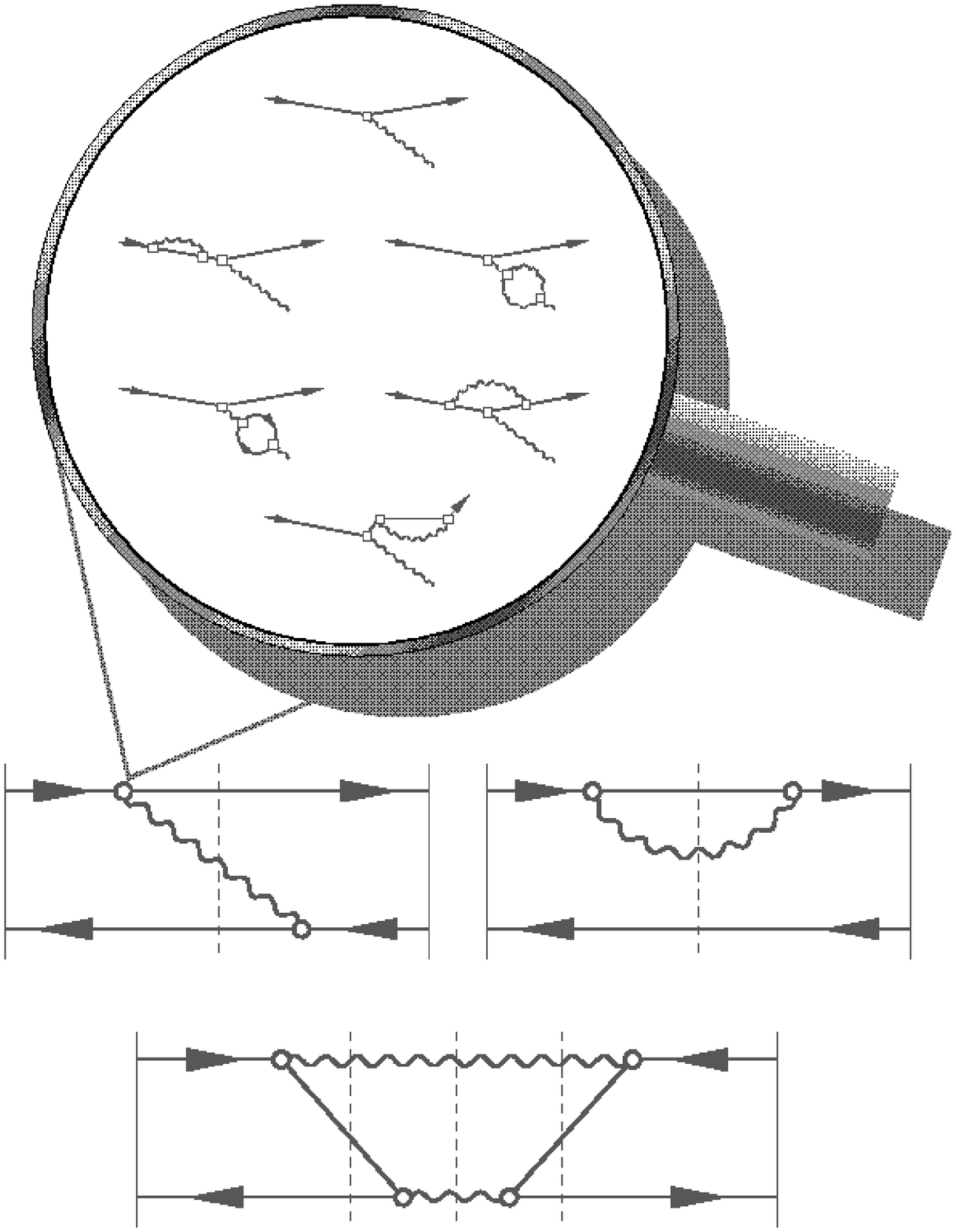} 
\end{minipage}
\end{figure}

\begin{figure}
\begin{minipage}[t]{72mm} \makebox[0mm]{}
 \epsfxsize=70mm\epsfbox{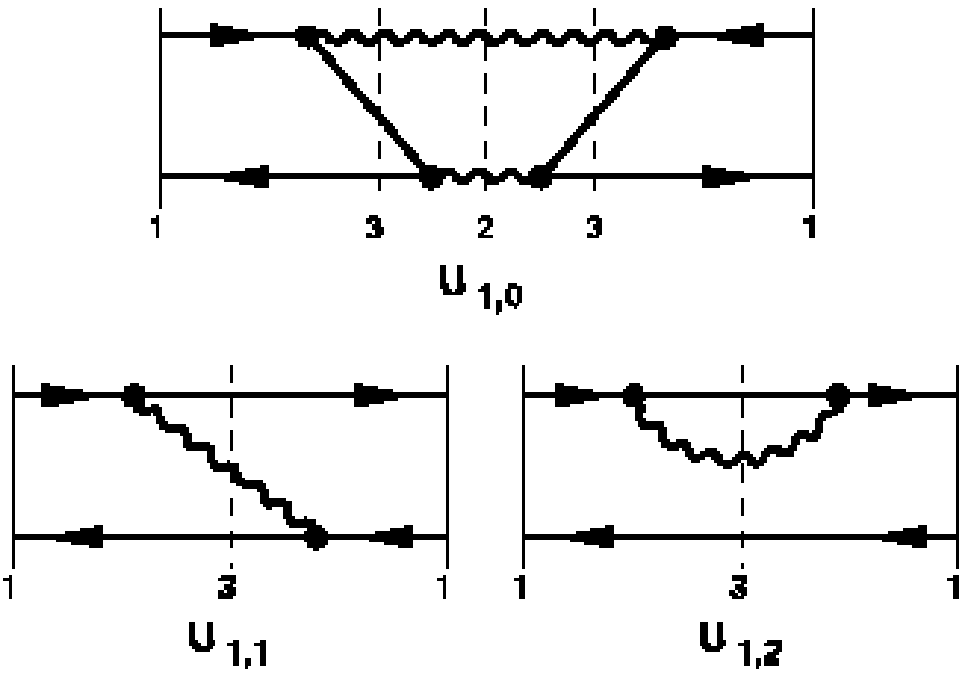}
\caption{ \label{fig:6_2} \sl
    The three graphs of the  spectator interaction 
    in the $q\bar q\,g$-space. 
    Note the role of the gluon  as a spec\-tator.
}\end{minipage}
\hfill
\begin{minipage}[t]{72mm} \makebox[0mm]{}
 \epsfxsize=70mm\epsfbox{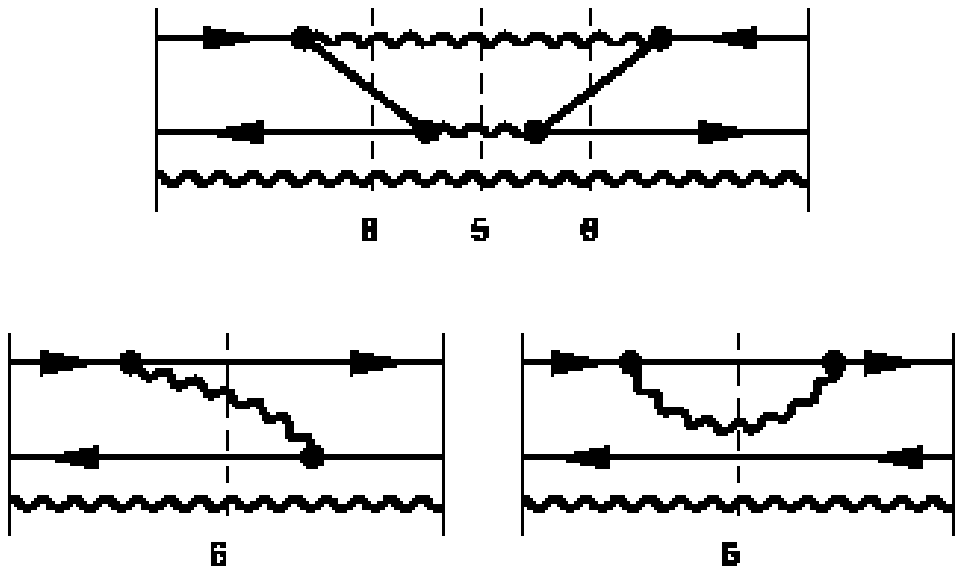}
 \caption{\label{fig:6_3} \sl
    Some six graphs of the  participant interaction 
    in the $q\bar q\,g$-space. 
}\end{minipage}
\end{figure}

Consider first the the effective interaction in the space of
one $q\bar q$-pair and one gluon as given 
by Eq.(\ref{aeq:620}).
The corresponding diagrams can be grouped into two 
topologically distinct classes, displayed 
in Figs.~\ref{fig:6_2}  and \ref{fig:6_3}.
The the diagrams in Fig.~\ref{fig:6_2} are obtained by
adding a non-interacting gluon line to the 
former diagrams in Fig.~\ref{fig:6_1}. 
The gluon does not change quantum numbers under impact
of the interaction and acts like a spectator.  Therefore, 
the graphs in Fig.~\ref{fig:6_2} will be refered to as the  
`spectator interaction' $\overline U _3$. 
In the graphs of Fig.~\ref{fig:6_3} the gluons are scattered
by the interaction, and correspondingly these graphs will be 
refered to as the `participant interaction' $\widetilde U _3$.  
Thus, $U_3 =  \overline U _3 + \widetilde U _3$.
The separation into a graph with only one interacting quark-pair
Fock-space sectors 
\begin {equation}  
   U_n =  \overline U _n + \widetilde U _n
,\label{aeq:626} \end {equation} 
except in those with only gluons.
More explicitly, the spectator and participant interactions in the lowest 
sectors with one quark-pair become  
\begin{eqnarray}
\begin{array} {lclclclclcl} 
     \overline  U_3 &=& V G _6 V  &+& V G_6 V  G_5 V G_6 V, &\quad&
     \widetilde U_3 &=& V G _4 V  &+& V G_6 V 
,\label{aeq:n50}\\  
     \overline  U_6 &=& V G _ {10} V &+& V G_{10} V  G_9 V G_{10} V , &\quad&
     \widetilde U_6 &=& V G _7     V &+& V G_{10} V
.\label{aeq:n51}
\end{array} 
\end {eqnarray}
The same operators can appear in both 
$\overline  U$ and $\widetilde U$, but they refer to different graphs. 
Since the Hamiltonian is additive in spectator and participant
interactions, $\overline U _n$ can be associated with 
its own resolvent 
\begin{eqnarray} 
   \overline G _n  =   {1\over \omega-T_n-\overline U_n }
,  \qquad \biggl( {\rm while\ } G _n\equiv{1\over \omega-H_n} 
     ={1\over \omega-T_n-\overline U_n-\widetilde U_n}\ \biggr)
.\label{aeq:641}\end{eqnarray} 
The relation of $\overline G _n$ to the full resolvent is 
$G_n=\overline G_n+\overline G_n\,\widetilde U_n\,G_n$, or 
\begin{equation} 
     G _n  = \overline G _n
     + \overline G _n \widetilde U _n\, \overline G _n 	
     + \overline G _n \widetilde U _n\, \overline G _n 
                      \widetilde U _n\, \overline G _n   + \dots 
,\label{eq:series}\end{equation}  
writing it as an infinite series. 
Note that the propagator $\overline G _n$ contains the 
interaction $\widetilde U _n$ in contrast to the 
free Tamm-Dancoff propagator $\widetilde G$ in Eq.(\ref{aeq:352}).
One deals here therefore with `perturbation theory  in medium'.
But since it is driven to all orders one better speaks of
`propagation in medium'.

The series can be identically written as
\begin{eqnarray} 
   G _n = C_n\,\overline G _n \,C_n^\dagger
, \qquad{\rm with}\quad
  C_n = \sqrt{{1\over 1 - \overline G _n \widetilde U _n} }
.\label{aeq:6520}\end{eqnarray}
One can verify this order by order in perturbation theory
or by the identity 
\begin{eqnarray} 
  (\omega - \overline H _n ) (1-\overline G _n \widetilde U _n ) =
  (\omega - \overline H _n ) 
  \left(1-\frac {1}{\omega - \overline H _n }\widetilde U _n \right) 
  = \omega - H _n 
.\label{aeq:6530}\end{eqnarray} 
The inverse gives $C_n^2 \,\overline G _n = G _n$, and with
the identity 
$C_n\,\overline G _n = \overline G _n\, C_n^\dagger$, 
one gets Eq.(\ref{aeq:6520}).
This remarkable property is peculiar to the method of
iterated resolvents.
Whenever a quark-pair-glue resolvent is sandwiched
in between two vertex interactions,
it allows to define a modified vertex
interaction $\overline V$, since
\begin  {equation} 
     V\,G _n \, V ^\dagger = 
     V\,C_n\,\overline G _n \,C_n^\dagger\, V ^\dagger
     = \overline VG_n\,\overline V ^\dagger     
,\ {\rm with }\quad
     \overline V = V C_n 
.\label{aeq:652}\end{equation} 
One can thus rewrite the sector Hamiltonians like for example
\begin {eqnarray} 
     \overline U_6 
&=&   \overline V \,\overline G _ {10} \overline V
                     + \overline V \,\overline G _{10} \overline V  
              G _9 \overline V \,\overline G _ {10} \overline V 
, \label {aeq:662} \\
     \overline U_3 
&=&   \overline V \,\overline G _6 \overline V 
                    + \overline V \,\overline G _6 \overline V  
              G _5\overline V \,\overline G _6 \overline V 
, \label {aeq:663} \\      
       U_1
&=&  \overline V \,\overline G _3 \overline V 
                    + \overline V \,\overline G _3 \overline V  
             G _2 \overline V \,\overline G _3 \overline V 
.\label{aeq:664} \end {eqnarray} 
Now they have all essentially the same structure, contrary
to Eqs.(\ref{aeq:610}) where they were different.
Note that $G_2$, $G_5$ and $G_9$ are pure gluon propagators.

Generally spoken, the rectangular block matrix $V$ 
is multiplied with the square matrix $C_n $. 
Below it will be shown that $C_n$ is approximately 
diagonal and independent of the spin.
Each vertex matrix element is multiplied 
with a simple function which depends 
on the momentum transfer $Q$ across the vertex. 
Equivalently one replaces the coupling constant $g$ 
by $\overline g= g C_n(Q)$, such that 
$C_n$ can be interpreted as a {\em coupling function}.

\subsection{The exact effective interaction}

The most important result of this section is that gauge theory 
particularly QCD has only two structurally different contributions
to the effective interaction in the $q\bar q$-space,
see Eq.(\ref{aeq:664}).
It scatters a quark with 
helicity $\lambda_q$ and four-momentum 
$p = (xP^+, x\vec P_{\!\perp} + \vec k_{\!\perp}, p^-)$
into a state with $\lambda_q^\prime$ and four-momentum 
${p^\prime} = (x^\prime P^+, x^\prime\vec P_{\!\perp} 
+ \vec k_{\!\perp}^\prime, {p^\prime}^-)$, 
and correspondingly the antiquark.
In the {\em continuum limit}, the resolvents are replaced by 
propagators and the matrix eigenvalue problem 
$H _{\rm eff}\vert \psi\rangle=M^2\vert \psi\rangle$ 
becomes an integral equation with a rather transparent structure:
\begin{eqnarray} 
    &&M_i^2\langle x,\vec k_{\!\perp}; \lambda_{q},
    \lambda_{\bar q}  \vert \psi _i\rangle 
    = 
    \left[ {\overline m^2_{q} + \vec k_{\!\perp}^2 \over x } +
    {\overline m^2_{\bar  q} + \vec k_{\!\perp}^2 \over 1-x }\right]
    \langle x,\vec k_{\!\perp}; \lambda_{q},
    \lambda_{\bar q}  \vert \psi _i\rangle 
\label{aeq:445}\\ &+&
    \sum _{ \lambda_q^\prime,\lambda_{\bar q}^\prime}
    \!\int\!dx^\prime d^2 \vec k_{\!\perp}^\prime\,
    \, R (x^\prime,k_{\perp}^\prime)     
    \,\langle x,\vec k_{\!\perp}; \lambda_{q}, \lambda_{\bar q}
    \vert U_{\rm OGE} 
    \vert x^\prime,\vec k_{\!\perp}^\prime; 
    \lambda_{q}^\prime, \lambda_{\bar q}^\prime\rangle\,
    \langle x^\prime,\vec k_{\!\perp}^\prime; 
    \lambda_{q}^\prime,\lambda_{\bar q}^\prime  
    \vert \psi _i\rangle  
\nonumber\\ &+&
    \sum _{ \lambda_q^\prime,\lambda_{\bar q}^\prime}
    \!\int\!dx^\prime d^2 \vec k_{\!\perp}^\prime\,
    \, R (x^\prime,k_{\perp}^\prime)     
    \,\langle x,\vec k_{\!\perp}; \lambda_{q}, \lambda_{\bar q}
    \vert U_{\rm TGA}  
    \vert x^\prime,\vec k_{\!\perp}^\prime; 
    \lambda_{q}^\prime, \lambda_{\bar q}^\prime\rangle\,
    \langle x^\prime,\vec k_{\!\perp}^\prime; 
    \lambda_{q}^\prime,\lambda_{\bar q}^\prime  
    \vert \psi _i\rangle 
.\nonumber\end {eqnarray}
The domain of integration is set by the cut-off function 
$ R $ given in Eq.(\ref{aeq:i1}). 
The effective one-gluon exchange
\begin{equation}
             U_{\rm OGE} = \overline V \overline G _3 \overline V
\end{equation}
conserves the flavor along the quark line. 
As illustrated in Fig.~\ref{fig:6_1} the vertex interaction $V$ 
creates a gluon and scatters the system virtually into the 
$q\bar q\,g$-space. 
As indicated in the figure by the vertical line with subscript `3', 
the three particles propagate there under impact of the full 
Hamiltonian before the gluon is  absorbed. 
The gluon can be absorbed either by the antiquark or by the 
quark. If it is absorbed by the quark, it contributes to the
effective quark mass $\overline  m$.  
The second term in Eq.(\ref{aeq:664}) is 
the effective two-gluon-annihilation interaction,
\begin{equation}
             U_{\rm TGA}  = \overline V \overline G _3 \overline V G _2 
	           \overline V \overline G _3 \overline V 
,\end{equation}
as represented by the graph 
$U_{1,0}$ in Fig.~\ref{fig:6_1}. 
The virtual  annihilation of the $q\bar q$-pair into two gluons 
can generate an interaction between different quark flavors.~--
The eigenvalue $M_i^2$ is one of the invaraint-mass squared 
eigenvalues of the full light-cone Hamiltonian, and  
the wavefunction 
$      \langle x,\vec k_{\!\perp}; \lambda_{q},
        \lambda_{\bar q}  \vert \psi _i\rangle$ 
gives the probability amplitudes for finding in the $q\bar q$-state
a flavored quark with momentum fraction $x$, intrinsic transverse 
momentum $\vec k_{\!\perp}$ and helicity $\lambda_{q}$,
and correspondingly an anti-quark with $1-x$, $-\vec k_{\!\perp}$
and $\lambda_{\bar q}$.  

\section {The breaking of the propagator hierarchy}

The content of the preceeding sections is exact but rather formal. 
In the sequel, rigor will be given up 
by the aim to obtain a solvable equation. 
In order to spot easier the essential approximations,
they will be dressed as `theorems', 
which can -- or cannot -- be proven later.

All sector Hamiltonians can be diagonalized 
on their own merit, for example
\begin{eqnarray} 
    M_{1;i} ^2
    \,\langle q;\bar q\vert\Psi_{1;i}\rangle &=&
    \sum _{q^\prime,\bar q^\prime}
    \langle q;\bar q\vert 
    H_1 
    \vert q^\prime;\bar q^\prime\rangle
    \,\langle q^\prime;\bar q^\prime
    \vert\Psi_{1;i} \rangle 
, \nonumber\label{aeq:6.60}\\
    M _{3;i} ^2 
    \,\langle q;\bar q;g\vert\Psi_{3;i}\rangle &=&
    \sum _{q^\prime,\bar q^\prime,g^\prime}
    \langle q;\bar q;g\vert
    \overline H_{3}
    \vert q^\prime;\bar q^\prime;g^\prime\rangle
    \,\langle q^\prime;\bar q^\prime;g^\prime
    \vert\Psi_{3;i}\rangle 
.\label{aeq:6.62}\end{eqnarray}
The spectra might be continuous, but the eigenvalues 
are denumerated for simplicity by $i$. 
By construction, one knows the relation between these two
sets of eigenvalues, by the following reason:
Because of the separation into spectators and participants
the gluon in the $q\bar q\,g$-equation is a non-interacting
particle. 
it moves freely relative to a $q\bar q$-bound state. 
If its four-momentum  is parametrized as 
$q^\mu=(yP^+,y\vec P_{\!\perp}+\vec q_{\!\perp},q_g^-)$ 
one has 
\begin  {equation}
    M_{3;i} ^2 
    = {M^2 + \vec q_{\!\perp}^{\,2} \over (1-y)} 
    + {\vec q_{\!\perp}^{\,2} \over y} 
.\end{equation}
Every $q\bar q$ bound state $M$ is a band-head in the $q\bar q\,g$
spectrum. With $\omega=M^2_{1,0}(\omega)\equiv M^2$
one gets
\begin{equation}
     \omega - M_{3;i} ^2 
     = M^2 - \left({M_{1;0}^2 + \vec q_{\!\perp}^{\,2} \over (1-y)} 
     + {\vec q_{\!\perp}^{\,2} \over y}\right)
     = -{y^2 M^2 + \vec q_{\!\perp}^{\,2} \over y(1-y)} 
.\label{aeq:n70}\end {equation}
Knowing the eigenvalues and the eigenfunctions of
the $q\bar q$ bound state one alos know the spectrum and the
eigenfunctions of $\overline H_3$.
Knowing the eigenfunctions, the exact resolvent can be calculated.
\begin{theorem}
   The exact propagators can be approximated by closure.
\end{theorem}
The substitution of the exact propagators by closure
is a widely used approximation in many-body
theory and in chemistry. 
In principle, the exact propagator is non-diagonal.
Performing closure, it becomes diagonal, which is
reflected by the Dirac-delta function 
$\langle q;\bar q;g\vert q^\prime;\bar q^\prime;g^\prime\rangle$
in the single particle momenta and helicities
\begin{eqnarray}
   \langle q;\bar q;g \vert \overline G _3
   \vert q ^\prime;\bar q^\prime;g ^\prime \rangle =
   \langle q;\bar q;g\vert q ^\prime;\bar q^\prime;g^\prime\rangle 
   \overline G _3 (q;\bar q;g)
.\label{aeq:6.69}\end{eqnarray} 
The Dirac-delta function is multiplied with the function 
\begin{eqnarray}
   \overline G _3 (q;\bar q;g) = 
   -{y(1-y)\over y^2 M ^2 +\vec q_{\!\perp}^{\,2}}
.\label{aeq:6.68}\end{eqnarray}
The in-medium propagator $\overline G _3$ ceases to be a 
functional of the higher resolvents. 
{\em The hierarchy of the iterated resolvents is broken}.

\begin{figure}
\begin{minipage}[t]{72mm} \makebox[0mm]{}
\centering
 \epsfxsize=50mm\epsfbox{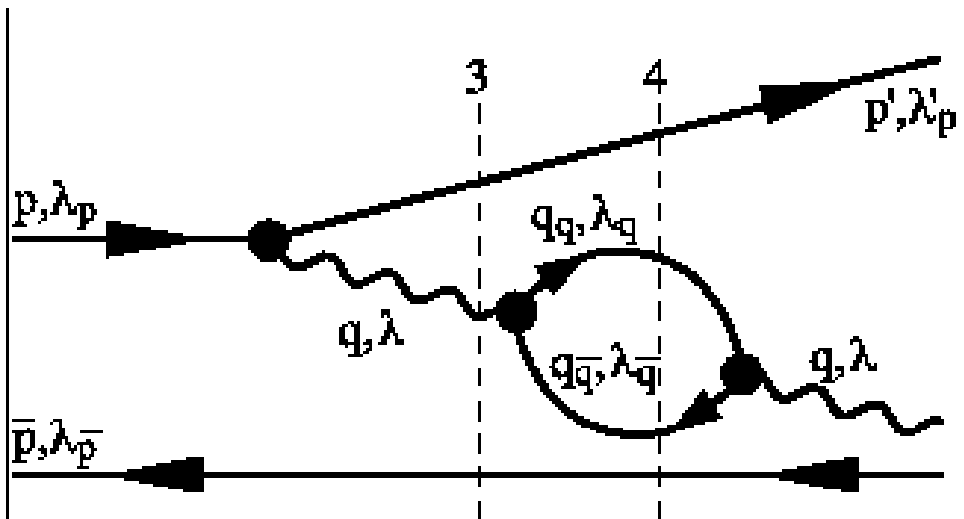}
 \caption{\label{fig:g_1} \sl 
    The $q\bar q$ vacuum polarization graph.
}\end{minipage}
\hfill
\begin{minipage}[t]{72mm} \makebox[0mm]{}
\centering
 \epsfxsize=50mm\epsfbox{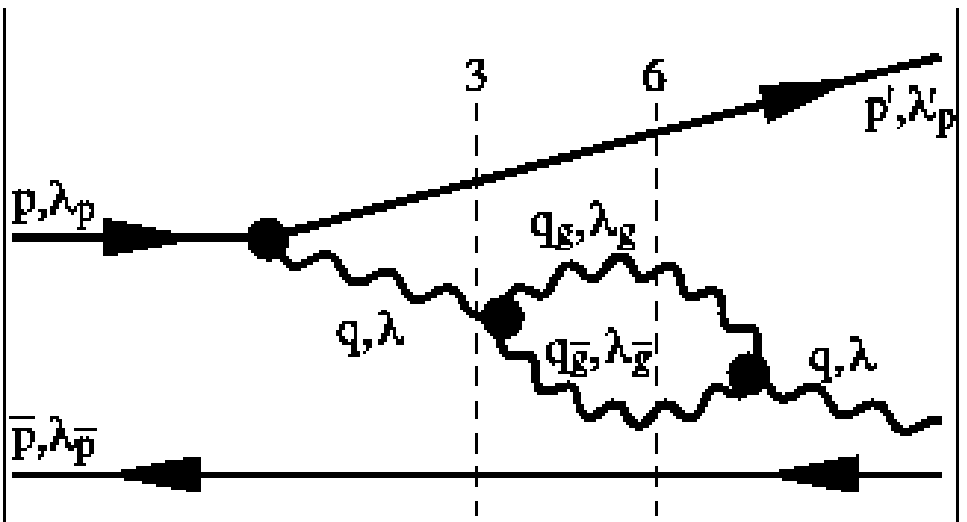}
\caption{\label{fig:g_2} \sl 
    The $gg$ vacuum polarization graph.
}\end{minipage}
\end{figure}

\begin{theorem}
In the solution, the interacting particles propagate free particles.
The in-medium propagators can be replaced by the free propagators.
\end{theorem}
The free propagator in the  $q\bar q g$-space
can be written 
\begin{eqnarray}
      G_{3,{\rm free}} &=& 
      {1\over P^+(p^--p^{\prime -} -q^-)} = 
      -{y\over Q^2} = -\frac {y} 
      {y^2(2\overline m)^2 +\vec q_{\!\perp}^{\,2}}
,\label{aeq:m72}\end{eqnarray}
where $Q^2=(p-p^\prime)^2$ is the four-momentum transfer 
along the quark line, in the notation of Fig.~\ref{fig:g_1}. 
For sufficiently small $y$ holds 
$(p-p^\prime)^2= -[y^2(2\overline m)^2 +\vec q_{\!\perp}^{\,2}]$.
If one substitutes $M\simeq 2\overline m$, which holds to 
rather good approximation, the momentum transfer in 
Eq.(\ref{aeq:m72}) is the same as in Eq.(\ref{aeq:6.68}). 

Similar considerations also hold all other 
propagators with at least one $q\bar q$-pair. 

Having replaced the in-medium-propagators $\overline G_n$
by the free propagators,
one can calculate  the graphs like in the 
usual (light-cone-) time-ordered perturbation theory.
The diagram $U_{1,2}$ in Fig.~\ref{fig:6_1} yields 
the effective quark mass 
\begin{equation}
      \overline m_f^2 =m_f^2 +m_f^2 {\alpha\over \pi}
      {n_c^2-1\over 2n_c}
      \ln{ {\Lambda^2\over m_{g}^2} }
.\label{aeq:i64}\end{equation}
Correspondingly one can calculate the coupling function 
\begin{equation}
      C_3 =  \sqrt{{1\over 1 - \overline G _3 \widetilde U _3}}
               =   1
               +  {1\over 2} \overline G _3 \widetilde U _3
               +  {3\over 8} \overline G _3 \widetilde U _3  \,
                           \overline G _3 \widetilde U _3 + \dots  
,\label{aeq:m74}\end{equation}
as defined by Eq.(\ref{aeq:6520}).
With $\widetilde U_3 = V G _4 V + V G_6 V $ from Eq.(\ref{aeq:n50}),
and thus $\widetilde U_3 = 
\overline V \overline G _4 \overline V + 
\overline V \overline G_6 \overline V $
from Eq.(\ref{aeq:652}),
the first two terms in the expansion are 
\begin{equation}
    \overline V 
    \simeq V + {1\over2}(
    V \overline G_3 \overline V\overline G_4 \overline V +
    V \overline G_3 \overline V\overline G_6 \overline V 
    ) 
    \simeq V + {1\over2} (
    V \overline G_3 V\overline G_4 V +
    V \overline G_3 V\overline G_6 V 
    ) 
.\label{aeq:m73}\end{equation}
In the last step $R_4=R_6=1$ was set. 
In general, they contribute terms of higher orders in $g$, 
which must be suppressed for consistency.
Figs.~\ref{fig:g_1} and \ref{fig:g_2} give only 2
representative graphs. 
According to the rules of time ordered perturbation theory,
there are about 22 different time-ordered and instantaneous diagrams.
All of them were re-calculated by Raufeisen \cite{rau97},
for all values of $Q$.
As shown to some detail in \cite{pau98} one gets for 
sufficiently large $\Lambda$
\begin{eqnarray}
      \overline G_3V(\overline G_6 + \overline G_4) V 
      &=& 
      {11\alpha n_c \over 12\pi} 
      \ln{\left({\Lambda^2\over 4m^2_g+Q^2}\right)}
      - {\alpha\over 6\pi}  \sum_f 
      \ln{\bigg({\Lambda^2\over 4m_f^2+Q^2}\bigg)}
,\nonumber\\ &=&
      \alpha\, b_0 \ln{(\Lambda^2/\kappa^2)} - b(Q)
,\quad b_0 = \frac {11n_c - 2 n_f} {12\pi}
,\end{eqnarray}
with an arbitrary $\kappa$ and a $b(Q)$
independent of $\Lambda$ given in Eq.(\ref{aeq:m96}).
One conculdes that the coupling function $C_3$ 
is identical with the familiar vertex correction.
In the limit where $Q$ is larger than all mass scales,
this result agrees with
the calculations of both, Thorn \cite{tho79} and Perry \cite{phz93},
having been done previously.

What happens if one substitutes the free propagators
in Eq.(\ref{aeq:m74}) by
the non-perturbative propagators
$\overline G_4= (\omega- \overline H_{q\bar q\,q\bar q})^{-1}$ and
$\overline G_6= (\omega- \overline H_{q\bar q\,gg})^{-1}$,
at least in an approximate fashion?~---
There are additional graphs.
In the fermion loop of vacuum polarization appear two 
graphs in addition to Fig.~\ref{fig:g_1}.
In one of them, a gluon is emitted and absorbed 
on the same quark line 
which changes the bare quark mass $m _f$
into the  physical quark mass $\overline m _f$.
In the other graph, the gluon is emitted from 
the quark and absorbed by the anti-quark which
represents an interaction. In consequence one has
a bound state with a physical mass scale $\mu_f$.
We replace therefore
$2m _f \Longrightarrow 2\overline m _f \Longrightarrow \mu _f$.
Similar considerations hold for the gluon loop
in Fig.~\ref{fig:g_2} and lead to the substitution
$2m _g \Longrightarrow 2\overline m _g \Longrightarrow \mu _g$.
Both $\mu_g$ and $\mu_f$ are interpreted as physical mass scales.
The physical gluon mass $\overline m _g$ vanishes of course 
due to gauge invariance. This is not in conflict
with {\it f.e.} Cornwall's suggestion of a finite effective 
gluon mass \cite{cor83} since one can define
$(\overline m _g)_{\rm eff}\equiv \mu_g/2$.

As a consequence of the approximation in Eq.(\ref{aeq:6.69}),
the propagators $\overline G_3$ and $\overline G_4$
are diagonal in Fock space, which in turn leads 
to diagonal products $\overline G_3V\overline G_4V$. 
They can be resumed therefore to all orders 
according to Eq.(\ref{aeq:m74}).
With $\overline \alpha \equiv {g^2} C_3^2 / {4\pi}$, one gets 
\begin{eqnarray}
   \overline \alpha(Q;\Lambda) &=& 
   \frac {1}{1/\alpha-b_0 \ln{(\Lambda^2/\kappa^2)} + b(Q)}     
      \quad{\rm with}
\nonumber\\
      b(Q) &=& 
      \frac {11n_c} {12\pi} 
      \ln{\Big(\frac {\kappa^2} {\mu_g^2+Q^2}\Big) }
      - {1\over 6\pi}  \sum_f 
      \ln{\Big(\frac {\kappa^2} {\mu_f^2+Q^2}\Big) }
.\label{aeq:m96}\end{eqnarray}
The effective fine structure constant
depends on the momentum transfer $Q$ across the vertex
and on the cut-off $\Lambda$.
Now, all the pieces are together which are needed for a further
discussion of the effective interaction as defined 
in Eq.(\ref{aeq:445}).
Finally, the the instantaneous interaction 
in the effective one-gluon-exchange interaction 
$U_{OGE}= \overline V \overline G_{3}\overline V $ 
is restored according to 
$\overline V \overline G_{3}\overline V \longrightarrow
 C_3^2(V \overline G_{3} V + S)$,
with the expression given in Eq.(\ref{eq:4s.15}).

\section{Status and discussion}

What have we reached?~-- 
The general expression for the effective Hamiltonian
in the $q\bar q$-sector was given in Eq.(\ref{aeq:445}).
If one restricts to consider mesons in which
the quark and the anti-quark have different flavors,
$f_{q}\neq f_{\bar q}$,
the two-gluon annihilation diagram can not get active 
and one remains with  
\begin{eqnarray} 
    M _n^2\langle x,\vec k_{\!\perp}; \lambda_{q},
    \lambda_{\bar q}  \vert \psi _n\rangle 
    = 
    \left[ {\overline m^2_{q}(\Lambda) + \vec k_{\!\perp}^2 \over x } +
    {\overline m^2_{\bar  q}(\Lambda) + \vec k_{\!\perp}^2 \over 1-x }
    \right]
    \langle x,\vec k_{\!\perp}; \lambda_{q},
    \lambda_{\bar q}  \vert \psi _n\rangle & &
\nonumber\\ +
    \sum _{ \lambda_q^\prime,\lambda_{\bar q}^\prime}
    \!\int\!dx^\prime d^2 \vec k_{\!\perp}^\prime\,
    \,\langle x,\vec k_{\!\perp}; \lambda_{q}, \lambda_{\bar q}
    \vert U_{\rm OGE} 
    \vert x^\prime,\vec k_{\!\perp}^\prime; 
    \lambda_{q}^\prime, \lambda_{\bar q}^\prime\rangle 
    \ \langle x^\prime,\vec k_{\!\perp}^\prime; 
    \lambda_{q}^\prime,\lambda_{\bar q}^\prime  
    \vert \psi _n\rangle &.& 
\label{new-aeq:445}\end {eqnarray}
The essential achievement of the two preceeding sections is
that the kernel of this integral equation can be written
down in the very explicit form
\begin{equation} 
    \langle
    \lambda_q,\lambda_{\bar q}\vert U_{\rm OGE} \vert 
    \lambda_q^\prime,\lambda_{\bar q}^\prime\rangle
    = - 
    {\overline\alpha(Q;\Lambda) \over 3\pi^2 Q  ^2} \,\langle
    \lambda_q,\lambda_{\bar q}\vert S\vert 
    \lambda_q^\prime,\lambda_{\bar q}^\prime\rangle
    \frac {R (x^\prime,k_{\perp}^\prime;\Lambda) }
    {\sqrt{ x(1-x) x'(1-x')} }
.\label{aeq:i65}\end{equation}
All many-body effects reside in the effective mass
$\overline m_{q}(\Lambda)$ and in the effective 
coupling constant $\overline \alpha (Q;\Lambda)$.
They are given in Eqs.(\ref{aeq:i64}) and (\ref{aeq:m96}), 
respectively.
The mean of the four-momentum transfers
of quark and anti-quark,
$Q^2 _{q} = -(k_q-k_q')^2 $ and 
$\overline Q^2 _{\bar q}= (k_{\bar q}-k_{\bar q}')^2$,
respectively, is denoted by $Q^2$. 
The spinor factor is 
\[
    \langle
    \lambda_{q},\lambda\vert S\vert 
    \lambda_{q}^\prime,\lambda^\prime\rangle =
    \left[ \overline u (k,\lambda)\gamma^\mu
    u(k^\prime,\lambda^\prime)\right]_q  \,
    \left[ \overline u (k,\lambda) 
    \gamma_\mu
    u(k^\prime,\lambda^\prime)\right] _{\bar q}
.\] 
The cut-off function $ R $ sets the domain of integration and
is defined in Eq.(\ref{aeq:i1}).
The approximations made have been carefully enumerated
in Section~7.

One has thus reached the goal of reducing the field theoretical 
many-body problem of $M ^2 = H _{\rm LC}$ to a one-body problem
with an effective interaction. For any value of the
Lagrangian coupling constant $g$, the flavor masses $m _f$,
and the cut-off $\Lambda$, one can generate 
the eigenvalues $ M_n^2$ on a computer, 
very much in analogy to QED \cite{kpw92,trp96}.
With the known eigenfunction in the $q\bar q$-sector
one an generate all higher Fock-space amplitudes,
according to the explicit formulas in Section~6.
It looks as if one has solved the problem.

But one has solved the problem only in a superficial
way, since the solutions depend on the unphysical
parameter $\Lambda$. One of the central issues in 
gauge-field theory is to remove this dependence
by a renormalization group analysis, see below.

The eigenvalues of the integral equation are
the invariant masses squared. Very often it is more
convenient to think in terms of an analogue
of a non-relativistic Hamiltonian $H_{S}$, 
\begin{equation}
   H_{\rm LC} = (\overline m _{\bar q}+\overline m _{q'})^2 +
       2(\overline m _{\bar q}+\overline m _{q'})\ H_{S}
,\label{peq:105}\end{equation}
whose eigenvalues $E=H_{S}$ are the more convential binding energies,
while the eigenfunctions are the same.
As shown in \cite{bpp97}, the so defined `Schr\"odinger operator'
has much in common with a non-relativistic Hamiltonian 
for the Coulomb problem.

\subsection{Renormalization group analysis}

All eigenvalues $M ^2_n$  depend thus
on the cut-off
$\Lambda$, the bare masses $m$ and the bare $\alpha$. 
In line with modern interpretation of quantum field theory 
\cite{wei95}, they are unphysical parameters and 
can be replaced by ($m _\Lambda,\alpha _\Lambda$),
by functions of $\Lambda$, such that the eigenvalues
are cut-off independent, {\it i.e.} 
\begin{equation}
    {d\over d\Lambda} M ^2_n (\Lambda,\alpha _\Lambda,m _\Lambda)= 0
.\label{ceq:8}\end{equation}
This fundamental equation of the renormalization group 
must hold for all eigenvalues.

The effective Hamiltonian $H=H_{\rm LC,eff}$ depends 
on $\Lambda$ only through $\overline m_{f}$, $\overline \alpha$ 
and the regulator function $R$. 
The renormalization group equations 
can be written therefore in the operator form
\begin{eqnarray}
       \delta \overline m_{f}\,
       {{\delta H \over \delta \overline m_{f}}}\vert \psi_n\rangle 
     + \delta \overline \alpha \, 
       {{\delta H \over \delta \overline \alpha}}\vert \psi_n\rangle 
     + \delta R\, {{\delta H \over \delta R}} \vert \psi_n\rangle 
     = 0 
.\label{ceq:9}\end{eqnarray}
The simultaneous variation of all three terms is
a very difficult problem.
But one can replace this equation by three independent ones,
\begin{eqnarray}
     \frac {d}{d\Lambda} \overline m_{f} &=& 0 
,\label{ceq:10}\\
     \frac {d}{d\Lambda} \overline \alpha &=& 0 
,\label{ceq:11}\\
     \frac {dR}{d\Lambda} 
     \frac {\delta H} {\delta R} \vert \psi_n\rangle &=& 0 
.\label{ceq:12}\end{eqnarray}
This is possible since one still solves Eq.(\ref{ceq:9}).

Eqs.(\ref{ceq:10}) and (\ref{ceq:11}) are two equations for
the two unknown functions $m_\Lambda$ and $\alpha_\Lambda$.
The first equation says that the effective flavor masses 
$\overline m_{f} = \overline m_{f} (\Lambda,\alpha_\Lambda,m_\Lambda)$
are renormalization group invariants. 
The numerical value of $\overline m_{f}$ must be fixed 
by experiment.~-- 
Eq.(\ref{ceq:11}) is then considered as an equation for
$\alpha_\Lambda$ at fixed values of $\overline m_{f}$.
Using the expression for $\overline \alpha (Q;\Lambda)$
given in Eq.(\ref{aeq:m96}) yields then \cite{pau98}
\begin{eqnarray}
   \alpha_\Lambda = \frac {1}{b_0 \ln{(\Lambda^2/\kappa^2)}} 
   ,\quad{\rm thus}\quad 
   \overline \alpha (Q) = \frac {1} {b(Q)}
,\end{eqnarray}
with $b(Q)$ given in Eq.(\ref{aeq:m96}).
All $\Lambda$-dependence cancels exactly in favor of the
renormalization group invariant $\kappa$,
which is sometimes called the QCD-scale $\Lambda_{\rm QCD}$. 
The scale $\kappa$ must be determind from experiment.~-- 
Once $\alpha_\Lambda$ is known, 
Eq.(\ref{ceq:10}) is an equation for $m_\Lambda$.
This function can be determined from Eq.(\ref{aeq:i64}),
but we renounce to do that here, since $m_\Lambda$
is needed nowhere in the present context.~-- 
Having fixed the functions $\alpha_\Lambda$ and $m_\Lambda$
one has exhausted all freedom provided by conventional
renormalization anaysis of field theory \cite{wei95}.
Since the cutoff-dependence was removed by renormalization, 
the cut-off can be driven to infinity, thus
\[
   R(x, \vec k _{\!\perp};\Lambda) =1
.\]
Formally, this solves Eq.(\ref{ceq:12}).

\subsection{The remaining divergence}

After renormalization, the kernel of the one-body equation
is independent of $\Lambda$. 
Let us discuss its relevant part
\[
     {\cal K} = \frac {1} {b(Q)}
     \ \frac {S} {Q^2}
.\]
At very small momentum transfers $Q ^2 \rightarrow 0$,
where $x\sim x'$ and $\vec k _{\!\perp} \sim \vec k _{\!\perp} '$,
the effective coupling constant locks into the finite
value $b(0)$, see Eq.(\ref{aeq:m96}).
The spinor function also goes to a constant,
$ S \sim 4 \overline m _{q} \overline m _{\bar  q}$,
see \cite{trp96} or \cite{bpp97}.
The remaining `Coulomb singularity' $Q ^{-2}$ is square-integrable
and can be dealt with by convenient numerical means \cite{kpw92,trp96}.
For very large momentum transfers $Q ^2 \rightarrow \infty$,
the behaviour is quite different. 
Both $S$ and $Q ^2$ tend to diverge, $S \propto (\vec k _{\!\perp} ') ^2$
and $Q ^2 \propto (\vec k _{\!\perp} ') ^2$,
but such that the ratio $S/Q^2$ tends to a {\em finite constant}.
Disregarding the very slow logarythmic increase of $b(Q)$,
the kernel of the integral equation is therefore essentially
a dimensionless constant
\[ 
   {\cal K} \sim constans
   ,\qquad{\rm for}\quad 
   Q ^2 \gg 0
.\]
One has sufficient evidence \cite{trp96} that
this behaviour creates all kinds of problems,
among them a diverging eigenvalue.

One can separate the kernel of the integral
equation identically into two pieces
${\cal K} = {\cal K}_1 + {\cal K}_2$, {\it i.e.}
\[
     {\cal K} = \frac {1} {b(Q)} \frac {S} {Q^2}
     \ \frac {\mu^2+Q^2} {\mu^2+Q^2} =
     \frac {1} {b(Q)} \frac {S} {Q^2 (1+ Q^2/\mu^2)} +
     \frac {S} {b(Q)(\mu^2+Q^2)} 
,\]
with an arbitrary mass parameter $\mu$.
The first part ${\cal K}_1$ is well behaved and vanishes at least like
$Q^{-2}$ in the limit of large $Q$.
The second part ${\cal K}_2$ is a strange object:
it is a constant almost everywhere, in particular
\[
   {\cal K}_2\sim constans,
   \qquad{\rm since}\quad
   \frac {S}{Q^2} \sim constans
   ,\qquad{\rm for}\quad 
   Q ^2 \gg 0
.\]
One should emphasize the following aspect.
The typical field theoretical divergences like the divergence of the
effective coupling constant lead to a {\em divergence} of the kernel,
but adding a {\em finite constant} to the kernel leads also to
a divergence.
Thus far one does not understand the reason for the latter.

\begin{table}[t]
\begin{minipage}[t]{72mm} \makebox[0mm]{}
\begin{tabular}{c|rrrrrr} 
   \rule[-1em]{0mm}{1em}
     & $\overline u$ & $\overline d$ 
     & $\overline s$ & $\overline c$ & $\overline b$ \\ \hline
   \rule[1em]{0mm}{0.5em}
 $u$ &      & 768  & 892  & 2007 & 5325 \\ 
 $d$ & 140  &      & 896  & 2010 & 5325 \\ 
 $s$ & 494  & 498  &      & 2110 &  --- \\ 
 $c$ & 1865 & 1869 & 1969 &      &  --- \\ 
 $b$ & 5278 & 5279 & 5375 &  --- &      \\ 
\end{tabular}
\end{minipage}
\hfill
\begin{minipage}[t]{72mm} \makebox[0mm]{}
\begin{tabular}{c|cccccc} 
   \rule[-1em]{0mm}{1em}
     & $\overline u$ & $\overline d$ 
     & $\overline s$ & $\overline c$ & $\overline b$ \\ \hline
   \rule[1em]{0mm}{0.5em}
 $u$ &       &$\rho^+$&$K^{*+} $&$\overline D^{*0}$&$B^{*+}$\\ 
 $d$ &$\pi^-$&        &$K^{*0}$&$D^{*-}$&$ B^{*0}$\\ 
 $s$ &$K^-$  &$\overline K ^0$&      &$D_s^{*-}$&$B_s^{*0}$\\ 
 $c$ &$D^0$  &$D^+$&$D_s^+$&      &$B_c^{*+}$\\ 
 $b$ &$B^-$  &$\overline B ^0$&$\overline B_s ^0$&$B_c^{-}$&      \\ 
\end{tabular}
\end{minipage}
\caption{ \sl
   The empirical masses of the flavor-off-diagonal physical mesons in MeV.
   Vector mesons are given in the upper, 
   pseudo-scalar mesons in the lower triangle.
   Their physical nomenclature is given on the right.
}\label{tab:meson-masses}
\end{table}
\begin{table}[t]
\begin{minipage}[t]{72mm} \makebox[0mm]{}
\begin{tabular}{c|rrrrrr} 
   \rule[-1em]{0mm}{1em}
     & $\overline u$ & $\overline d$ 
     & $\overline s$ & $\overline c$ & $\overline b$ \\ \hline
   \rule[1em]{0mm}{0.5em}
 $u$ &          & $^*$768 &  902 & 2012 & 5331 \\ 
 $d$ & $^*$140  &         &  902 & 2012 & 5331 \\ 
 $s$ & $^*$494  &     494 &      & 2155 & 5476 \\ 
 $c$ & $^*$1865 &    1865 & 2108 &      & 6617 \\ 
 $b$ & $^*$5278 &    5278 & 5423 & 6573 &       
\end{tabular}
\end{minipage}
\hfill
\begin{minipage}[t]{72mm} \makebox[0mm]{}
\begin{tabular}{c|rrrrrr} 
   \rule[-1em]{0mm}{1em}
     & $\overline u$ & $\overline d$ 
     & $\overline s$ & $\overline c$ & $\overline b$ \\ \hline
   \rule[1em]{0mm}{0.5em}
 $u$ &          & $^*$768 & 1002 & 2301 & 5696 \\ 
 $d$ & $^*$140  &         & 1002 & 2301 & 5696 \\ 
 $s$ & $^*$494  &     494 &      & 2535 & 5829 \\ 
 $c$ & $^*$1865 &    1865 & 2102 &      & 7227 \\ 
 $b$ & $^*$5278 &    5278 & 5512 & 6811 &      \\
\end{tabular}
\end{minipage}
\caption{ \sl
   The calculated masses of the flavor-off-diagonal mesons in MeV, 
   as obtained from a fit to Eq.(\protect{\ref{peq:104}}) on the
   left and to Eq.(\protect{\ref{peq:103}}) on the right.
   Vector mesons are given in the upper, pseudo-scalar mesons
   in the lower triangle.
   The $^*$upperscript marks mesons where the shift and mass 
   parameters have be fitted to. 
}\label{tab:mesons-mass-calc}
\end{table}
Perhaps one can get further insight by studying the 
following eigenvalue in (usual) three-momentum space ($\vec k$)
\begin{equation}
   E \psi(\vec k) = \frac {\vec k^{\,2}}{2m} \psi(\vec k) -
   \frac {\alpha} {\pi^2} \int \!d^3k'
   \left[\frac {1}{(\vec k - \vec k')^2} + \frac {1}{\mu^2}\right]
   \psi(\vec k ') 
.\label{peq:106}\end{equation}
In the spirit of Eq.(\ref{peq:105}),
the right-hand side is a conventional Schr\"odinger
equation for the Coulomb problem with the
reduced mass $1/m = 1/m_{q} + 1/m_{\bar q}$, but 
where an arbitrary constant has been added in the kernel.
In Heidelberg, we are presently working on the question
whether its eigenvalue $E$ diverges.
Preliminary studies with the Fourier transform of 
Eq.(\ref{peq:106}), where the constant works its way into a
weird Dirac-delta function $\delta^{(3)}(\vec x)$ as a potential, 
\[
   E \phi(\vec x) = -\frac {\vec \nabla^{\,2}}{2m} \phi(\vec x) -
   \left[\frac {\alpha}{|\vec x|} + 
   \frac {\alpha}{\mu^2}\delta^{(3)}(\vec x)\right] \phi(\vec x ') 
,\]
indicate indeed that the eigenvalue of the $1s$-state
in particular diverges.

If these preliminary results materialize,
one must regularize and subsequently renormalize Eq.(\ref{peq:106}).
The energy of the $1s$-state will then be essentially
a renormalization constant $E_{1s}= \overline s$.
Given that to be true, all of us have been riding 
the wrong horse.
It is not the regular part of the interaction kernel ${\cal K}_1$,
but it is the `constant' ${\cal K}_2$ which determines the 
mass in the first place.
Since ${\cal K}_2$ involves a renormalization
and since the singlet can have an energy shift ($\overline s_-$)
different from the triplet ($\overline s_+$),
one has at most two additional renormalization constants, 
subject to be determined by experiment.

\subsection{A mass formula}

Is there some physical evidence for the above considerations?~--
Inspired by Eq.(\ref{peq:105}), one can seek the masses
of the physical mesons in the form \cite{Pau99}
\begin{equation}
   M^2= (\overline m _{\bar q}+\overline m _{q'})^2 +
       2(\overline m _{\bar q}+\overline m _{q'})\ \overline s_\pm
,\label{peq:103}\end{equation}
and fit the energy shifts $\overline s_\pm$ and the physical quark 
masses $\overline m _q$ to the physical meson masses,
compiled in Table~\ref{tab:meson-masses} 
from the data of the particle data group \cite{PDG98} which
do not include yet the topped mesons. 

The following procedure was applied.
First, the up and the down mass were chosen equal. 
Then the empirical masses of $\pi^+$ and $\rho^+$ were used 
to determine the mass shifts for the singlet and the triplet.
The remaining quark masses are obtained  
from the pseudo-scalar mesons with an up quark,
which exhausts all freedom in determining physical parameters. 
The remaining 13 off-diagonal pseudo-scalar and 
vector meson masses are calculated straightforwardly 
and compiled in Table~\ref{tab:mesons-mass-calc}.
In comparing the so obtained numbers with the experiment,
one notes with great surprise that the discrepancy
exceeds only rarely an estimated error of 10\%.
The resulting quark masses are given in Table~\ref{tab:quarks}. 

Actually, there is no reason why one should stick to 
Eq.(\ref{peq:103}). 
Equivalently, one can replace $\overline s _\pm$ by 
$\overline s _\pm \,a/(\overline m _{\bar q}+\overline m _{q'})$.
Choosing $a= \overline m _{\bar u}+\overline m _{d} = 700$~MeV,
one gets
\begin{equation}
   M^2= (\overline m _{\bar q}+\overline m _{q'})^2 +
       2(\overline m _{\bar u}+\overline m _{d})\ \overline s_\pm
.\label{peq:104}\end{equation}
A form like this was suggested to me by Dae Sung Hwang
during the lectures.
Redoing the fit for the heavy quarks gives the results
compiled in Table~\ref{tab:mesons-mass-calc}.
The discrepancy with the experimental numbers is now 
on the level of 1\%.
This excellent agreement should not be over-emphasized. 
In any case it does not falsify the above considerations.
\begin{table}[t]
\begin{center}
\begin{tabular}{c||rrrrr||rr} 
   \rule[-1em]{0mm}{1em}
   method & 
   $\overline m_u$ &
   $\overline m_d$ &
   $\overline m_s$ &
   $\overline m_c$ &
   $\overline m_b$ &
   $\overline s_-$ &
   $\overline s_+$ \\ \hline
   \rule[1em]{0mm}{0.5em}
   Eq.(\protect{\ref{peq:103}}) & 
   350 & 350 & 583 & 1881 & 5275 & -336 & 71 \\
   \rule[1em]{0mm}{0.5em}
   Eq.(\protect{\ref{peq:104}}) & 
   350 & 350 & 495 & 1637 & 4972 & -336 & 71
\end{tabular}
\end{center}
\caption{ \sl
   The quark mass parameters $\overline m_q$ and 
   the energy shifts $\overline s_\pm$, 
   all in MeV, as obtained from a fit to 
   the physical meson masses.
}\label{tab:quarks}
\end{table}

\section{A short summary}

In these lecture notes the attempt was made to phrase and discuss
a relativistic quantum theory such as QCD from the point of view of
Dirac's front form of hamiltonian dynamics.
Sometimes refered to as light-cone quantization, it has
the goal to describe and understand the bound-state structure
of hadrons in terms of their constituents from a covariant theory.
As the lectures show, this goal has not been reached yet.

As reviewed in \cite{bpp97}, there might be alternative roads
to reach the same goal, like for example the Bethe-Salpeter equations,
lattice gauge theory, Hamiltonian flow equations, just to name a few.
Each of these approaches have their own inherent adventages or
disadventages. 
By reasons of space their discussion is left out in these notes.
By the same reason the light-front renormalizations approach of Wilson
and collaborators \cite{wwh94,glw98} is omitted. 
The many articles have thus far not provided conclusive evidence 
for success, apart from the fact that virtually every Langrangian 
symmetry has been violated in the approach. 
It is much too early to draw definitive conclusions.

The front form is however useful to study the structure
of arbitrary field theories. The theories considered are often 
not physical, but are selected to help in the understanding 
of a particular non-perturbative phenomenon.
The relatively simple vacuum properties  of light-front field theories
underly many of these `analytical' approaches.
The relative simplicity of the light-cone vacuum provides a firm 
starting point to attack many non perturbative issues. 
As we have seen, the problems in two dimensions,
in one space and one time dimension,
not only are they tractable from the outset
but in many cases like the Schwinger model can they be solved
numerically by `discretized light-cone quantization (DLCQ)'
to almost arbitrary accuracy.
This solution gives a unique insight and understanding. 
The Schwinger particle indeed has the simple parton
structure that one hopes to see in QCD.

Unfortunately the same cannot be said for the physical world in
3 space and 1 time dimensions.
The essential problem is that the number of degrees of freedom
needed to specify each Fock state even in a discrete basis 
grows much too quickly.
As discussed in this review, the basic
procedure is to diagonalize the full light-cone Hamiltonian 
in the free light-cone Hamiltonian basis.  
The eigenvalues are the invariant mass squared of the discrete 
and continuum eigenstates of the spectrum.  
The  projection of the eigenstate on the free Fock basis 
are the light-cone wavefunctions and provide a rigorous 
relativistic many-body representation in terms of its particle 
degrees of freedom. 
Given the light-cone wavefunction one can compute the structure
functions and distribution amplitudes.  
More generally, the light-cone wavefunctions provide the 
interpolation between hadron scattering amplitudes and 
the underlying parton subprocesses. 
The method of iterated resolvents can be a useful intermediary
step to generate these wavefunctions.
The unique property of light-cone quantization that makes the  
calculations of light cone wavefunctions particularly useful 
is that they are independent of the reference frame and 
that the same wavefunction can be use in many different problems. 

Finally, let us highlight the intrinsic advantages  of light-cone field theory:
\begin{itemize}
\item
The light-cone wavefunctions are independent of the
momentum of the bound state --
only relative momentum coordinates appear.
\item
The vacuum state is simple and in many cases trivial.
\item
Fermions and fermion derivatives
are treated exactly; there is no fermion doubling problem.
\item
The minimum number of physical degrees of freedom are used because of
the light-cone gauge. No Gupta-Bleuler or Faddeev-Popov
ghosts occur and unitarity is explicit.
\item
The output is the full color-singlet spectrum of the theory,
both bound states and continuum, together with their respective
wavefunctions.
\end{itemize}

\end{document}